\documentclass[aps,pre,amsmath,amssymb,superscriptaddress,showpacs,showkeys]{revtex4-2}

\usepackage{graphicx}
\usepackage{dcolumn}
\usepackage{bm}
\usepackage{CJKutf8}
\usepackage{xcolor}
\usepackage{enumitem}
\usepackage{stmaryrd}

\newcommand{\tensorarrow}{ \overset{ \lower0.5em\hbox{$\smash{\scriptscriptstyle\leftrightarrow}$}}}

\newcommand{\bbNabla}{%
  \nabla\mkern-12mu\nabla%
}
\newcommand{\bbDelta}{%
  \Delta\mkern-12mu\Delta%
}

\begin{document}

\title{Onsager-variational formulation of diffuse-domain methods for \\ 
computational modeling of microscale fluid-structure interactions} 



\author{Xinpeng Xu (\begin{CJK*}{UTF8}{gbsn}徐新鹏\end{CJK*})}\thanks{Corresponding author}
\email{xu.xinpeng@gtiit.edu.cn}
\affiliation{Department of Physics and MATEC Key Lab, Guangdong Technion --- Israel Institute of Technology, 241 Daxue Road, Shantou, Guangdong 515063, China}
\affiliation{Technion --- Israel Institute of Technology, Haifa 3200003, Israel}


\date{\today}

\begin{abstract}
Direct numerical simulation (DNS) of microscale fluid--structure interactions (mFSIs) in multicomponent and multiphase flow systems requires numerical methods that can represent moving boundaries together with fields constrained to evolving interfaces and surfaces. Diffuse-domain methods (DDMs) address this geometric
difficulty by replacing sharp surfaces with diffuse volumetric representations on regular computational domains. In this work, we formulate DDMs within Onsager's variational principle (OVP). Rather than extending sharp-interface governing equations and boundary conditions term by term, we embed sharp-surface free-energy and dissipation functionals into the bulk through a diffuse surface delta density and derive the corresponding diffuse equations from the Rayleighian. The resulting framework distinguishes balance-law fields, internal nonconserved order parameters, and kinematic or constitutive rate variables. It also makes explicit a moving-surface kinematic distinction: conserved surface densities are transported by the full material surface velocity, whereas explicitly tangential vector and tensor internal nonconserved variables require projected objective/co-rotational rates in their admissible tangential state spaces. For scalar transport on rigid and deformable interfaces, and for interfacial hydrodynamics near rigid walls, the formulation recovers established DDM models with known sharp-interface limits. The same variational construction provides thermodynamically consistent finite-$\epsilon$ diffuse-domain models for multicomponent deformable vesicles with surface viscosity, tangential slip, and finite areal compressibility, and for active shells carrying chemical and tangential vector order. These results give a unified framework for constructing thermodynamically consistent passive diffuse-domain models of interfacial and surface dynamics, and for incorporating active stresses through active work power. They also identify the geometric, state-space, and asymptotic issues that remain in fully coupled vector- and tensor-valued surface theories. The framework is relevant to soft condensed matter, microfluidic interfaces, biological membranes, and morphogenetic surface
dynamics. 
\end{abstract}


\maketitle


\section{Introduction} \label{sec:introduction}  
\begin{figure}[htbp]
\centering
\includegraphics[width=0.8\columnwidth]{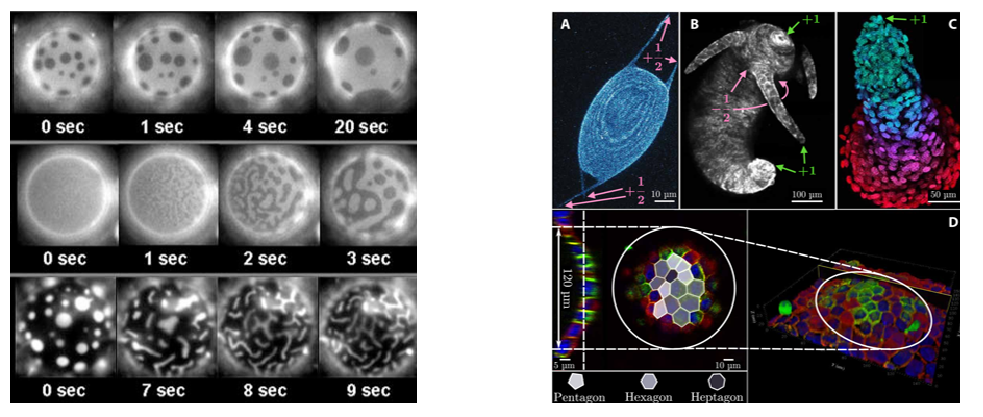}
\caption{Experimental motivation for modeling complex dynamics on curved surfaces. Left: liquid--liquid phase separation on a vesicle membrane, illustrating curvature-mediated interfacial dynamics such as domain coarsening, spinodal decomposition, and viscous fingering (adapted from Ref.~\cite{VeatchKeller2003}). Right: active polar or nematic dynamics and topological defects on deformable epithelial surfaces, illustrating defect-mediated morphogenesis in systems including microtubule--kinesin assemblies encapsulated in lipid vesicles, regenerating \textit{Hydra}, collectively migrating myoblast monolayers, and cultured MDCK GII epithelial sheets (adapted from Ref.~\cite{hoffmann2022theory}).}
\label{Fig:Experiment}
\end{figure}

Microscale fluid–structure interactions (mFSIs) in multicomponent and multiphase flow systems occur whenever the bulk flow is coupled either to (moving) solid boundaries or to physical processes that are kinematically or dynamically constrained to interfaces and surfaces. At characteristic length scales on the order of $1$–$100\,\mu\mathrm{m}$, two broad classes of mFSIs problems may be distinguished: (i) dynamics occurring \emph{in the vicinity of} interfaces or surfaces, where fluids interact with solid structures that may translate, deform, or undergo morphological evolution; and (ii) dynamics \emph{intrinsically confined to} interfaces or surfaces, in which the dominant physical phenomena are localized on two-dimensional manifolds embedded in three-dimensional space, such as lipid bilayer membranes, vesicular interfaces, and active biological shells [Fig.~\ref{Fig:Experiment}]~\cite{Kirby2010Book,Helfrich1973,ArroyoDesimone2009PRE,Dziuk2013}. At these scales, the large surface-area-to-volume ratio makes interfacial effects dominant. Interfacial and line tension, wettability, slip, curvature elasticity, surface viscosity, and surface transport can all couple strongly to the surrounding bulk hydrodynamics. Continuum descriptions therefore often combine bulk hydrodynamics, Helfrich-type curvature elasticity, surface hydrodynamics, and surface phase transition/separation dynamics~\cite{Helfrich1973,ArroyoDesimone2009PRE,Dziuk2013}. 
Analytical progress is limited except in special geometries and weak-coupling
limits, so direct numerical simulation (DNS) is often needed to test continuum theories and explore regimes in which geometry, transport, and hydrodynamics are strongly coupled. The principal difficulty is geometric: the relevant interfaces are often highly curved, time dependent, or topologically nontrivial. This has motivated both discretizations of surface partial differential equations and diffuse or implicit surface representations~\cite{Dziuk2013,RatzVoigt2006,ElliottStinner2009,LiLowengrubRatzVoigt2009}. Recent experiments and continuum theories---from phase separation on curved vesicles to active orientational dynamics on epithelial shells---further underscore the need for numerical frameworks that treat bulk flow, interfacial physics, and geometry in a unified way~\cite{Keber2014,hoffmann2022theory,wang2023patterning,Salbreux2017,Khoromskaia2023}.

From a computational perspective, DNS strategies for mFSI fall broadly into two classes~\cite{Hou2012}. (i) \emph{Partitioned methods} solve the fluid, solid, and interfacial subproblems on separate computational domains and couple them through prescribed interfacial conditions by resolving and tracking interfaces explicitly. Representative examples include immersed-boundary methods~\cite{Griffith2020IBM}, arbitrary Lagrangian--Eulerian formulations~\cite{Feng2009ALE}, and particle-based approaches such as smoothed particle hydrodynamics and dissipative particle dynamics~\cite{Shadloo2016SPH,Hoogerbrugge1992DPD}. Their main advantage is modularity, since existing solvers can often be reused. Their main limitation is geometric complexity: explicit interface tracking or carefully designed particle interactions becomes increasingly difficult in multiphase flows with evolving contact lines, rich surface physics, or interfaces that deform, merge, and reorganize dynamically. (ii) \emph{Monolithic methods}, by contrast, embed fluids, solids, and interfaces into a single regular computational domain and impose interfacial effects implicitly through auxiliary fields. Examples include level-set and phase-field formulations~\cite{Jacqmin2000,Mokbel2018,hong2021hybrid}, fictitious-domain methods~\cite{Glowinski2001FDM}, smoothed-profile or fluid-particle-dynamics approaches~\cite{Yamamoto2005SPM,Yamamoto2021SPM,tanaka2000simulation}, and diffuse-domain methods (DDMs)~\cite{Levine2003,Levine2005a,Levine2005b,Levine2010,Levine2013,RatzVoigt2006,LiLowengrubRatzVoigt2009}. Among these approaches, DDM is particularly appealing, as it is applicable to both classes of mFSI problems---dynamics near interfaces or surfaces as well as those occurring directly on them---while operating on fixed, regular computational domains and representing complex interfacial geometries through continuous phase fields.

In DDM, integrals over sharp surfaces are replaced by bulk integrals weighted by a diffuse surface delta density $\delta_\epsilon $. This idea originates in diffuse-interface approximations of surface measures and in numerical methods designed to avoid explicit interface tracking. In soft-matter and biological applications, this approach provides a natural framework for coupling membranes, interfaces, and surface order parameters to surrounding bulk fields by solving a unified set of partial differential equations defined over a fixed, regular computational domain. Early biological phase-field applications used such ideas in models of cell motility and intracellular dynamics~\cite{Levine2003, Levine2005a,Levine2005b,Levine2010,Levine2013}. The mathematical and numerical foundations of DDM were developed subsequently by R\"atz, Voigt, Li, Lowengrub, Guo, and collaborators, who clarified scalar surface transport, moving-surface advection--diffusion, diffuse imposition of boundary conditions, and the associated sharp-interface limits~\cite{RatzVoigt2006,ElliottStinner2009,DeckelnickStyles2018,LiLowengrubRatzVoigt2009,Lervag2014AnalysisDDM}. Since then, DDMs and closely related formulations have been applied to multicomponent vesicles and deformable membranes~\cite{WangDu2008MultiComponentVesicles,LowengrubRatzVoigt2009Vesicles,AlandEgererLowengrubVoigt2014,WenValizadehRabczukZhuang2024}, transport and adsorption on deformable interfaces~\cite{TeigenLiLowengrubWangVoigt2009DeformableInterface}, two-phase flows in complex geometries~\cite{AlandLowengrubVoigt2010}, vector-valued PDEs on surfaces~\cite{NestlerVoigt2024}, elastic and viscoelastic surfaces in fluids~\cite{KloppeAland2024}, and biological growth on evolving domains~\cite{ChenLowengrub2015TumorDiffuse}. More recently, we introduced the diffuse-resistance-domain (DRD) method, a DDM variant in which smoothly varying resistance coefficients enforce effective boundary conditions such as impermeability, slip, and dynamic contact-angle behavior across diffuse layers~\cite{GaoLiXuDRD2025}. These developments illustrate the breadth of the diffuse-domain paradigm, but they also expose a persistent limitation: many models are still constructed case by case, with free energies, dissipation mechanisms, and localized interfacial conditions combined in ways that are physically plausible but not systematically derived.

This lack of a unified thermodynamic construction becomes particularly restrictive in coupled problems involving surface viscous flow, chemical transport, tangential slip, elastic deformation, and active orientational dynamics. Three related difficulties recur. First, moving-surface kinematics are subtle in diffuse representations: a conserved surface density is tied to the evolving surface measure, whereas a tangential internal variable lives in the tangent bundle and should evolve through an objective rate. Second, many diffuse-domain models are assembled directly at the level of partial differential equations and boundary conditions, so thermodynamic consistency of coupled bulk/surface systems must be checked \emph{a posteriori} rather than built in from the outset. Third, although sharp-interface asymptotics are understood for several canonical DDM settings, much less is known for coupled problems involving surface hydrodynamics, tangential slip, finite areal compressibility, and active vector or tensor fields on deformable surfaces~\cite{KloppeAland2024,Salbreux2017,NitschkeVoigt2024ActiveNematodynamics}. At the numerical level, the interplay between diffuse-layer thickness, spatial resolution, and control of bulk extensions---especially for projected tangential fields---also remains poorly characterized. A central need is therefore a framework that derives diffuse-domain models from the same physical ingredients that define sharp-interface theories: geometry, kinematic and dynamic constraints, free energy, and dissipation.

Our objective here is to formulate field theories on moving surfaces while computing them in a fixed, regular three-dimensional domain. To do so, the diffuse embedding must preserve surface transport, local balance laws, and the energy-dissipation structure. We address this by formulating DDM within Onsager's variational principle (OVP)~\cite{Qian2006JFM,Doi1992,Doi2011OVP,Doi2013,Doi2015,Doi2021,Xu2017,Xu2021,Xu2022}. Instead of extending the governing equations and boundary conditions term by term, we embed sharp-surface free-energy and dissipation functionals into the bulk via a diffuse-surface delta density and derive the corresponding diffuse equations by minimizing the Rayleighian. This construction yields a unified treatment of balance-law fields, internal nonconserved order parameters, and kinematic or constitutive rate variables. In particular, it makes explicit a moving-surface kinematic distinction: conserved surface densities are transported by the full material surface velocity, whereas explicitly tangential vector and tensor internal variables evolve through projected objective/co-rotational rates in their admissible tangent state spaces.
For scalar transport on rigid and deformable interfaces, and for interfacial hydrodynamics near rigid walls, the resulting equations recover established DDM formulations with known sharp-interface limits. This recovery provides an internal consistency check on the variational embedding. We then use the same construction to formulate coupled diffuse-domain models for multicomponent deformable vesicles with surface viscosity, tangential slip, and finite areal compressibility, and for active polar shells with morphogen transport, phase separation, tangential flow, curvature coupling, and polarization. These latter models should be understood as thermodynamically consistent diffuse-Rayleighian formulations with formal sharp-interface correspondence at the level of the embedded free energy, dissipation functional, and leading-order surface kinematics. A complete matched-asymptotic derivation of all coupled field equations, together with systematic finite-$\epsilon$ numerical exploration, is left for future work.

\textcolor{black}{The main contributions of this work are threefold. First, we formulate DDMs from an embedded Rayleighian by placing surface free-energy and dissipation functionals in a fixed, regular, bulk domain through a diffuse surface density. This gives a unified OVP construction (as introduced in Sec.~\ref{sec:DDM}) for balance-law fields, internal nonconserved order parameters, and kinematic or constitutive rate variables, and it makes explicit the different moving-surface kinematics of conserved surface densities and explicitly tangential vector or tensor internal variables. 
Second, for scalar phase separation on fixed surfaces in Sec.~\ref{sec:SurfPSD-RigidSurf}, scalar transport on deformable interfaces in Sec.~\ref{sec:SurfPSD-DeformSurf}, and two-phase hydrodynamics near rigid walls in  Sec.~\ref{sec:MPF-2Phase}, the formulation recovers established diffuse-domain or diffuse-interface models with known sharp-interface consistency~\cite{RatzVoigt2006,LiLowengrubRatzVoigt2009,ElliottStinner2009,LowengrubRatzVoigt2009Vesicles,AlandLowengrubVoigt2010}. These examples serve mainly as reformulations and internal consistency checks: the same equations follow from embedded free-energy and dissipation functionals rather than from term-by-term assembly of diffuse equations and boundary conditions. Third, applying the same Rayleighian construction in Secs.~\ref{sec:MPF-vesicle} and \ref{sec:ActPolar} leads to variationally consistent finite-$\epsilon$ diffuse-domain models for extensible multicomponent vesicles and active polar shells. These models incorporate surface viscosity, tangential slip, finite areal compressibility, morphogen transport, curvature--composition or curvature--polarization coupling, and active surface stresses. To our knowledge, this combination has not previously been derived within a single diffuse-interface or phase-field framework. We emphasize, however, that these latter models are finite-$\epsilon$ variational constructions with sharp-interface correspondence at the level of the free energy, dissipation functional, constraints, and leading-order surface kinematics; a full matched-asymptotic derivation of all coupled field equations and a systematic numerical convergence study are left for future work.}
This positioning also clarifies the relation of the present work to recent OVP-based domain-embedding studies by Yu, Qian, Wang, and co-workers for Allen--Cahn and Cahn--Hilliard dynamics in arbitrary bulk domains~\cite{YuQianWang2025,YuWangZhangQian2025}. These studies share the same variational philosophy of deriving embedded phase-field equations from Onsager's principle. The present work addresses a different class of problems: coupled bulk--surface hydrodynamics, moving-surface geometry, scalar chemical transport on surfaces, and tangential vector- or tensor-field dynamics on deformable interfaces.

The remainder of this paper is organized as follows. Sec.~\ref{sec:DDM} introduces the diffuse-domain embedding of surface fields and the general variational construction. Sec.~\ref{sec:SurfPSD} applies this framework to phase separation on rigid and deformable interfaces. Sec.~\ref{sec:MPF} extends it to interfacial hydrodynamics in multiphase flows, including both dynamics near rigid boundaries and dynamics on multicomponent deformable vesicles. Sec.~\ref{sec:ActPolar} further develops the theory for active polar tissue shells, coupling morphogen transport, phase separation, tangential viscous flow, polarization, surface deformation, and bulk hydrodynamics. Sec.~\ref{sec:conclusion} summarizes the main results and discusses open problems and possible extensions.

\section{Diffuse-domain embedding of surface fields}\label{sec:DDM}

Let $\Gamma(t)$ be a smooth, oriented, closed surface embedded in a fixed, regular bulk domain $\Omega$ with boundary $\partial\Omega$; see Fig.~\ref{Fig:Schematic}(a). In this section, we introduce the diffuse-domain embedding of surface fields that are originally defined on $\Gamma(t)$. The goal
is to formulate the geometric, kinematic, and variational ingredients used in the OVP-based constructions below. For canonical scalar balance laws, this construction reduces to standard DDM formulations with known sharp-interface
consistency~\cite{RatzVoigt2006,LiLowengrubRatzVoigt2009,ElliottStinner2009,DeckelnickStyles2018}. For the coupled vector- and tensor-valued models developed later, the same construction should be understood as a diffuse Rayleighian-level variational embedding: under the assumptions stated below, the embedded free energy, dissipation functional, constraints, and Rayleighian reduce to their sharp-surface counterparts, and the leading-order surface kinematics are consistent with the corresponding sharp-surface OVP formulation. This statement is distinct from,
and does not replace, a complete matched-asymptotic derivation of all resulting finite-$\epsilon$ field equations~\cite{ArroyoDesimone2009PRE,Keber2014,Salbreux2017,Khoromskaia2023,AlandWohlgemuth2023,NitschkeVoigt2024ActiveNematodynamics}. The construction relies on three ingredients: a diffuse approximation of the surface measure, embedded balance laws for surface-conserved quantities, and embedded free-energy and dissipation functionals defined on the appropriate surface state spaces.

\textcolor{black}{\textbf{Notation guide} --- For convenience, we summarize the notation used repeatedly below. The parameter $\epsilon$ denotes the diffuse-surface thickness, whereas parameters such as $\epsilon_c$ or $\epsilon_m$ denote widths of composition or molecular transition layers on the surface.
\begin{description}[leftmargin=3.3cm,labelwidth=3.0cm,style=nextline]
\item[$\Gamma(t)$, $\Omega$] Sharp moving surface and fixed bulk computational domain.
\item[$\psi(\boldsymbol x,t)$] Phase-field variable representing the diffuse interface. Unless specified otherwise, one assumes $\psi \simeq 1$ in the interior region and $\psi \simeq 0$ in the exterior region, where applicable.
\item[{$\delta_\epsilon[\psi]$}] Normalized diffuse surface delta density used to replace $\int_\Gamma(\cdot)\,dA$ by $\int_\Omega\delta_\epsilon(\cdot)\,dV$.
\item[$\hat{\boldsymbol n}_\epsilon$, $\boldsymbol P_\epsilon$] Diffuse normal and tangential projector, with $\boldsymbol P_\epsilon=\boldsymbol I-\hat{\boldsymbol n}_\epsilon\otimes\hat{\boldsymbol n}_\epsilon$.
\item[$\boldsymbol\kappa_\epsilon$, ${\mathcal H}_\epsilon$] Diffuse curvature tensor and mean curvature, with ${\mathcal H}_\epsilon=\frac12\nabla_{\mathrm s}\cdot\hat{\boldsymbol n}_\epsilon$.
\item[$\boldsymbol{a}$, $\boldsymbol{b}$] Generic balance-law field and internal nonconserved order-parameter field, respectively.
\item[$\boldsymbol J_{\mathrm a}$, $\boldsymbol{k}_{\mathrm a}$] Nonconvective flux and source associated with a balance-law field $\boldsymbol{a}$.
\item[$\boldsymbol{\hat\mu}_{\mathrm a}$, $\boldsymbol h_{\mathrm b}$] Variational conjugates: chemical potential of $\boldsymbol{a}$ and molecular field of $\boldsymbol{b}$.
\item[$\boldsymbol V=\boldsymbol V_\parallel+V_n\hat{\boldsymbol n}_\epsilon$] Material surface velocity decomposed into tangential and normal parts.
\item[$\mathbb D_{\mathrm s}(\boldsymbol V)$] Full material surface rate-of-deformation tensor; $\mathbb D_{\mathrm s}(\boldsymbol V)=\mathbb D_{\mathrm s}(\boldsymbol V_\parallel)+V_n\boldsymbol\kappa_\epsilon$.
\item[$\nabla_{\mathrm s}$, $\bbNabla_{\mathrm s}$] Projected derivative and intrinsic projected surface gradient; for tensor fields $\bbNabla_{\mathrm s}$ projects both derivative and state slots.
\item[subscript $\Gamma$] Corresponding sharp-interface quantity, for example $\boldsymbol P_\Gamma$, $\nabla_\Gamma$, and $\mathbb D_\Gamma$.
\end{description}}

\subsection{Geometric embedding and diffuse surface measure}

\begin{figure}[htbp]
\centering
\includegraphics[width=0.9\columnwidth]{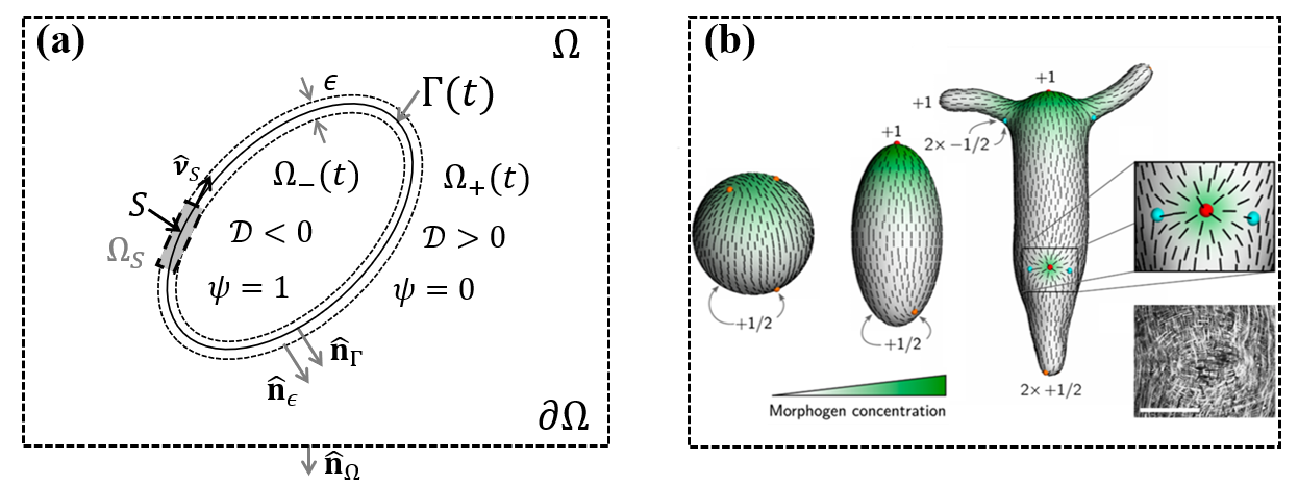}
\caption {Schematic illustration of the diffuse-domain embedding and its biological motivation from deformable active surfaces. (a) Diffuse-domain representation of an irregular, closed, time-evolving surface $\Gamma(t)$: the computational domain $\Omega=\Omega_- \cup \Gamma\cup \Omega_+$ (with boundary $\partial \Omega$) is fixed and regular, and the surface $\Gamma(t)$ is represented by a diffuse interface of thickness $\epsilon$, across which the phase field varies smoothly from $\psi = 1$ in the interior region $\Omega_-$ (with $\mathcal D<0$) to $\psi = 0$ in the exterior region $\Omega_+$ (with $\mathcal D>0$). The vectors $\hat{\boldsymbol n}_\Gamma$ and $\hat{\boldsymbol n}_\epsilon$ denote the sharp and diffuse outward unit normals, respectively. The set $S(t)\subset \Gamma(t)$ denotes a material surface patch, and $\Omega_S(t) \subset \Omega$ is the bulk region whose intersection with the diffuse interfacial layer projects onto $S(t)$. The vector $\hat{\boldsymbol{\nu}}_S$ denotes the outward co-normal unit vector along the boundary curve $\partial S(t)$. (b) Active nematic dynamics on a deformable surface during \textit{Hydra} regeneration: nematic orientational order is coupled to active actomyosin contractility, morphogen gradients, topological defect dynamics, and surface shape remodeling, thereby illustrating the integrated roles of active orientational order, mechano-chemical signaling, and surface deformation in biological morphogenesis (adapted from Wang \emph{et al}.~\cite{wang2023patterning}).
\label{Fig:Schematic}}
\end{figure}

\textbf{Diffuse interface, unit normal, and tangential projector} --- We represent the evolving closed surface $\Gamma(t)$ by a smooth phase field $\psi(\boldsymbol{x},t)\in [0,1]$ with an interfacial layer of thickness $O(\epsilon)$, such that $\psi\approx 1$ in the interior region $\Omega_-$, $\psi\approx 0$ in the exterior region $\Omega_+$, and the isosurface $\psi=\tfrac12$ approximates $\Gamma(t)$; see Fig.~\ref{Fig:Schematic}(a).  The outward diffuse unit normal and tangential projector are
\begin{equation}
\hat{\boldsymbol n}_\epsilon\equiv-\nabla\psi/|\nabla\psi|,
\qquad
\boldsymbol P_\epsilon\equiv\boldsymbol I-\hat{\boldsymbol n}_\epsilon\otimes\hat{\boldsymbol n}_\epsilon .
\end{equation}
On the sharp surface $\Gamma(t)$, the corresponding quantities are $\hat{\boldsymbol n}_\Gamma\equiv\nabla {\mathcal D}$ and $\boldsymbol P_\Gamma=\boldsymbol I-\hat{\boldsymbol n}_\Gamma\otimes\hat{\boldsymbol n}_\Gamma$, where $\mathcal D$ is the signed distance to $\Gamma(t)$~\cite{Aris1962Book,Federer1969GMT}; see Appendix~\ref{sec:surface-geometry}.  If $\psi(\boldsymbol x,t)=\tilde\psi(\mathcal D(\boldsymbol x,t)/\epsilon)$ with $\tilde\psi'<0$ (\emph{e.g.}, $\psi(\boldsymbol{x},t) = \frac{1}{2}\left[1-\tanh\left(3\mathcal{D}(\boldsymbol{x},t)/\epsilon\right)\right]$), then $\hat{\boldsymbol n}_\epsilon=\hat{\boldsymbol n}_\Gamma$ wherever the signed-distance representation is smooth.

\textbf{Diffuse surface delta density} $\delta_\epsilon $ --- Surface integrals along $\Gamma$ are embedded into $\Omega$ by a nonnegative diffuse surface delta density $\delta_\epsilon[\psi] \ge 0$ localized in the $O(\epsilon)$ interfacial layer and normalized along a signed-distance normal line, via $\int_{-\infty}^{\infty}\delta_\epsilon \,d\mathcal D=1$.  Standard choices are
\begin{equation}\label{Eq:DDM-surfDelta}
B_1=|\nabla\psi|,\qquad
B_2=2\epsilon^{-1}G(\psi),\qquad
B_3=\epsilon|\nabla\psi|^2, \qquad
B_4=\frac12\epsilon|\nabla\psi|^2+\epsilon^{-1}G(\psi),
\end{equation}
with $G(\psi)=18\psi^2(1-\psi)^2$~\cite{RatzVoigt2006,LiLowengrubRatzVoigt2009,ElliottStinner2009}. \textcolor{black}{For the standard one-dimensional equilibrium profile, $\psi(\boldsymbol{x},t) = \frac{1}{2}\left[1-\tanh\left(3\mathcal{D}(\boldsymbol{x},t)/\epsilon\right)\right]$, these choices have the same normal-line normalization. Away from such a profile they should be interpreted as different finite-$\epsilon$ regularizations which, in the limit $\epsilon\to 0$, converge asymptotically equivalent to the same sharp surface measure, while nevertheless remaining pointwise distinct functions for any finite $\epsilon$. In practice, $B_1$ is the most direct geometric choice because of its coarea interpretation. $B_2$ depends only on the phase-field value and is less sensitive to numerical differentiation of $\psi$, provided that the profile remains close to the prescribed one-dimensional equilibrium profile. The choices $B_3$ and $B_4$ are natural in phase-field models because they are tied to the gradient and double-well contributions to the diffuse-interface energy. In particular, $B_3$ is convenient in some diffuse-wall wetting formulations; and $B_4$ is preferred for the deformable-surface models in Secs.~\ref{sec:SurfPSD-DeformSurf}--\ref{sec:ActPolar} because its time derivative explicitly captures the normal-transport terms. 
At finite $\epsilon$, the choice of $\delta_\epsilon$ determines the specific form of the embedded equations and must be used consistently across the free energy, dissipation functional, and balance laws. Although we use the notation $\delta_\epsilon$ for a generic normalized diffuse surface density, it should be understood that specific choices are made based on the physical and numerical requirements of the model. } 

\textbf{Two foundational identities} --- Two sharp-surface consistency relations are used repeatedly.  

\begin{enumerate}[label=(\roman*)]
\item \emph{Coarea consistency.} Let $\boldsymbol f$ be a smooth extension of
a surface field $\boldsymbol f_\Gamma$ in a tubular neighborhood of $\Gamma(t)$, and assume
that $\boldsymbol f(\boldsymbol X+\mathcal D\hat{\boldsymbol n}_\Gamma,t)
\to\boldsymbol f_\Gamma(\boldsymbol X,t)$ as $\epsilon\to0$ uniformly on the
normal scale $|\mathcal D|=O(\epsilon)$. Then
\begin{equation}\label{Eq:DDM-coarea}
\lim_{\epsilon\to0}
\int_{\Omega_S(t)}
\boldsymbol f(\boldsymbol x,t)\,
\delta_\epsilon (\boldsymbol x,t)\,dV
=\int_{S(t)} \boldsymbol f_\Gamma(\boldsymbol X,t)\,dA.
\end{equation}
Here, $S(t)\subset\Gamma(t)$ is a material surface patch, and $\Omega_S(t)\subset\Omega$ is any material bulk region whose intersection with the diffuse layer projects to $S(t)$. For $\delta_\epsilon =|\nabla\psi|$, this follows directly from the coarea theorem, $\int_{\Omega_S}
\boldsymbol f|\nabla\psi|\,dV = \int_0^1
(\int_{\psi=u}\boldsymbol f\,dA_u)du$, together with convergence of the level-set measures to the measure on
$\Gamma(t)$. 
The same sharp-interface limit holds for the normalized asymptotically equivalent densities in Eq.~\eqref{Eq:DDM-surfDelta} under the
corresponding one-dimensional profile assumptions.

\item \emph{Tangential divergence identity.} Let $\boldsymbol F$ be a tensor
field whose last, flux-carrying slot is tangential,
$\boldsymbol F\cdot\hat{\boldsymbol n}_\epsilon=\boldsymbol0$. If
$\hat{\boldsymbol n}_\epsilon=\nabla\mathcal D$ and the normal field is constant
along normal lines, so that
$(\hat{\boldsymbol n}_\epsilon\cdot\nabla)\hat{\boldsymbol n}_\epsilon=\boldsymbol0$,
then
\begin{subequations}\label{Eq:DDM-tandiv}
\begin{equation}\label{Eq:DDM-tandiva}
\nabla\!\cdot\boldsymbol F =
\nabla_{\mathrm s}\!\cdot\boldsymbol F,
\end{equation}
\begin{equation}\label{Eq:DDM-tandivb}
\nabla\!\cdot(\delta_\epsilon \boldsymbol F)
= \delta_\epsilon \,\nabla_{\mathrm s}\!\cdot\boldsymbol F \quad\hbox{if }\delta_\epsilon=\delta_\epsilon(\mathcal D).
\end{equation}
\end{subequations}
For generic $\delta_\epsilon [\psi]$ and non-equilibrium phase-field
profiles, these identities should be interpreted as leading-order sharp-interface relations rather than exact finite-$\epsilon$ identities.
\end{enumerate}

\textbf{Diffuse curvature and material surface rate of deformation} ---
The diffuse curvature tensor, mean curvature, and the deviatoric curvature tensor are defined by
\begin{equation}
\boldsymbol \kappa_\epsilon \equiv \bbNabla_{\mathrm s} \hat{\boldsymbol n}_\epsilon = \boldsymbol P_\epsilon\cdot(\nabla\hat{\boldsymbol n}_\epsilon)\cdot \boldsymbol P_\epsilon, 
\qquad
\mathcal H_\epsilon \equiv \frac12\,\operatorname{tr}\boldsymbol \kappa_\epsilon = \frac12\,\nabla_{\mathrm s}\!\cdot\!\hat{\boldsymbol n}_\epsilon,
\qquad
\tilde{\boldsymbol \kappa}_\epsilon \equiv \boldsymbol \kappa_\epsilon-\mathcal H_\epsilon\boldsymbol P_\epsilon.
\label{Eq:DDM-diffuse-curvature}
\end{equation}
Here, the left projection in $\boldsymbol P_\epsilon \cdot (\nabla\hat{\boldsymbol n}_\epsilon) \cdot \boldsymbol P_\epsilon$ is redundant at the continuous level because $|\hat{\boldsymbol n}_\epsilon|=1$ implies $\hat{\boldsymbol n}_\epsilon\cdot\partial_j\hat{\boldsymbol n}_\epsilon=0$ for each derivative direction $j$, and hence $\boldsymbol \kappa_\epsilon =\bbNabla_{\mathrm s} \hat{\boldsymbol n}_\epsilon = \nabla_{\mathrm s}\hat{\boldsymbol n}_\epsilon$. We nevertheless keep both projectors to make explicit that $\boldsymbol \kappa_\epsilon$ is a fully tangential second-rank tensor.
In general, $\boldsymbol \kappa_\epsilon\neq\nabla\hat{\boldsymbol n}_\epsilon$ at
finite $\epsilon$; they coincide when the diffuse normal is constant along diffuse
normal lines, as for a signed-distance profile
$\psi=\tilde\psi(\mathcal D/\epsilon)$. 
On the sharp surface $\Gamma(t)$, the corresponding curvature tensor and mean
curvature are denoted by $\boldsymbol \kappa_\Gamma \equiv \boldsymbol P_\Gamma\cdot(\nabla\hat{\boldsymbol n}_\Gamma)\cdot\boldsymbol P_\Gamma$, $\mathcal H_\Gamma \equiv \frac12\,\operatorname{tr}\boldsymbol \kappa_\Gamma=\frac12\,\nabla_\Gamma\!\cdot\hat{\boldsymbol n}_\Gamma$. 
With the sign convention adopted here, a sphere of radius $R$ with outward normal has $\mathcal H_\Gamma=1/R>0$. 

The material surface velocity is decomposed as
\begin{equation}
\boldsymbol V
=
\boldsymbol V_\parallel+V_n\hat{\boldsymbol n}_\epsilon,
\qquad
\boldsymbol V_\parallel=\boldsymbol P_\epsilon\cdot\boldsymbol V,
\qquad
V_n=\boldsymbol V\cdot\hat{\boldsymbol n}_\epsilon .
\label{Eq:DDM-velocity-decomposition}
\end{equation}
The full material surface rate-of-deformation tensor is
\begin{equation}
\mathbb{D}_{\mathrm s}(\boldsymbol V) \equiv
\frac12\left[\bbNabla_{\mathrm{s}}\boldsymbol V+(\bbNabla_{\mathrm{s}}\boldsymbol V)^{\mathrm T}
\right]=\boldsymbol P_\epsilon\cdot \boldsymbol D(\boldsymbol V) \cdot\boldsymbol P_\epsilon,
\qquad
\boldsymbol D(\boldsymbol V) \equiv \frac12\left[
\nabla\boldsymbol V+(\nabla\boldsymbol V)^{\mathrm T}
\right].
\label{Eq:DDM-material-rate}
\end{equation}
Using Eq.~\eqref{Eq:DDM-velocity-decomposition}, this tensor decomposes as
\begin{equation}
\mathbb{D}_{\mathrm s}(\boldsymbol V)=
\mathbb{D}_{\mathrm s}(\boldsymbol V_\parallel)
+
V_n\boldsymbol \kappa_\epsilon,
\label{Eq:DDM-material-rate-decomposition}
\end{equation}
where $\mathbb{D}_{\mathrm s}(\boldsymbol V_\parallel)
\equiv \frac12\left[\bbNabla_{\mathrm{s}}\boldsymbol V_\parallel+(\bbNabla_{\mathrm{s}}\boldsymbol V_\parallel)^{\mathrm T}
\right]=
\boldsymbol P_\epsilon\cdot
\frac12\left[
\nabla\boldsymbol V_\parallel+
(\nabla\boldsymbol V_\parallel)^{\mathrm T}
\right]\cdot\boldsymbol P_\epsilon$. 
Thus, $\mathbb{D}_{\mathrm s}(\boldsymbol V_\parallel)$ measures lateral in-surface rearrangement, whereas $\mathbb{D}_{\mathrm s}(\boldsymbol V)$ measures the full metric rate of a moving material surface. The two tensors coincide only when $V_n=0$. Which tensor enters the surface-viscous dissipation is therefore a constitutive choice: a full Boussinesq--Scriven metric-rate model uses $\mathbb{D}_{\mathrm s}(\boldsymbol V)$, whereas a reduced lateral-rearrangement model uses $\mathbb{D}_{\mathrm s}(\boldsymbol V_\parallel)$ and omits the curvature-induced metric-rate contribution.

\subsection{Surface fields and embedded balance laws}

\textcolor{black}{A surface field $\boldsymbol f_\Gamma$ is represented in the diffuse layer by an ambient extension $\boldsymbol f$, but the extension away from $\Gamma(t)$ and the admissible surface state space are distinct choices.  Scalar surface fields have no tensor-slot tangentiality.  For an explicitly tangential vector or tensor field, however, admissible variations and variational conjugates must be projected onto the same tangential state space.  We write $\boldsymbol f_\parallel=\boldsymbol P_\epsilon^{(r)}\!\cdot\boldsymbol f$ for the slotwise tangential projection of a rank-$r$ field.  In the scalar R\"atz--Voigt embedding used below, the gradient operator is the full ambient gradient $\mathcal G=\nabla$, which provides finite-$\epsilon$ normal regularization.  For intrinsic tangential vector or tensor distortion energies, $\mathcal G$ is instead the fully projected gradient $\bbNabla_{\mathrm s}$.  The component definitions of $\boldsymbol P_\epsilon^{(r)}$, $\nabla_{\mathrm s}$, $\bbNabla_{\mathrm s}$, and the associated projected divergences are given in Appendix~\ref{app:tangential-state-spaces}.}

\textbf{Three operational classes of surface fields} --- For the variational construction, it is useful to distinguish three operational classes of surface quantities according to the mathematical form of their evolution equations:
\begin{enumerate}[label=(\roman*)]
\item \emph{Balance-law fields} $\boldsymbol a_\Gamma$, whose dynamics are posed through local balance equations with flux and possibly source terms. Prototypical examples include surface densities of extensive quantities, defined per unit current surface area, and internal conserved order-parameter fields, \emph{e.g.}, surface concentrations.

\item \emph{Internal non-conserved order-parameter fields} $\boldsymbol b_\Gamma$, whose dynamics are posed through material or objective rates rather than through conservative balances. They characterize the local material state. Typical examples are a polarization vector $\boldsymbol p_\Gamma$, a director field, or a $\boldsymbol Q_\Gamma$ tensor.

\item \emph{Kinematic or constitutive rate variables}, such as surface velocities $\boldsymbol V_\Gamma$, nonconvective fluxes, and local source terms.
\end{enumerate}  
Only classes (i) and (ii) are thermodynamic state variables in the free-energy
functional. Class (iii) variables enter through kinematic relations, constraints,
and the dissipation functional. The diffuse-domain construction below is formulated first for class-(i) variables, for which the embedded conservative balance is unambiguous.
Class-(ii) variables are then treated through the choice of a projected material
or objective dissipative rate.

\textbf{Embedded balance laws for balance-law fields} --- Let $\boldsymbol{a}_\Gamma$ denote a balance-law field of arbitrary tensorial rank, endowed with a nonconvective surface flux $\boldsymbol{J}_{\mathrm{a}\Gamma}$ and source $\boldsymbol{k}_{\mathrm{a}\Gamma}$. Let $\boldsymbol{a}$, $\boldsymbol{J}_{\mathrm a}$, and $\boldsymbol{k}_{\mathrm a}$ represent their corresponding extensions into the diffuse layer. The associated conservative embedded variable is given by $\delta_\epsilon \boldsymbol{a}$. For a material diffuse control volume $\Omega_S(t)$ whose intersection with the diffuse layer projects onto the surface patch $S(t)$, the embedded integral balance equation is
\begin{subequations}\label{Eq:DDM-bal}
\begin{equation}\label{Eq:DDM-bal-global}
\frac{d}{dt} \int_{\Omega_S(t)}\delta_\epsilon  \boldsymbol a\,dV = -\int_{\partial\Omega_S(t)}
(\delta_\epsilon \boldsymbol J_{\mathrm a})\cdot
\hat{\boldsymbol n}_{\Omega_S}\,dA + \int_{\Omega_S(t)} \delta_\epsilon \boldsymbol k_{\mathrm a}\,dV.
\end{equation}
Using the bulk Reynolds transport theorem and Gauss’ divergence theorem, the local form of the embedded balance equation is
\begin{equation}\label{Eq:DDM-bal-local}
\partial_t(\delta_\epsilon \boldsymbol{a})+\nabla\!\cdot\left[(\delta_\epsilon \boldsymbol{a})\otimes \boldsymbol V\right]
=-\nabla\!\cdot(\delta_\epsilon\boldsymbol J_{\mathrm{a}})+\delta_\epsilon \boldsymbol{k}_{\mathrm{a}},
\end{equation}
\end{subequations}
where the last index of $\boldsymbol J_{\mathrm{a}}$ carries the flux direction.  
In the geometrically faithful tangential embedding, this flux slot is tangential with $\boldsymbol J_{\mathrm a}\cdot\hat{\boldsymbol n}_\epsilon=\boldsymbol0$, so that Eq.~\eqref{Eq:DDM-tandiv} reduces the bulk divergence to the embedded surface divergence in the sharp-interface limit.  Equations~\eqref{Eq:DDM-bal} yield, as $\epsilon\to 0$, the sharp surface balance equations
\begin{subequations}\label{Eq:DDM-SI-bal}
\begin{equation}\label{Eq:DDM-SI-bal-global}
\frac{d}{dt}\int_{S(t)} \boldsymbol{a}_\Gamma\,dA =-\int_{\partial S(t)}\boldsymbol J_{\mathrm{a}\Gamma}\cdot \hat{\boldsymbol{\nu}}_S\,d\ell +\int_{S(t)} \boldsymbol k_{\mathrm{a}\Gamma}\,dA,
\end{equation}
\begin{equation}\label{Eq:DDM-SI-bal-local}
\dot{\boldsymbol{a}}_\Gamma+\boldsymbol{a}_\Gamma\,(\nabla_\Gamma\!\cdot\boldsymbol V_\Gamma)
=-\nabla_\Gamma\!\cdot\boldsymbol J_{\mathrm{a}\Gamma}+ \boldsymbol k_{\mathrm{a}\Gamma},
\end{equation}
\end{subequations}
where $\hat{\boldsymbol{\nu}}_S$ is the outward co-normal vector along the boundary curve $\partial S(t)$, 
$\dot{\boldsymbol{a}}_\Gamma=\partial_t\boldsymbol{a}_\Gamma+\boldsymbol V_{\Gamma\parallel}\cdot\nabla_\Gamma \boldsymbol{a}_\Gamma$, $\boldsymbol V_\Gamma=\boldsymbol V|_{\Gamma}$, $\nabla_\Gamma\!\cdot\boldsymbol V_\Gamma=\nabla_\Gamma\!\cdot\boldsymbol V_{\Gamma\parallel}+2{\mathcal H}_\Gamma V_{\Gamma n}$, and ${\mathcal H}_\Gamma \equiv \tfrac12\,\nabla_\Gamma\!\cdot\hat{\boldsymbol n}_\Gamma$ is the mean curvature (so, a sphere of radius $R$ has ${\mathcal H}_\Gamma=1/R>0$). For scalar surface fields, $\nabla_\Gamma\!\cdot\boldsymbol J_{\mathrm{a}\Gamma}$
is the usual surface divergence of a tangential flux. For ambient Cartesian vector or tensor components, it is applied componentwise to the flux slot. If $\boldsymbol a_\Gamma$ is constrained to a fully tangential tensor state space, then Eq.~\eqref{Eq:DDM-SI-bal-local} is understood after projection onto the same tangential state space, \emph{i.e.}, $\nabla_\Gamma\!\cdot\boldsymbol J_{\mathrm{a}\Gamma}$ is replaced by the intrinsic projected covariant divergence $\bbNabla_\Gamma\!\cdot\boldsymbol J_{\mathrm{a}\Gamma}$.

\textcolor{black}{\textbf{Dynamic equations and dissipative rates for internal non-conserved order-parameter fields} --- For a class-(ii) internal variable—namely, a nonconserved order parameter $\boldsymbol{b}_\Gamma$ together with its ambient extension $\boldsymbol{b}$—the appropriate dissipative variable is not the temporal derivative $\partial_t(\delta_\epsilon \boldsymbol{b})$, but rather a material (objective) rate of $\boldsymbol{b}$ itself, denoted by $\dot{\mathbb B}$, which is weighted by $\delta_\epsilon$ solely within the dissipation functional. This distinction is implemented by formulating separate balance equations for the carrier density $\rho_\Gamma$ and for the carrier-weighted internal variable $\boldsymbol{m}_\Gamma \equiv \rho_\Gamma \boldsymbol{b}_\Gamma$. The derivation of these balance laws, together with the explicit forms of the projected co-rotational rates $\dot{\mathbb B}$ employed for vector- and tensor-valued fields, is presented in Appendix~\ref{app:internal-nonconserved-rates}.}

\subsection{Fixed-surface OVP variational construction}

We now state the OVP construction in the setting in which it is used as a template for the examples below.  In this subsection, the diffuse surface is fixed, so $\boldsymbol V=0$ and $\psi$, $\delta_\epsilon$, and $\boldsymbol P_\epsilon$ are time independent.  Moving-surface models require additional $\psi$-dependent terms; these are derived case by case in Secs.~\ref{sec:SurfPSD-DeformSurf}, \ref{sec:MPF-vesicle}, and \ref{sec:ActPolar}, with supporting algebra in Appendix~\ref{sec:App-ActPolar-deriv}.

\textbf{Embedded free energy, dissipation functional, and Rayleighian} --- Let $F_s$ be a sharp-surface free-energy density per unit current area. We assume $F_{\mathrm s}$ depends on fully tangential balance-law fields $\boldsymbol a_\Gamma$, fully tangential internal nonconserved order parameters $\boldsymbol b_\Gamma$, and selected surface-gradient variables. The embedded free energy is written as
\begin{equation}\label{Eq:DDM-F}
\mathcal F[\boldsymbol a_\parallel,\boldsymbol b_\parallel,\psi] = \int_\Omega \delta_\epsilon [\psi] F_{\mathrm s}(
\boldsymbol a_\parallel,\boldsymbol b_\parallel,
\mathcal G_{\mathrm a}\boldsymbol a_\parallel,
\mathcal G_{\mathrm b}\boldsymbol b_\parallel )\,dV.
\end{equation}
Here, $\boldsymbol a_\parallel=\boldsymbol P_\epsilon^{(r_{\mathrm a})}\cdot\boldsymbol a$
and $\boldsymbol b_\parallel=\boldsymbol P_\epsilon^{(r_{\mathrm b})}\cdot\boldsymbol b$
for fully tangential tensor fields, with $\boldsymbol P_\epsilon^{(0)}=1$ for scalar fields. 
Tangentiality may be imposed strongly by using these
projected variables as the state variables, or weakly by anchoring or penalty terms. The operators $\mathcal G_{\mathrm a}$ and $\mathcal G_{\mathrm b}$ denote the chosen gradient embeddings. 
For scalar surface fields, the isotropic Rätz--Voigt embedding uses the full ambient gradient, $\mathcal G=\nabla$, which yields weighted isotropic operators of the form $\nabla\cdot(\delta_\epsilon \nabla\cdot)$ and provides
finite-$\epsilon$ normal regularization~\cite{RatzVoigt2006,LiLowengrubRatzVoigt2009}.
For explicitly tangential vector or tensor state fields, an intrinsic surface distortion energy should instead be built from the intrinsic projected gradient $\mathcal G=\bbNabla_{\mathrm s}$, with the state-space projection imposed separately from the derivative projection~\cite{NestlerVoigt2024,KloppeAland2024,AlandWohlgemuth2023}. 

For fixed surfaces, using $\partial_t(\mathcal G_{\boldsymbol{\alpha}} \boldsymbol{\alpha})=\mathcal G_{\boldsymbol{\alpha}}(\partial_t\boldsymbol{\alpha})$ with $\boldsymbol \alpha=\boldsymbol a_\parallel,\boldsymbol b_\parallel$ and Eq.~\eqref{Eq:DDM-bal-local} for the fixed-surface balance-law of $\boldsymbol a_{\parallel}$, with $\boldsymbol J_{\mathrm a}$ and $\boldsymbol k_{\mathrm a}$ valued in the same admissible tangential state space as $\boldsymbol a_\parallel$, integrating by parts, and dropping outer-boundary terms, the free-energy rate can be written as
\begin{equation}\label{Eq:DDM-dFdt}
\dot{\mathcal F}
=\int_\Omega \delta_\epsilon \left(\hat{\boldsymbol\mu}_{\mathrm a} :\partial_t \boldsymbol a_{\parallel}
- \boldsymbol h_{\mathrm b}:\dot{\mathbb B}\right)\,dV
=\int_\Omega
\delta_\epsilon \left(\boldsymbol J_{\mathrm a}:\mathcal G_{\mathrm a}\hat{\boldsymbol\mu}_{\mathrm a}+\hat{\boldsymbol\mu}_{\mathrm a}:\boldsymbol k_{\mathrm a}-\boldsymbol h_{\mathrm b}:\dot{\mathbb B}\right)\,dV.
\end{equation} 
Here $\dot{\mathcal F}$ denotes the time derivative of the integrated free energy,
whereas overdots on local fields denote the corresponding material or objective
rates specified below.
The symbol $\dot{\mathbb B}$ denotes the dissipative rate associated with the internal nonconserved order parameter $\boldsymbol  b_\parallel$. Its precise form is constitutive and depends on the tensorial character of
$\boldsymbol b_\parallel$, the motion of the surface, and the imposed tangential state space; on a fixed surface it reduces to $\partial_t\boldsymbol b_\parallel$. 
The generalized chemical potential for the balance-law field $\boldsymbol a$ is defined by
\begin{equation}\label{Eq:DDM-mu}
\delta_\epsilon \hat{\boldsymbol\mu}_{\mathrm a} \equiv
\frac{\delta \mathcal F}{\delta \boldsymbol a_{\parallel}} =\boldsymbol P_\epsilon^{(r_{\mathrm a})} \cdot (\delta_\epsilon \,\partial_{\boldsymbol a_{\parallel}}F_{\mathrm s})
-\mathcal D_{\mathcal G_{\mathrm a}}^{\ast}(\delta_\epsilon \boldsymbol\pi_{\mathrm a}),
\qquad
\boldsymbol\pi_{\mathrm a} \equiv
\partial_{\mathcal G_{\mathrm a}\boldsymbol a_{\parallel}}F_{\mathrm s},
\end{equation}
where $\mathcal D_{\mathcal G_{\mathrm a}}^{\ast}$ is the divergence operator adjoint to $\mathcal G_{\mathrm a}$, defined by
$\int_\Omega \delta_\epsilon  \boldsymbol\pi_{\mathrm a}:\mathcal G_{\mathrm a}\boldsymbol\eta\,dV = -\int_\Omega D_{\mathcal G_{\mathrm a}}^{\ast}(\delta_\epsilon \boldsymbol\pi_{\mathrm a}):\boldsymbol\eta\,dV$
for admissible variations $\boldsymbol\eta$ and assuming that boundary terms vanish. In the isotropic R\"atz--Voigt embedding for scalar surface fields $a(\boldsymbol x,t)$, we take $\mathcal G_{\mathrm a} a=\nabla a$ and $\mathcal D_{\mathcal G_{\mathrm a}}^{\ast}(\delta_\epsilon \boldsymbol\pi_{\mathrm a})=\nabla\!\cdot\!(\delta_\epsilon \boldsymbol\pi_{\mathrm a})$. In this case $\hat\mu_{\mathrm a}$ is a scalar and $\boldsymbol\pi_{\mathrm a}$ is an ambient bulk vector; at finite $\epsilon$ it need not be tangential, although under the closest-point extension $\hat{\boldsymbol n}_\epsilon\!\cdot\!\nabla a=0$ it becomes tangential to leading order in the diffuse layer.
For the intrinsic projected/tangential embedding of an explicitly tangential vector or tensor field $\boldsymbol a_{\parallel}(\boldsymbol x,t)$, we take $\mathcal G_{\mathrm a}\boldsymbol a_{\parallel}=\bbNabla_{\mathrm s}\boldsymbol a_{\parallel}$ and $\mathcal D_{\mathcal G_{\mathrm a}}^{\ast}(\delta_\epsilon \boldsymbol\pi_{\mathrm a})=\nabla\cdot (\delta_\epsilon  \boldsymbol P_{\epsilon}^{(r_{\mathrm a})} \cdot \boldsymbol \pi_{\mathrm a} \cdot \boldsymbol P_\epsilon)$, equivalently $\delta_\epsilon  \nabla\cdot(\boldsymbol P_{\epsilon}^{(r_{\mathrm a})} \cdot \boldsymbol \pi_{\mathrm a} \cdot \boldsymbol P_\epsilon)$ by Eq.~\eqref{Eq:DDM-tandivb}. Here $\boldsymbol\pi_{\mathrm a}$ carries one additional derivative slot, and this added slot is right-tangential by construction. If $\boldsymbol a_{\parallel}$ is imposed to be fully tangential, then $\boldsymbol\pi_{\mathrm a}$ inherits the same tangentiality in the original slots. The tensor $\hat{\boldsymbol\mu}_{\mathrm a}$ has the same rank as $\boldsymbol a_{\parallel}$; only its component in the admissible slotwise tangential subspace couples to admissible variations of $\boldsymbol a_{\parallel}$. Accordingly, when tangentiality is imposed strongly, $\hat{\boldsymbol\mu}_{\mathrm a}$ is identified with that projected component.
The generalized molecular field for the internal non-conserved order parameter $\boldsymbol b$ is defined by
\begin{equation}\label{Eq:DDM-h}
\delta_\epsilon \boldsymbol h_{\mathrm b} \equiv
-\frac{\delta \mathcal F}{\delta \boldsymbol b_{\parallel}} =
-\boldsymbol P_\epsilon^{(r_{\mathrm b})}\cdot (\delta_\epsilon \,\partial_{\boldsymbol b_{\parallel}}F_{\mathrm s})
+\mathcal D_{\mathcal G_{\mathrm b}}^{\ast}(\delta_\epsilon \boldsymbol\pi_{\mathrm b}),
\qquad
\boldsymbol\pi_{\mathrm b} \equiv \partial_{\mathcal G_{\mathrm b}\boldsymbol b_{\parallel}}F_{\mathrm s}.
\end{equation}
Similar considerations apply to $\boldsymbol{b}$: for a scalar quantity $b$, the corresponding molecular field $h_{\mathrm b}$ is a scalar, while $\boldsymbol\pi_{\mathrm b}$ is an ambient bulk vector. For an explicitly tangential vector or tensor field $\boldsymbol b_{\parallel}$, or more generally for a strongly tangential vector or tensor field, only the projection of the conjugate field $\boldsymbol h_{\mathrm b}$ onto the admissible tangential state space is thermodynamically conjugate to admissible variations. By construction, the additional derivative slot of $\boldsymbol\pi_{\mathrm b}$ is right-tangential, while $\boldsymbol{h}_{\mathrm b}$ retains the tensorial rank of $\boldsymbol{b}_\parallel$. If tangentiality is enforced weakly via penalty terms, $\boldsymbol{h}_{\mathrm b}$ may exhibit non-vanishing normal components at finite $\epsilon$; in such cases, its projected tangential component must be employed in the constitutive relations.

Under the assumption that the dissipative mechanisms consist of density fluxes $\boldsymbol J_{\mathrm a}$, density sources $\boldsymbol k_{\mathrm a}$, and the dissipative rate $\dot{\mathbb B}$ of the internal non-conserved order parameter, we write the quadratic dissipation functional as
\begin{equation}\label{Eq:DDM-Phi}
\Phi[\boldsymbol J_{\mathrm a},\boldsymbol k_{\mathrm a},\dot{\mathbb B}]= \int_\Omega
\delta_\epsilon \left(
\frac{|\boldsymbol J_{\mathrm a}|^2}{2M_{\mathrm a}}+
\frac{|\boldsymbol k_{\mathrm a}|^2}{2\Gamma_{\mathrm a}}
+\frac{\gamma_{\mathrm b}}{2}\big|\dot{\mathbb B}\big|^2
\right)dV,
\end{equation}
where $M_{\mathrm a},\Gamma_{\mathrm a}>0$ are mobility coefficients and $\gamma_{\mathrm b}>0$ is a kinetic coefficient. Minimizing the embedded Rayleighian $\mathcal R=\dot{\mathcal F}+\Phi$ with respect to $\boldsymbol J_{\mathrm a}$, $\boldsymbol k_{\mathrm a}$, and $\dot{\mathbb B}$ yields the Onsager constitutive relations
\begin{equation}\label{Eq:DDM-Onsager}
\boldsymbol J_{\mathrm a}=-M_{\mathrm a}\,\mathcal G_{\mathrm a}\hat{\boldsymbol\mu}_{\mathrm a},
\qquad
\boldsymbol k_{\mathrm a}=-\Gamma_{\mathrm a}\hat{\boldsymbol\mu}_{\mathrm a},
\qquad
\gamma_{\mathrm b}\dot{\mathbb B}=\boldsymbol h_{\mathrm b}.
\end{equation}
For the fixed-surface case considered here, the last equation of Eq.~\eqref{Eq:DDM-Onsager} simply gives $\gamma_{\mathrm b}\partial_t\boldsymbol b_{\parallel}=\boldsymbol h_{\mathrm b}$. Substituting the first two relations in Eq.~\eqref{Eq:DDM-Onsager} into the fixed-$\psi$ balance law~\eqref{Eq:DDM-bal-local} with $\boldsymbol V=\boldsymbol0$ gives
\begin{equation}\label{Eq:DDM-bal-local2}
\delta_\epsilon \partial_t\boldsymbol a_\parallel = \nabla\!\cdot\left(\delta_\epsilon  M_{\mathrm a}\mathcal G_{\mathrm a}\hat{\boldsymbol\mu}_{\mathrm a} \right) -
\delta_\epsilon \Gamma_{\mathrm a} \hat{\boldsymbol\mu}_{\mathrm a},
\end{equation}
where the divergence acts on the flux slot generated by $\mathcal G_{\mathrm a} \hat{\boldsymbol\mu}_{\mathrm a}$. If this flux slot is tangential and the assumptions of Eq.~\eqref{Eq:DDM-tandiv} hold, the first term may be written as a weighted surface divergence in the diffuse layer. For a scalar field with the projected surface-gradient embedding $\mathcal G_{\mathrm a}=\nabla_{\mathrm s}$ and constant $M_{\mathrm a}$, Eq.~\eqref{Eq:DDM-bal-local2} reduces to $\partial_t a = M_{\mathrm a} \Delta_{\mathrm s}\hat\mu_{\mathrm a}-\Gamma_{\mathrm a}\hat\mu_{\mathrm a}$ inside the diffuse layer. For the isotropic scalar embedding $\mathcal G_{\mathrm a}=\nabla$, used later for Rätz--Voigt-type scalar models, Eq.~\eqref{Eq:DDM-bal-local2} should instead be kept in the weighted bulk form $\delta_\epsilon \partial_t a = \nabla\!\cdot(\delta_\epsilon  M_{\mathrm a}\nabla\hat\mu_{\mathrm a}) - \delta_\epsilon \Gamma_{\mathrm a}\hat\mu_{\mathrm a}$. For tangential vector or tensor fields, the corresponding operator is the projected surface divergence of the intrinsic projected gradient; it can not be generally identified with a componentwise ambient Laplacian. In addition, substituting Eq.~\eqref{Eq:DDM-Onsager} into Eq.~\eqref{Eq:DDM-dFdt} yields the passive fixed-surface energy (dissipation) law
\begin{equation}\label{Eq:DDM-energy-fixed}
\dot{\mathcal F}
=-\int_\Omega \delta_\epsilon  \left(\Gamma_{\mathrm a}|\hat{\boldsymbol\mu}_{\mathrm a}|^2+M_{\mathrm a}|\mathcal G_{\mathrm a} \hat{\boldsymbol\mu}_{\mathrm a}|^2+ \gamma_{\mathrm b} \big|\dot{\mathbb B} \big|^2\right)dV
=-2\Phi\le 0.
\end{equation}


\textbf{Rayleighian-level sharp-interface reduction} --- Under the smoothness, normalization, and normal-extension assumptions stated above, Eq.~\eqref{Eq:DDM-coarea} implies the formal convergence, as $\epsilon\to0$, of the embedded free energy, dissipation functional, and Rayleighian to
\begin{equation}\label{Eq:DDM-SI-FPhi-general}
\mathcal F_\Gamma[\boldsymbol a_{\Gamma\parallel},\boldsymbol b_{\Gamma\parallel}]=
\int_\Gamma F_{\mathrm s}(\boldsymbol a_{\Gamma\parallel},\boldsymbol b_{\Gamma\parallel},\bbNabla_{\Gamma}\boldsymbol a_{\Gamma\parallel},\bbNabla_{\Gamma}\boldsymbol b_{\Gamma\parallel})\,dA,
\qquad
\Phi_\Gamma[\boldsymbol J_{\mathrm{a}\Gamma},\boldsymbol k_{\mathrm{a}\Gamma},\dot{\mathbb B}_\Gamma]
= \int_\Gamma\left(
\frac{|\boldsymbol J_{\mathrm{a}\Gamma}|^2}{2M_{\mathrm a}}+
\frac{|\boldsymbol k_{\mathrm{a}\Gamma}|^2}{2\Gamma_{\mathrm a}}+
\frac{\gamma_{\mathrm b}}{2}|\dot{\mathbb B}_\Gamma|^2
\right)dA,
\end{equation}
and $\mathcal R_\Gamma=\dot{\mathcal F}_\Gamma+\Phi_\Gamma$. For a fixed surface, the sharp-surface free-energy rate is
\begin{equation}\label{Eq:DDM-SI-dFdt-general}
\dot{\mathcal F}_\Gamma
=\int_\Gamma\left(\hat{\boldsymbol\mu}_{\mathrm{a}\Gamma}:\partial_t\boldsymbol a_{\Gamma\parallel}-\boldsymbol h_{b\Gamma}:\dot{\mathbb B}_\Gamma\right)\,dA,
\end{equation}
with the corresponding generalized sharp-surface chemical potential and molecular field given by
\begin{subequations}
\begin{equation}\label{Eq:DDM-SI-muh}
\hat{\boldsymbol\mu}_{\mathrm{a}\Gamma}\equiv \frac{\delta \mathcal F_\Gamma}{\delta \boldsymbol a_\Gamma}=
\partial_{\boldsymbol a_{\Gamma\parallel}}F_{\mathrm s}-\mathcal D_{\bbNabla_\Gamma}^{\ast}\boldsymbol\pi_{\mathrm{a}\Gamma},
\qquad
\boldsymbol\pi_{\mathrm{a}\Gamma}\equiv\partial_{\mathcal \bbNabla_\Gamma\boldsymbol a_{\Gamma\parallel}}F_{\mathrm s},
\end{equation}
\begin{equation}\label{Eq:DDM-SI-h}
\boldsymbol h_{b\Gamma}\equiv
-\frac{\delta \mathcal F_\Gamma}{\delta \boldsymbol b_{\Gamma\parallel}}=
-\partial_{\boldsymbol b_{\Gamma\parallel}}F_{\mathrm s}+\mathcal D_{\bbNabla_\Gamma}^{\ast}\boldsymbol\pi_{b\Gamma},
\qquad
\boldsymbol\pi_{b\Gamma}\equiv\partial_{\bbNabla_\Gamma\boldsymbol b_{\Gamma\parallel}}F_{\mathrm s}.
\end{equation}
\end{subequations}
Minimizing the sharp-surface Rayleighian together with the fixed-surface balance
law gives
\begin{equation}\label{Eq:DDM-SI-reduced-general}
\partial_t\boldsymbol a_{\Gamma\parallel} =
\bbNabla_\Gamma\!\cdot \left(M_{\mathrm a}\bbNabla_\Gamma\hat{\boldsymbol\mu}_{\mathrm{a}\Gamma}
\right) - \Gamma_{\mathrm a}\hat{\boldsymbol\mu}_{\mathrm{a}\Gamma},
\qquad
\gamma_{\mathrm b}\dot{\mathbb B}_\Gamma=\boldsymbol h_{b\Gamma},
\end{equation}
where the divergence acts on the flux slot. For scalar $a_\Gamma$, this reduces to the usual surface divergence
$\nabla_\Gamma\!\cdot\boldsymbol J_{\mathrm{a}\Gamma}$. In the fixed-surface case, $\dot{\mathbb B}_\Gamma=\partial_t\boldsymbol b_{\Gamma\parallel}$, so the second relation reduces to $\gamma_{\mathrm b}\partial_t\boldsymbol b_{\Gamma\parallel}=\boldsymbol h_{b\Gamma}$. Moreover, the embedded energy dissipation law~\eqref{Eq:DDM-energy-fixed} converges to the sharp-surface energy law as $\epsilon\to 0$: $\dot{\mathcal F}_\Gamma=-2\Phi_\Gamma\le 0$. For a scalar field with $\bbNabla_\Gamma \to\nabla_\Gamma$ and constant $M_{\mathrm a}$, the first equation reduces to $\partial_t a_\Gamma = M_{\mathrm a}\Delta_\Gamma\hat\mu_{\mathrm{a}\Gamma} - \Gamma_{\mathrm a}\hat\mu_{\mathrm{a}\Gamma}$. Thus, the scalar Laplace--Beltrami form is a special case, not the generic tensorial statement.

\textcolor{black}{We emphasize that the reduction established here represents a sharp-interface correspondence at the level of the Rayleighian, rather than a formal matched-asymptotic expansion of the finite-$\epsilon$ field equations. This approach ensures that the embedded free energy, dissipation functional, and kinematic constraints consistently recover their sharp-surface counterparts within the Onsager variational framework. Matched-asymptotic analysis serves a complementary role: for a specified diffuse regularization, it can verify equation-level convergence, identify boundary-layer corrections, and quantify finite-$\epsilon$ errors. Thus, while the variational construction provides thermodynamically consistent finite-$\epsilon$ models and their sharp-interface variational counterpart, asymptotic and numerical analyses remain essential for validating the resulting diffuse equations.}

\textcolor{black}{Building on this foundation, we summarize the systematic construction of the variational DDM models employed throughout this work. The procedure consists of four steps: 
(i) Identify the role of the field variable: balance-law field, internal nonconserved order parameter, or kinematic/constitutive rate variable;
(ii) Specify the admissible state space, incorporating slotwise projections for tangential vector or tensor fields; (iii) Select the appropriate gradient embedding, such as $\nabla$ for scalar fields or $\bbNabla_{\mathrm s}$ for tangential vector/tensor distortion energies; and (iv) Formulate the embedded free energy and dissipation functional using a consistent measure weight $\delta_\epsilon$, with conjugate forces and dissipative rates projected onto the admissible state space. Optional finite-$\epsilon$ normal-extension controls for numerical robustness are described in Appendix~\ref{app:finiteepsilon-reg}. Crucially, this construction maintains a clear separation between state-space projections, derivative-slot projections, and the projection of thermodynamic conjugates or fluxes---a distinction that is strictly preserved in all subsequent model developments.}

\section{Surface phase separation dynamics (single scalar field)}\label{sec:SurfPSD}

\subsection{Phase separation on fixed rigid surfaces}\label{sec:SurfPSD-RigidSurf}

We now specialize the general diffuse-domain framework of Sec.~\ref{sec:DDM} to a single scalar surface composition field $c_\Gamma$ on a fixed, rigid, closed surface $\Gamma$ embedded in the regular computational domain $\Omega$ [Fig.~\ref{Fig:Schematic}(a)]. The field $c_\Gamma$ may be
interpreted either as a conserved surface composition in the Cahn--Hilliard case or as a nonconserved scalar order parameter in the Allen--Cahn case.

\textbf{State variables and constraints} --- Following the embedding procedure described in Sec.~\ref{sec:DDM}, we represent $c_\Gamma$ by a smooth bulk extension $c(\boldsymbol x,t)$ in the diffuse layer. Since the surface is fixed, the phase field $\psi$, the diffuse surface density $\delta_\epsilon $, and the projector $\boldsymbol P_\epsilon$ are all time independent. If $c_\Gamma$ is a conserved surface composition, the corresponding fixed-surface balance law is
\begin{equation}\label{Eq:SurfPSD-RigidSurf-bal}
\partial_t\!\left(\delta_\epsilon  c\right)
=-\nabla\!\cdot\! \left(\delta_\epsilon  \boldsymbol J_c\right)
+\delta_\epsilon  k_c,
\end{equation}
where the diffuse-layer flux $\boldsymbol J_c$ need not be strictly tangential. 

\textbf{Free energy functional} --- In this subsection, we use OVP to derive the thermodynamically consistent governing equations~\cite{Qian2006JFM,Doi1992,Doi2011OVP,Doi2013,Doi2015,Doi2021,Xu2017,Xu2021,Xu2022}. Following Sec.~\ref{sec:DDM}, we adopt the isotropic R\"atz--Voigt scalar embedding, $\mathcal G_c=\nabla$, and take the
embedded free energy to be
\begin{equation}\label{Eq:SurfPSD-RigidSurf-Fc}
\mathcal{F}_c[c,\psi]
=\gamma_{\mathrm L}\int_\Omega \delta_\epsilon 
\left[\epsilon_c^{-1} f(c)+\frac{\epsilon_c}{2}|\nabla c|^2\right]dV,
\end{equation}
where $f(c)=18c^2(1-c)^2$. With this normalization, $\gamma_{\mathrm L}$ is the sharp line tension associated with the one-dimensional composition interface in the subsequent $\epsilon_c\to0$ limit. The parameter $\epsilon_c$ sets the width of the composition transition on the surface and is independent of the diffuse-surface thickness $\epsilon$.

Because $\psi$ and hence $\delta_\epsilon $ are fixed in this subsection, the free-energy rate can be written as $\dot{\mathcal F}_c
=\int_\Omega \delta_\epsilon \hat\mu_c\,\partial_t c\,dV
=\int_\Omega \hat\mu_c\,\partial_t(\delta_\epsilon  c)\,dV$. 
The weighted chemical potential is the scalar $\mathcal G_c=\nabla$ specialization of Eq.~\eqref{Eq:DDM-mu}:
\begin{equation}\label{Eq:SurfPSD-RigidSurf-muc}
\delta_\epsilon \hat{\mu}_c =
\gamma_{\mathrm L}\left[\epsilon_c^{-1}\delta_\epsilon  f'(c)
-\epsilon_c\nabla\!\cdot(\delta_\epsilon \nabla c) \right],
\end{equation}
with $f'(c)=df/dc$. The integrations by parts assume either that the diffuse layer does not intersect $\partial\Omega$, or that the natural weighted boundary condition $\hat{\boldsymbol n}_{\partial\Omega}\!\cdot\delta_\epsilon \nabla c=0$ is imposed. Pointwise division by $\delta_\epsilon $ is only meaningful inside the diffuse layer, or after introducing the small regularization discussed in Sec.~\ref{sec:DDM}.

\textbf{Dissipation functional} --- While the physically relevant membrane concentration $c$ is conserved in the applications below, it is useful to record both the nonconserved Allen--Cahn and the conserved Cahn--Hilliard limits for later reference~\cite{AllenCahn1979,CahnHilliard1958}. The embedded dissipation functional depends on the specific concentration dynamics as follows.
\begin{enumerate}[label=(\roman*)]
\item Allen--Cahn surface dynamics (internal nonconserved scalar order parameter $c$). For the Allen--Cahn case, $c$ is treated as a scalar internal nonconserved order parameter, \emph{i.e.}, a class-(ii) field in the notation of
Sec.~\ref{sec:DDM}, rather than as a conserved surface concentration. Since the surface is fixed, the dissipative rate is simply $\dot{\mathbb B}_c=\partial_t c$. The molecular field conjugate to this rate is $\delta_\epsilon  h_c =
- {\delta\mathcal F_c}/{\delta c} = -\delta_\epsilon \hat\mu_c $, where $\hat\mu_c$ is defined in Eq.~\eqref{Eq:SurfPSD-RigidSurf-muc}. The dissipation functional is
\begin{equation}\label{Eq:SurfPSD-RigidSurf-ACPhi}
\Phi[\partial_t c]
=\int_\Omega \delta_\epsilon  \left[
\frac{(\partial_t c)^2}{2\tilde{\Gamma}_c}\right]\,dV,
\end{equation}
where $\tilde{\Gamma}_c>0$ is a local kinetic mobility. 

\textbf{Rayleighian and dynamic equations} --- Minimizing
$\mathcal R=\dot{\mathcal F}_c+\Phi$ with respect to
$\partial_t c$ gives 
\begin{equation}\label{Eq:SurfPSD-RigidSurf-DynEqn1}
\delta_\epsilon \partial_t c = -\tilde{\Gamma}_c\delta_\epsilon \hat{\mu}_c =
-\Gamma_c\left[\epsilon_c^{-1}\delta_\epsilon  f'(c)
-\epsilon_c\nabla\!\cdot(\delta_\epsilon \nabla c)\right],
\end{equation}
where $\Gamma_c\equiv\gamma_{\mathrm L}\tilde{\Gamma}_c$. 
Thus, on a fixed rigid surface, the class-(ii) Allen--Cahn construction is algebraically identical to the source-relaxation form obtained from Eq.~\eqref{Eq:SurfPSD-RigidSurf-bal} with $J_c=0$. The class-(ii) interpretation is nevertheless preferable here, because $c$ is not a conserved surface density in the Allen--Cahn model.
 
\item Cahn--Hilliard and viscous Cahn--Hilliard surface dynamics (conserved $c$). Setting $k_c=0$ in Eq.~\eqref{Eq:SurfPSD-RigidSurf-bal} gives the conserved weighted balance
\begin{equation}\label{Eq:SurfPSD-RigidSurf-CHct}
\delta_\epsilon \partial_t c =
-\nabla\!\cdot\!\left(\delta_\epsilon \boldsymbol J_c\right).
\end{equation}
With the no-flux condition $\hat{\boldsymbol n}_{\partial\Omega}\!\cdot \delta_\epsilon \boldsymbol J_c=0$, this equation conserves $\int_\Omega\delta_\epsilon  c\,dV$. If only flux dissipation is retained, one obtains the standard conserved Cahn--Hilliard reduction of Sec.~\ref{sec:DDM}.  Following the rigid-surface diffuse-domain formulation of R\"atz and Voigt~\cite{RatzVoigt2006}, we additionally introduce the viscous Cahn--Hilliard (VCH) regularization. In this approach, the conserved dynamics is augmented by a local penalization term acting on $\partial_t c$:
\begin{equation}\label{Eq:SurfPSD-RigidSurf-CHPhi}
\Phi[\boldsymbol J_c,\partial_t c] =\int_\Omega \delta_\epsilon 
\left[\frac{|\boldsymbol J_c|^2}{2M_c} +\frac{(\partial_t c)^2}{2\tilde{\Gamma}_c}\right]dV.
\end{equation}
Here $M_c>0$ and $\tilde{\Gamma}_c>0$ are taken as constants. The second term is a viscous, or pseudo-parabolic, relaxation of the chemical-potential response that mimics viscoelastic effects. It is one minimal Onsager-consistent regularization within the present choice of rates; more detailed viscoelastic descriptions would require additional internal variables and Onsager couplings~\cite{Doi1992,Doi2021}.

\textbf{Rayleighian and dynamic equations} --- In the conserved case, $\partial_t c$ and $\boldsymbol J_c$ are not independent rates; they are linked by Eq.~\eqref{Eq:SurfPSD-RigidSurf-CHct}. The constrained Rayleighian should therefore be written as
\begin{equation}\label{Eq:SurfPSD-RigidSurf-RayleighianVCH}
\mathcal R[\boldsymbol J_c,\partial_t c,\lambda_c]=
\dot{\mathcal F}_c+\Phi+\int_\Omega \lambda_c\left[
\delta_\epsilon \partial_t c +\nabla\!\cdot(\delta_\epsilon  \boldsymbol J_c)\right]dV,
\end{equation}
where $\lambda_c$ is a Lagrange multiplier. Variation with respect to $\partial_t c$ and $\boldsymbol J_c$ gives $\lambda_c=-\hat{\mu}_c-\tilde{\Gamma}_c^{-1}\partial_t c$ and $\boldsymbol J_c=-M_c\nabla\!\left(\hat{\mu}_c+\tilde{\Gamma}_c^{-1}\partial_t c\right)$. Thus, the physical viscous chemical potential is $\bar{\mu}_c^{\,\mathrm{phys}}=\hat{\mu}_c+\tilde{\Gamma}_c^{-1}\partial_t c$. For comparison with the usual R\"atz--Voigt notation, we introduce the rescaled chemical potential $\bar{\mu}_c\equiv\bar{\mu}_c^{\,\mathrm{phys}}/\gamma_{\mathrm L}$. Then the VCH system becomes
\begin{subequations}\label{Eq:SurfPSD-RigidSurf-DynEqn}
\begin{equation}\label{Eq:SurfPSD-RigidSurf-DynEqn-a}
\delta_\epsilon \partial_t c=
\nu_c\nabla\!\cdot\!\left(\delta_\epsilon \nabla\bar{\mu}_c\right),
\end{equation}
\begin{equation}\label{Eq:SurfPSD-RigidSurf-DynEqn-b}
\delta_\epsilon \bar{\mu}_c=
\epsilon_c^{-1}\delta_\epsilon  f'(c)
-\epsilon_c\nabla\!\cdot(\delta_\epsilon \nabla c)
+\alpha_c\epsilon_c\delta_\epsilon \partial_t c .
\end{equation}
\end{subequations}
Here $\nu_c\equiv\gamma_{\mathrm L}M_c$, $\Gamma_c\equiv\gamma_{\mathrm L}\tilde{\Gamma}_c$, and $\alpha_c\equiv\Gamma_c^{-1}\epsilon_c^{-1}$.
Equations~\eqref{Eq:SurfPSD-RigidSurf-DynEqn-a}--%
\eqref{Eq:SurfPSD-RigidSurf-DynEqn-b} are the weighted isotropic diffuse-domain VCH equations of R\"atz--Voigt type~\cite{RatzVoigt2006}, up to notation and coefficient scaling. The boundary conditions are $\hat{\boldsymbol n}_{\partial\Omega}\!\cdot
\delta_\epsilon \nabla c=0$ from the free-energy variation and
$\hat{\boldsymbol n}_{\partial\Omega}\!\cdot
\delta_\epsilon \nabla\bar{\mu}_c=0$ for constant $M_c$, equivalently $\hat{\boldsymbol n}_{\partial\Omega}\!\cdot
\delta_\epsilon \boldsymbol J_c=0$.
\end{enumerate}

For the fixed rigid surface considered here, both reductions possess the expected passive Onsager dissipation law under the stated boundary conditions. In the Allen--Cahn case, $\dot{\mathcal F}_c= -\int_\Omega\delta_\epsilon \tilde{\Gamma}_c|\hat{\mu}_c|^2\,dV
=-2\Phi\le 0$. 
In the VCH case, $\dot{\mathcal F}_c=-\int_\Omega \delta_\epsilon 
\left[ {M_c}^{-1}{|\boldsymbol J_c|^2}+{\tilde{\Gamma}_c^{-1}}{(\partial_t c)^2}\right]dV=-2\Phi\le 0$. 
The sharp-surface limit invoked here is the diffuse-domain limit $\epsilon\to0$ at fixed composition-interface width $\epsilon_c$. Under the coarea and closest-point-extension assumptions of Sec.~\ref{sec:DDM}, Eq.~\eqref{Eq:SurfPSD-RigidSurf-DynEqn} reduces to the surface VCH system
\begin{equation}
\partial_t c_\Gamma =
\nu_c\Delta_\Gamma\bar{\mu}_{c\Gamma},\qquad
\bar{\mu}_{c\Gamma}=\epsilon_c^{-1}f'(c_\Gamma)
-\epsilon_c\Delta_\Gamma c_\Gamma+\Gamma_c^{-1}\partial_t c_\Gamma.
\end{equation}
A further limit $\epsilon_c\to 0$ would be a separate sharp-composition-interface limit associated with line tension. Finally, the use of $|\nabla c|^2$ in Eq.~\eqref{Eq:SurfPSD-RigidSurf-Fc} is deliberate: it is the scalar R\"atz--Voigt embedding $\mathcal G_c=\nabla$ adopted in Sec.~\ref{sec:DDM}, generates the weighted bulk operator $\nabla\!\cdot(\delta_\epsilon \nabla\,\cdot)$, and provides finite-$\epsilon$ normal-extension regularization. Therefore, no additional weak normal diffusion is needed for this scalar fixed-surface model~\cite{RatzVoigt2006,LiLowengrubRatzVoigt2009}. 

\subsection{Phase separation on deformable vesicle surfaces} \label{sec:SurfPSD-DeformSurf}

\textbf{State variables and constraints} --- We now extend the fixed-surface scalar formulation of Sec.~\ref{sec:SurfPSD-RigidSurf} to a deformable
multicomponent vesicle~\cite{Seifert1993CurvatureSegregation,
JulicherLipowsky1993DomainBudding,JulicherLipowsky1996ShapeTransformations,
Taniguchi1996ShapeDeformation}, whose geometry is represented by a
time-dependent bulk phase field $\psi(\boldsymbol x,t)$. Once $\psi$ is
dynamic, the fixed-surface identities used in
Eqs.~\eqref{Eq:DDM-dFdt} and \eqref{Eq:DDM-bal-local2} no longer apply
directly, because the diffuse surface density $\delta_\epsilon [\psi]$ also
evolves.

In this subsection, we consider a reduced nonhydrodynamic deformable-interface DDM model for scalar surface phase separation. No independent hydrodynamic membrane velocity is solved for. Instead, the normal motion of the diffuse
surface is inferred kinematically from the evolution of the level sets of $\psi$, while the tangential material velocity is left unspecified and is set to zero in this reduced model. The fundamental conserved concentration
variable is the diffuse areal quantity $\delta_\epsilon [\psi]c$, rather than the bulk extension $c$ alone. This distinction is essential: for a moving surface, a conserved surface density is transported by the
surface velocity, whereas $c$ itself is only a diffuse-layer extension of the surface concentration. The fully hydrodynamic version, in which the material velocity is determined from momentum balance, is introduced later in
Sec.~\ref{sec:MPF-vesicle}.

For the membrane concentration $c$, we keep the isotropic R\"atz--Voigt scalar embedding used in Secs.~\ref{sec:DDM} and
\ref{sec:SurfPSD-RigidSurf}, namely $\mathcal G_c=\nabla$. Thus, the finite-$\epsilon$ concentration flux is the diffuse-layer flux generated by the weighted bulk operator $\nabla\!\cdot(\delta_\epsilon \nabla\,\cdot)$; it becomes the tangential surface flux only in the sharp-surface limit under
the usual closest-point-extension assumptions. By contrast,  $\psi$ is the bulk diffuse-interface variable representing the vesicle geometry and evolves through an Allen--Cahn- or Cahn--Hilliard-type bulk metric.

\textbf{Free energy functional} --- The total embedded free energy is
\begin{equation}
\mathcal F[c,\psi]=\mathcal F_c[c,\psi]+\mathcal F_\psi[c,\psi],
\end{equation}
where $\mathcal F_c$ is still given by Eq.~\eqref{Eq:SurfPSD-RigidSurf-Fc}. For the vesicle-shape contribution we use the diffuse-interface bending/tension  functional~\cite{Helfrich1973,WangDu2008MultiComponentVesicles,LowengrubRatzVoigt2009Vesicles}
\begin{equation}\label{Eq:SurfPSD-DeformSurf-Fpsi}
\mathcal{F}_\psi[c,\psi] = \int_{\Omega} \left\{ \frac{K_{\mathrm b}(c)}{2\epsilon}Q_\psi^2
+ \gamma_\psi(c)\delta_\epsilon [\psi]\right\}dV,
\end{equation}
with $Q_\psi \equiv -\epsilon^{-1}G'(\psi) +6\psi(1-\psi)\kappa_0(c) +\epsilon\nabla^2\psi$. 
Here $\epsilon$ is the diffuse-surface thickness of the vesicle phase field, whereas $\epsilon_c$ in $\mathcal F_c$ is the width of the composition transition on the membrane. The functions $K_{\mathrm b}(c)$, $\kappa_0(c)$, and $\gamma_\psi(c)$ denote, respectively, the effective bending-rigidity parameter, spontaneous curvature, and energetic areal tension density of the reduced vesicle model.
The same diffuse surface density $\delta_\epsilon [\psi]$ appears in the tension term of $\mathcal F_\psi$ and in the concentration free energy $\mathcal F_c$. In the material-balance formulation below, we take $\delta_\epsilon [\psi]= B_4(\psi,\nabla\psi)$, because this choice makes the normal transport associated with the evolving phase
field explicit at finite $\epsilon$.
Since $K_{\mathrm b}$, $\kappa_0$, and $\gamma_\psi$ may depend on $c$, the concentration chemical potential receives contributions from both $\mathcal F_c$ and $\mathcal F_\psi$:
\begin{equation}\label{Eq:SurfPSD-DeformSurf-muc}
\delta_\epsilon [\psi]\hat\mu_c \equiv \frac{\delta \mathcal F}{\delta c} = \gamma_{\mathrm L}\left[ \epsilon_c^{-1}\delta_\epsilon [\psi]f'(c)-\epsilon_c\nabla\!\cdot\!\left(\delta_\epsilon [\psi]\nabla c\right)\right]+ \frac{\delta \mathcal F_\psi}{\delta c}.
\end{equation}
At fixed $\psi$, ${\delta \mathcal F_\psi}/{\delta c}
=\frac12 \epsilon^{-1}{K_{\mathrm b}'(c)} Q_\psi^2+\epsilon^{-1}{K_{\mathrm b}(c)} Q_\psi\,6\psi(1-\psi)\kappa_0'(c)+\gamma_\psi'(c) \delta_\epsilon [\psi]$. As in Sec.~\ref{sec:SurfPSD-RigidSurf}, pointwise division by $\delta_\epsilon [\psi]$ is meaningful only inside the diffuse layer, or after introducing the small regularization described in Sec.~\ref{sec:DDM}.

\textbf{Dissipation functional} --- For a conserved membrane composition, the fundamental diffuse conserved variable is $\delta_\epsilon [\psi]c$. 
The motion of the diffuse surface is determined by the level-set kinematics of $\psi$. Thus, inside the diffuse layer, the associated normal level-set velocity is
\begin{equation}\label{Eq:SurfPSD-DeformSurf-Vpsi}
\boldsymbol V_\psi = V_{\psi n}\hat{\boldsymbol n}_\epsilon=-\frac{\partial_t\psi}{|\nabla\psi|^2}\nabla\psi.
\end{equation}
Equation~\eqref{Eq:SurfPSD-DeformSurf-Vpsi} is used only in the diffuse layer; in computations, the denominator should be regularized where $|\nabla\psi|$ is small.
The material balance-law form consistent with Sec.~\ref{sec:DDM} is then
\begin{equation}\label{Eq:SurfPSD-DeformSurf-material-balance}
\partial_t\!\left(\delta_\epsilon [\psi]c\right)
+\nabla\!\cdot\!\left(\delta_\epsilon [\psi]c\,\boldsymbol V_\psi \right)=-\nabla\!\cdot\!\left( \delta_\epsilon [\psi]\boldsymbol J_c\right),
\end{equation}
where $\boldsymbol J_c$ is the non-convective concentration flux. This equation is the finite-$\epsilon$ counterpart of the sharp moving-surface balance $\dot c_\Gamma+c_\Gamma\nabla_\Gamma\!\cdot\boldsymbol V_{\Gamma}=-\nabla_\Gamma\!\cdot\boldsymbol J_{c\Gamma}$. 
For purely normal motion, this becomes $\partial_t^\circ c_\Gamma+ 2{\mathcal H}_\Gamma V_{\Gamma n}c_\Gamma=-\nabla_\Gamma\!\cdot\boldsymbol J_{c\Gamma}$, where
$\partial_t^\circ c_\Gamma$ denotes the normal time derivative. 
The reduced Eulerian DDM balance equation
\begin{equation}\label{Eq:SurfPSD-DeformSurf-ct}
\partial_t(\delta_\epsilon [\psi]c)= -\nabla\!\cdot(\delta_\epsilon [\psi]\boldsymbol J_c^E)
\end{equation}
is recovered from Eq.~\eqref{Eq:SurfPSD-DeformSurf-material-balance} only if $\delta_\epsilon \boldsymbol J_c^E$ is understood as the total Eulerian flux, $\delta_\epsilon \boldsymbol J_c^E=\delta_\epsilon \boldsymbol J_c+\delta_\epsilon  c\boldsymbol V_\psi$. This equation conserves $\int_\Omega\delta_\epsilon [\psi]c\,dV$ under the weighted no-flux condition $\hat{\boldsymbol n}_\Omega\!\cdot
\delta_\epsilon [\psi]\boldsymbol J_c^E=0$. It is the Eulerian deformable-interface counterpart of the fixed-surface balance
Eq.~\eqref{Eq:SurfPSD-RigidSurf-CHct}. The Onsager dissipation, however, is associated with the nonconvective part $\boldsymbol J_c$.

Following the classical deformable-interface construction
\cite{TeigenLiLowengrubWangVoigt2009DeformableInterface,
LowengrubRatzVoigt2009Vesicles}, we choose $\delta_\epsilon [\psi]= B_4(\psi,\nabla\psi)=
\frac{\epsilon}{2}|\nabla\psi|^2+\epsilon^{-1}G(\psi)$, 
for which the identity $\delta_\epsilon  c\boldsymbol V_\psi  \simeq -\epsilon c\nabla\psi\,\partial_t\psi$ holds for the standard one-dimensional interfacial profile with $\delta_\epsilon \simeq \epsilon|\nabla\psi|^2$. 
Expanding Eq.~\eqref{Eq:SurfPSD-DeformSurf-ct} then gives 
\begin{equation}\label{Eq:SurfPSD-DeformSurf-ct-expanded}
\delta_\epsilon \partial_t c + A_4 (c,\psi)\partial_t\psi = - \nabla\!\cdot\! \left(\delta_\epsilon  \boldsymbol J_c\right),
\end{equation}
where $A_4 (c,\psi) \equiv -\epsilon\nabla\!\cdot(c\nabla\psi)+\epsilon^{-1}cG'(\psi)$and $\delta_\epsilon  \boldsymbol J_c =\delta_\epsilon (\boldsymbol J_c^E - c\boldsymbol V_\psi) \simeq \delta_\epsilon \boldsymbol J_c^E+\epsilon c\nabla\psi\,\partial_t\psi$. 
Using Eq.~\eqref{Eq:SurfPSD-DeformSurf-ct-expanded} to eliminate $\partial_t c$, the free-energy rate becomes
\begin{equation}\label{Eq:SurfPSD-DeformSurf-Fdot1}
\dot{\mathcal F} = \int_\Omega \left[\delta_\epsilon  \boldsymbol J_c\!\cdot\nabla\hat\mu_c+
\bar{\mu}_\psi\,\partial_t\psi\right]dV.
\end{equation} 
where the chemical potential $\hat\mu_c$ is defined in Eq.~\eqref{Eq:SurfPSD-DeformSurf-muc} and the shifted (vesicle) geometry chemical potential associated with $\psi$ is defined by
\begin{equation}
\bar{\mu}_\psi\equiv \left. \frac{\delta\mathcal F}{\delta\psi}\right|_{\delta_\epsilon  c}
=\left.\frac{\delta\mathcal F}{\delta\psi}\right|_c - A_4 (c,\psi)\hat\mu_c,  
\end{equation}
at fixed, conserved concentration $\delta_\epsilon  c$.
\begin{enumerate}[label=(\roman*)]
\item \emph{Allen--Cahn dynamics for nonconserved $\psi$.}
We first consider a nonconserved relaxational dynamics for the vesicle phase field $\psi$. This case is useful as a standard deformable-interface phase-field/DDM model~\cite{DuLiu2005,WangDu2008MultiComponentVesicles,Campelo2006dynamic,Campelo2021gaussian}, but it does not conserve the enclosed volume of a closed vesicle unless an additional volume constraint or penalty is imposed. The resulting sharp-interface geometry metric leads to a purely local normal drag law~\cite{CaetanoElliottGrasselliPoiatti2026Biomembrane,BarrettGarckeNuernberg2018GradientFlowBiomembranes}. 
In this case, the dissipation functional is
\begin{equation}\label{Eq:SurfPSD-DeformSurf-Phi1}
\Phi[\boldsymbol J_c,\partial_t\psi]=
\int_\Omega\left[\delta_\epsilon \frac{| \boldsymbol J_c|^2}{2M_c}+\frac{(\partial_t\psi)^2}{2\Gamma_\psi}\right]dV,
\end{equation}
with constant mobility coefficients $M_c,\,\Gamma_\psi>0$. 

\textbf{Rayleighian and dynamic equations} --- Using Eqs.~\eqref{Eq:SurfPSD-RigidSurf-Fc}, \eqref{Eq:SurfPSD-DeformSurf-Fpsi}, \eqref{Eq:SurfPSD-DeformSurf-ct-expanded}, \eqref{Eq:SurfPSD-DeformSurf-Phi1}, and minimizing the Rayleighian $\mathcal R[\boldsymbol J_c,\partial_t\psi]
= \dot{\mathcal F} + \Phi[\boldsymbol J_c,\partial_t\psi]$ gives
\begin{subequations}\label{Eq:SurfPSD-DeformSurf-DynEqn1}
\begin{equation}\label{Eq:SurfPSD-DeformSurf-DynEqn1ct}
\delta_\epsilon \partial_t c + A_4 (c,\psi)\partial_t\psi =\nabla\!\cdot\! \left(M_c\delta_\epsilon \nabla\hat\mu_c\right),
\end{equation}
\begin{equation}\label{Eq:SurfPSD-DeformSurf-DynEqn1phit}
\partial_t\psi = -\Gamma_\psi\left[\left.\frac{\delta\mathcal F}{\delta\psi}\right|_c-A_4 (c,\psi) \hat\mu_c\right].
\end{equation}
Equivalently, introducing the rescaled coefficients $\beta_c\equiv \epsilon M_c$ and $\beta_\psi\equiv \epsilon\Gamma_\psi$, Eqs.~\eqref{Eq:SurfPSD-DeformSurf-DynEqn1ct} and \eqref{Eq:SurfPSD-DeformSurf-DynEqn1phit} can be written in the same scaled form as the classical deformable-interface DDM system~\cite{LowengrubRatzVoigt2009Vesicles}.  
\end{subequations}

\item \emph{Cahn--Hilliard dynamics for conserved $\psi$.} We next consider conserved bulk dynamics for the geometry phase field, 
\begin{equation}\label{Eq:SurfPSD-DeformSurf-phit}
\partial_t\psi=-\nabla\!\cdot\boldsymbol J_\psi.
\end{equation}
Under the no-flux condition $\hat{\boldsymbol n}_\Omega\!\cdot\boldsymbol J_\psi=0$, this dynamics
conserves $\int_\Omega\psi\,dV$, which is the diffuse analogue of enclosed volume conservation for a closed impermeable vesicle. The resulting sharp-interface geometry metric leads to a volume-preserving normal relaxation law and is therefore different from the local Allen--Cahn metric above~\cite{CaetanoElliottGrasselliPoiatti2026Biomembrane,BarrettGarckeNuernberg2018GradientFlowBiomembranes}.
In this case, the dissipation functional is chosen as
\begin{equation}\label{Eq:SurfPSD-DeformSurf-Phi2}
\Phi[\boldsymbol J_c,\boldsymbol J_\psi]=
\int_\Omega\left[\delta_\epsilon  \frac{|\boldsymbol J_c|^2}{2M_c}+\frac{|\boldsymbol J_\psi|^2}{2M_\psi}\right]dV,
\end{equation}
with constant mobility coefficients, $M_c,\,M_\psi>0$. 

\textbf{Rayleighian and dynamic equations} --- Using Eqs.~\eqref{Eq:SurfPSD-RigidSurf-Fc}, \eqref{Eq:SurfPSD-DeformSurf-Fpsi}, \eqref{Eq:SurfPSD-DeformSurf-phit}--\eqref{Eq:SurfPSD-DeformSurf-Phi2}, minimizing the Rayleighian gives
\begin{subequations}\label{Eq:SurfPSD-DeformSurf-DynEqn2}
\begin{equation}\label{Eq:SurfPSD-DeformSurf-DynEqn2ct}
\delta_\epsilon \partial_t c+A_4 (c,\psi)\partial_t\psi=
\nabla\!\cdot\!\left(M_c\delta_\epsilon \nabla\hat\mu_c\right),
\end{equation}
\begin{equation}\label{Eq:SurfPSD-DeformSurf-DynEqn2phit}
\partial_t\psi = M_\psi\nabla^2 \left[ \left.\frac{\delta\mathcal F}{\delta\psi}\right|_c-
A_4 (c,\psi)\hat\mu_c\right].
\end{equation}
\end{subequations} 
Equivalently, with $\beta_c\equiv \epsilon M_c$ and $\tilde{\beta}_\psi\equiv \epsilon M_\psi$, Eq.~\eqref{Eq:SurfPSD-DeformSurf-DynEqn2phit} may be written in the rescaled form $\epsilon\partial_t\psi-\tilde{\beta}_\psi\nabla^2(\cdots)=0$. 
\end{enumerate}
Note that, in the above formulation, the integrations by parts assume either periodic boundary conditions, or that the diffuse layer remains away from $\partial\Omega$, or boundary conditions that remove all outer-boundary terms. For the concentration field, this includes the weighted natural condition $\hat{\boldsymbol n}_\Omega\!\cdot
\delta_\epsilon \nabla c=0$ and the weighted no-flux condition
$\hat{\boldsymbol n}_\Omega\!\cdot M_c\delta_\epsilon \nabla\hat\mu_c=0$. For the conserved-$\psi$ case, one also imposes $\hat{\boldsymbol n}_\Omega\!\cdot
M_\psi\nabla\left.({\delta\mathcal F}/{\delta\psi}\right|_{\delta_\epsilon  c})=0$. Because $\mathcal F_\psi$ contains $\nabla^2\psi$, the natural boundary conditions for $\psi$ require, in addition, conditions that eliminate the fourth-order bending boundary terms; for example, in an auxiliary-field implementation with $Q_\psi$ defined above, one may impose $\hat{\boldsymbol n}_\Omega\!\cdot\nabla\psi=0$ together with the corresponding no-boundary-work condition on $K_{\mathrm b}(c)Q_\psi$, or an equivalent periodic/split-form set of boundary conditions.

Finally, Eqs.~\eqref{Eq:SurfPSD-DeformSurf-DynEqn1} and
\eqref{Eq:SurfPSD-DeformSurf-DynEqn2} are reduced time-dependent-$\psi$ extensions of the fixed-surface isotropic scalar DDM model of Sec.~\ref{sec:SurfPSD-RigidSurf}. In both cases, the membrane concentration is still embedded through the full-gradient scalar R\"atz--Voigt operator $\nabla\!\cdot(\delta_\epsilon [\psi]\nabla\,\cdot)$. The finite-$\epsilon$ flux is therefore a diffuse-layer flux; its sharp-surface limit is the usual tangential surface flux under the coarea and closest-point-extension assumptions of Sec.~\ref{sec:DDM}. 
The explicit $c$--$\psi$ coupling terms have two origins: the
$c$-dependence of the total free energy, including
$K_{\mathrm b}(c)$, $\kappa_0(c)$, and $\gamma_\psi(c)$, and the fact that the conserved concentration variable is $\delta_\epsilon [\psi]c$, not $c$ itself. Equivalently, the geometry force appearing in the $\psi$-equation is the variational derivative at fixed $\delta_\epsilon  c$. 
The two geometry metrics considered above differ at finite $\epsilon$: the Allen--Cahn-$\psi$ model gives a local $L^2$ relaxation of the vesicle phase field, whereas the Cahn--Hilliard-$\psi$ model gives an $H^{-1}$ relaxation and conserves $\int_\Omega\psi\,dV$. In both cases, however, the normal level-set velocity is determined kinematically by
Eq.~\eqref{Eq:SurfPSD-DeformSurf-Vpsi}. The tangential velocity is not specified in this reduced non-hydrodynamic setting.
Using $\delta_\epsilon = B_4$ in both cases makes the concentration balance compatible with the material balance-law structure of Sec.~\ref{sec:DDM}: the conserved surface amount is $\delta_\epsilon  c$, the normal transport is carried by
$\boldsymbol V_\psi$, and the Onsager dissipation is assigned to the non-convective flux $\boldsymbol J_c$. 
Under the stated boundary conditions, both systems satisfy a passive Onsager dissipation law for the total free energy. For the Allen--Cahn-$\psi$ case, $\dot{\mathcal F} = -\int_\Omega \left[\delta_\epsilon  M_c|\nabla\hat\mu_c|^2+ \Gamma_\psi|\bar{\mu}_\psi|^2\right]dV=-2\Phi\le0$. For the conserved Cahn-Hilliard-$\psi$ case, $\dot{\mathcal F}=-\int_\Omega\left[\delta_\epsilon  M_c|\nabla\hat\mu_c|^2 + M_\psi|\nabla\bar{\mu}_\psi|^2\right]dV = -2\Phi\le0$. 
A complete matched-asymptotic derivation of the fully coupled sharp-interface limit is not claimed here; the intended correspondence is the standard sharp-surface diffuse-domain limit for scalar surface transport together with the chosen phase-field metric for vesicle-shape relaxation~\cite{RatzVoigt2006,LiLowengrubRatzVoigt2009,TeigenLiLowengrubWangVoigt2009DeformableInterface,LowengrubRatzVoigt2009Vesicles,CaetanoElliottGrasselliPoiatti2026Biomembrane,WinterLiuZiepkeDadunashviliFrey2025PhaseSeparation}.

\section{Interfacial hydrodynamics in multiphase flows (scalar \& vector fields)}\label{sec:MPF} 

\subsection{Two-phase hydrodynamics near and on rigid surfaces} \label{sec:MPF-2Phase} 

\begin{figure}[htbp]  
  \centering
  \includegraphics[width=0.8\columnwidth]{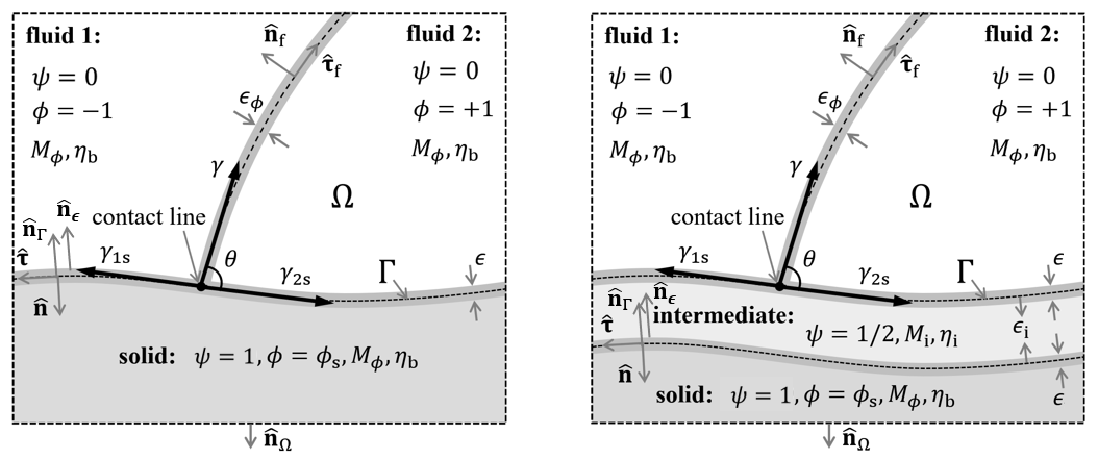}
  \caption {Schematic illustration of a moving contact line in immiscible two-phase flow on a rigid solid surface. (left) Diffuse-domain method (DDM): the two fluids are distinguished by the phase field $\phi=\pm1$, while the rigid wall is represented by the diffuse wall phase field $\psi$. The contact line forms with contact angle $\theta$ and interfacial tensions $\gamma$, $\gamma_{1s}$, and $\gamma_{2s}$. (right) Diffuse-resistance-domain (DRD) method (a variant of DDM): an intermediate thin layer centered near $\psi=1/2$ with low viscosity $\eta_i$ ($<\eta_{\mathrm f}\ll\eta_{\mathrm s}$) and mobility $M_i$ ($<M_{\mathrm f}$, and $M_{\mathrm s}\ll M_{\mathrm f}$) is introduced to regularize the contact-line dynamics of Qian--Wang--Sheng type~\cite{Qian2006JFM,Qian2008JFM}. The intermediate layer has a small thickness $\epsilon_i$ (exaggerated for illustration). }
  \label{Fig:TwoPhaseFlow}
\end{figure}

We now consider immiscible two-phase hydrodynamics near a fixed rigid solid boundary, as sketched in Fig.~\ref{Fig:TwoPhaseFlow}. This subsection is not a pure surface-field specialization of Sec.~\ref{sec:DDM}; rather, it is a mixed bulk/surface embedding. 

\textbf{State variables and constraints} --- The binary-fluid phase field $\phi(\boldsymbol r,t)\in[-1,+1]$ is a bulk order parameter defined in the fluid region and extended to the fixed computational domain. The rigid solid is represented by a fixed wall phase field $\psi(\boldsymbol r)\in[0,1]$. Here, we take the convention that $\psi\simeq 1$ in the solid region and $\psi\simeq 0$ in the fluid region, so that $1-\psi$ is the diffuse fluid mask and $\delta_\epsilon [\psi]$ is the diffuse fluid--solid surface measure. This convention is the rigid-wall specialization of the orientation convention introduced in Sec.~\ref{sec:DDM}, with the solid treated as the interior region. The labels in Fig.~\ref{Fig:TwoPhaseFlow} should therefore be read as fluid: $\psi=0$ and solid: $\psi=1$. 
Unlike the scalar fields in Sec.~\ref{sec:SurfPSD}, $\phi$ is not a surface-confined state variable. The full gradient $\nabla\phi$ in the Cahn--Hilliard energy below is the physical bulk gradient of the fluid--fluid phase field. No tangential state-space projection is imposed on $\phi$; the wall contribution is instead localized by $\delta_\epsilon [\psi]$.

\textbf{Free energy functional} --- The total free energy is therefore taken as a masked bulk Cahn--Hilliard energy plus a wall energy localized by $\delta_\epsilon [\psi]$~\cite{Jacqmin2000,Qian2006JFM,AlandLowengrubVoigt2010,AlandWohlgemuth2023,GaoLiXuDRD2025}:
\begin{equation}\label{Eq:theory-DRD-F}
\mathcal F[\phi(\boldsymbol r,t),\psi(\boldsymbol r)]=
\frac{3\sqrt{2}\gamma_\phi}{4}
\int_\Omega \left[(1-\psi) \left(
\epsilon_\phi^{-1}f_{\mathrm b}(\phi)
+\frac{\epsilon_\phi}{2}|\nabla\phi|^2
\right) +\delta_\epsilon [\psi]f_{\mathrm s}(\phi) \right]dV.
\end{equation}
Here $f_{\mathrm b}(\phi)=\tfrac14(1-\phi^2)^2$ is the standard double-well bulk free-energy density and $\gamma_\phi$ is the fluid-fluid interfacial tension parameter. For the fluid-solid interfacial energy density $f_{\mathrm s}(\phi)$, two common choices are the quadratic wetting potential $f_{\mathrm s}(\phi)=\tfrac12\alpha(\phi-h_1)^2-h_2\phi$
and the contact-angle form $f_{\mathrm s}(\phi)=-\tfrac{\sqrt2}{6}\cos\theta_{\mathrm s} (\phi^3-3\phi)$,
with $\alpha$, $h_1$, $h_2$, and $\theta_{\mathrm s}$ constant wetting parameters~\cite{Jacqmin2000,Qian2006JFM, Qian2008JFM,GaoLiXuDRD2025,YuQianWang2025}. In the present subsection, $\delta_\epsilon [\psi]$ may be any normalized diffuse surface density from Sec.~\ref{sec:DDM}; in practice one often takes $B_3=\epsilon|\nabla\psi|^2$, where $\epsilon$ now denotes the thickness of the diffuse fluid-solid layer. 
As in Secs.~\ref{sec:DDM} and \ref{sec:SurfPSD}, pointwise division by the mask or by $\delta_\epsilon [\psi]$ requires a numerical floor if the fields are evaluated away from their physical support~\cite{RatzVoigt2006}. In computations, one may replace $\delta_\epsilon [\psi]\rightarrow \delta_\epsilon [\psi]+\delta_{\mathrm reg}$ and $(1-\psi) \rightarrow (1-\psi)+\bar{\delta}_{\mathrm reg}\psi$, where $\delta_{\mathrm reg}$ has the same physical dimension as $\delta_\epsilon $ and is small compared with $O(\epsilon^{-1})$, while $0<\bar{\delta}_{\mathrm reg}\ll 1$ is dimensionless. These regularizations should be understood as finite-$\epsilon$ numerical devices, not as changes to the intended sharp-wall model. 

The rate of change of the total free energy is given by $\dot{\mathcal F}=\int_\Omega (1-\psi)\hat\mu_\phi\,\partial_t\phi\,dV$
for fixed $\psi$, in direct analogy with Eq.~\eqref{Eq:DDM-dFdt}, where the masked chemical potential reads
\begin{equation}\label{Eq:theory-DRD-muEq}
(1-\psi)\,\hat\mu_\phi\equiv\frac{\delta\mathcal F}{\delta\phi}
= \frac{3\sqrt{2}\gamma_\phi}{4}
\left[
\epsilon_\phi^{-1}(1-\psi)f'_{\mathrm b}(\phi)
-\epsilon_\phi\nabla\!\cdot\!\left((1-\psi)\nabla\phi\right)
+\delta_\epsilon [\psi]f'_{\mathrm s}(\phi)
\right],
\end{equation}
with $f'_{\mathrm{b,s}}(\phi)\equiv d f_{\mathrm{b,s}}/d\phi$, in analogy with Eq.\eqref{Eq:SurfPSD-RigidSurf-muc}, as the scalar $\mathcal G=\nabla$ specialization of Eq.~\eqref{Eq:DDM-mu}. The quantity $\hat\mu_\phi$ is therefore best regarded as a masked chemical potential. Pointwise use of $\hat\mu_\phi$ inside the solid region is meaningful only after the regularization described above.

In addition, the bulk velocity field $\boldsymbol v(\boldsymbol x,t)$ is constrained by the masked incompressibility condition~\cite{LiLowengrubRatzVoigt2009,AlandLowengrubVoigt2010,AlandWohlgemuth2023,GaoLiXuDRD2025}
\begin{equation}\label{Eq:theory-DDM-incomp}
\nabla\!\cdot[(1-\psi)\boldsymbol v]=0.
\end{equation}
In the fluid bulk, where $1-\psi\simeq 1$ and $\nabla\psi=\boldsymbol 0$, Eq.~\eqref{Eq:theory-DDM-incomp} reduces to $\nabla\!\cdot\boldsymbol v=0$. Across the diffuse wall layer it suppresses the normal component of the fluid flux and yields the sharp-wall impermeability condition $\boldsymbol v\cdot\hat{\boldsymbol n}_\epsilon=0$ in the limit $\epsilon\to0$ near the rigid solid wall (with $\psi\simeq 1$). Tangential slip is not imposed by Eq.~\eqref{Eq:theory-DDM-incomp}; it is controlled separately by the wall-friction term in the dissipation functional.


\subsubsection{Diffuse-domain method (DDM)}

In the DDM [Fig.~\ref{Fig:TwoPhaseFlow}(a)], the conserved phase variable is the masked bulk quantity $(1-\psi)\phi$, in direct analogy with the conservative embedded variable $\delta_\epsilon \boldsymbol a$ in Sec.~\ref{sec:DDM}. Since $\psi$ is fixed,
\begin{equation}\label{Eq:theory-DDM-phi}
\partial_t[(1-\psi)\phi]
=
-\nabla\!\cdot[(1-\psi)\phi\boldsymbol v]
-\nabla\!\cdot[(1-\psi)\boldsymbol J_\phi],
\end{equation}
where $\boldsymbol J_\phi$ is the diffusive flux. Equation~\eqref{Eq:theory-DDM-phi} is the direct bulk analogue of the conservative balance law in Eq.~\eqref{Eq:DDM-bal-local}.

\textbf{Dissipation functional} --- The dissipative mechanisms retained in DDM are bulk viscous flow and phase diffusion in the fluid region, together with wall slip and local relaxational dissipation at the diffuse fluid-solid interface. We therefore choose the dissipation functional as
\begin{equation}\label{Eq:theory-DDM-Phi}
\Phi[\boldsymbol v,\boldsymbol J_\phi,\partial_t\phi]=
\int_\Omega\left[(1-\psi)\left(\eta_{\mathrm b}\,\boldsymbol D:\boldsymbol D +\frac{|\boldsymbol J_\phi|^2}{2M_\phi}\right)
+\delta_\epsilon [\psi]\left(\frac{\beta}{2}|{\bf P}_\epsilon\cdot\boldsymbol v|^2+\frac{(\dot\phi_{\mathrm w})^2}{2\Gamma_\phi}\right)\right]dV.
\end{equation}
Here, $\boldsymbol D=\tfrac12(\nabla\boldsymbol v+\nabla\boldsymbol v^{\mathrm T})$; $\eta_{\mathrm b}$ is the bulk viscosity, $M_\phi$ is the bulk phase mobility; $\beta$ is the tangential wall-friction coefficient; $\dot\phi_{\mathrm w} \equiv \partial_t\phi+({\bf P}_\epsilon\cdot\boldsymbol v)\cdot\nabla_s\phi$ is the tangential material rate along the wall,  $\nabla_s\phi\equiv {\bf P}_\epsilon\cdot\nabla\phi$, and $\Gamma_\phi$ is the local wall-relaxation mobility. The use of $\dot\phi_{\mathrm w}$ in the last term corresponds to a wall relaxation law written in the fixed-wall frame.

\textbf{Rayleighian and dynamic equations} --- As in the conserved scalar formulation of Sec.~\ref{sec:SurfPSD-RigidSurf}, $\partial_t\phi$ and $\boldsymbol J_\phi$ are linked by the masked phase balance. Using Eq.~\eqref{Eq:theory-DDM-phi} and enforcing the masked incompressibility constraint~\eqref{Eq:theory-DDM-incomp}, we follow OVP to minimize the Rayleighian 
$\mathcal R[\boldsymbol v,\boldsymbol J_\phi,\partial_t\phi]=
\dot{\mathcal F}+\Phi-\int_\Omega P\,\nabla\!\cdot[(1-\psi)\boldsymbol v]dV$, we obtain 
\begin{subequations}\label{Eq:theory-DDM-Dyn}
\begin{equation}\label{Eq:theory-DDM-Dyn-Stokes}
\rho(1-\psi)\left(\partial_t \boldsymbol v+\boldsymbol v\!\cdot\nabla \boldsymbol v\right)= -(1-\psi)\nabla P
+\nabla\!\cdot\!\left[(1-\psi)\eta_{\mathrm b}\left(\nabla \boldsymbol v+\nabla \boldsymbol v^{\mathrm T}\right)\right]
+(1-\psi)\bar\mu_\phi\nabla\phi
-\delta_\epsilon [\psi]\beta\,\boldsymbol P_\epsilon \cdot \boldsymbol v-\delta_\epsilon [\psi]\Gamma_\phi^{-1}
\dot\phi_{\mathrm w}\nabla_s\phi,
\end{equation}
\begin{equation}\label{Eq:theory-DDM-Dyn-phi}
\partial_t[(1-\psi)\phi]+\nabla\!\cdot[(1-\psi)\phi\boldsymbol v]=\nabla\!\cdot\!\left[(1-\psi)M_\phi\nabla\bar\mu_\phi\right].
\end{equation}
\end{subequations}
Here, $P$ is the pressure-like Lagrange multiplier enforcing
Eq.~\eqref{Eq:theory-DDM-incomp}, up to the usual isotropic pressure shift used to write the capillary force as
$(1-\psi)\bar\mu_\phi\nabla\phi$. The modified chemical potential $\bar\mu_\phi$ is  
\begin{equation}\label{Eq:theory-DDM-muT}
(1-\psi)\bar\mu_\phi \equiv (1-\psi)\hat\mu_\phi
+\delta_\epsilon [\psi]\Gamma_\phi^{-1}\dot\phi_{\mathrm w}.
\end{equation} 
Note that because $\dot\phi_{\mathrm w}$ depends on the tangential velocity, variation of the Rayleighian produces the additional wall-localized force $-\delta_\epsilon [\psi] \Gamma_\phi^{-1}\dot\phi_{\mathrm w}\nabla_s\phi$ in Eq.~\eqref{Eq:theory-DDM-Dyn-Stokes}. In the sharp-wall limit of $\epsilon\to 0$, this term is the diffuse-domain counterpart of the uncompensated Young-stress force in the generalized Navier boundary condition~\cite{Qian2006JFM}. The OVP derivation gives the overdamped Stokes form; the inertial term in Eq.~\eqref{Eq:theory-DDM-Dyn-Stokes} is appended in the standard equal-density Navier--Stokes approximation.

Equations~\eqref{Eq:theory-DDM-Dyn} are supplemented by the masked incompressibility equation~\eqref{Eq:theory-DDM-incomp} and by standard outer-boundary conditions on $\partial\Omega$, for example,
$\boldsymbol v=\boldsymbol 0, \quad
\hat{\boldsymbol n}_\Omega\!\cdot\nabla\phi=0, \quad
\hat{\boldsymbol n}_\Omega\!\cdot[(1-\psi)\boldsymbol J_\phi]=0$, with $\hat{\boldsymbol n}_\Omega$ the outward unit normal of the computational domain. The limit $\beta\to\infty$ gives the tangential no-slip condition, together with impermeability from Eq.~\eqref{Eq:theory-DDM-incomp}. Since $\Gamma_\phi$ is a mobility in Eq.~\eqref{Eq:theory-DDM-Phi}, fast local wall relaxation corresponds to $\Gamma_\phi\to\infty$, whereas $\Gamma_\phi\to0$ suppresses the wall relaxation rate. In particular, for constant $\rho$, Eqs.~\eqref{Eq:theory-DDM-Dyn} reduce to the familiar rigid-wall DDM form~\cite{AlandLowengrubVoigt2010}. 

\subsubsection{Diffuse-Resistance-Domain (DRD) method}

The DRD formulation [a variant of DDM, Fig.~\ref{Fig:TwoPhaseFlow}(b)] keeps the same conserved masked phase variable $(1-\psi)\phi$, but represents the rigid wall through smoothly varying resistance (transport) coefficients rather than through explicit wall-localized friction and relaxation terms~\cite{GaoLiXuDRD2025}. The convective flux is still masked by $(1-\psi)$, while the diffusive flux is controlled by a mobility $M_\phi(\phi,\psi)$ that becomes small in the solid region. Thus, the DRD phase balance takes the same form as DDM in Eq.~\eqref{Eq:theory-DDM-phi}:
\begin{equation*}\label{Eq:theory-DRD-phi}
\partial_t[(1-\psi)\phi]=
-\nabla\!\cdot[(1-\psi)\phi\boldsymbol v]
-\nabla\!\cdot\boldsymbol J_\phi.
\end{equation*}

\textbf{Dissipation functional} --- The dissipation functional is taken in the simpler isotropic form
\begin{equation}\label{Eq:theory-DRD-Phi}
\Phi[\boldsymbol v,\boldsymbol J_\phi]
=\int_\Omega
\left[
\eta_{\mathrm b}(\phi,\psi)\,\boldsymbol D:\boldsymbol D
+\frac{|\boldsymbol J_\phi|^2}{2M_\phi(\phi,\psi)}
\right]dV.
\end{equation}
Instead of imposing explicit wall-localized friction and relaxation terms as in DDM, DRD encodes the rigid boundary through a large viscosity $\eta_{\mathrm b}(\phi,\psi)$ and a small phase mobility $M_\phi(\phi,\psi)$ in the solid region, equivalently through large resistance coefficients $\eta_{\mathrm b}$ and $M_\phi^{-1}$. With the present convention $\psi=0$ in the fluid and $\psi=1$ in the solid, two forms of interpolations over different regions can be taken:
\begin{subequations}\label{Eq:theory-Dprofile}
\begin{equation}\label{Eq:theory-Dprofile12}
a(\phi,\psi)= a_{\mathrm f}(\phi)+\big[a_{\mathrm s}-a_{\mathrm f}(\phi)\big]\psi,
\end{equation}
\begin{equation}\label{Eq:theory-Dprofile3}
a(\phi,\psi)=
\begin{cases}
a_{\mathrm f}(\phi)+2\psi\left(a_{\mathrm i}-a_{\mathrm f}(\phi)\right), & 0\le \psi\le \tfrac12,\\
2a_{\mathrm i}-a_{\mathrm s}+2\psi\left(a_{\mathrm s}-a_{\mathrm i}\right), & \tfrac12<\psi\le 1,
\end{cases}
\end{equation}
\end{subequations}
with $a$ denoting either $\eta_{\mathrm b}$ or $M_\phi$; if one interpolates the inverse mobility instead, the same formulas are applied to $M_\phi^{-1}$.
Here $\eta_{\mathrm s}\gg \eta_{\mathrm f}(\phi)$ and $M_{\mathrm s}\ll M_{\mathrm f}(\phi)$, while the second profile introduces an intermediate layer with values $\eta_i$ and $M_i$ near $\psi=\tfrac12$. This non-monotonic choice is useful when contact-line slip and wall relaxation are important~\cite{Qian2006JFM,Qian2008JFM,GaoLiXuDRD2025}.

\textbf{Rayleighian and dynamic equations} --- Using the constrained balance~\eqref{Eq:theory-DDM-phi} and minimizing the Rayleighian $\mathcal R[\boldsymbol v,\boldsymbol J_\phi]=\dot{\mathcal F}+\Phi-\int_\Omega P\,\nabla\!\cdot[(1-\psi)\boldsymbol v]dV$
with respect to $\boldsymbol v$ and $\boldsymbol J_\phi$ gives
\begin{subequations}\label{Eq:theory-DRD-Dyn}
\begin{equation}\label{Eq:theory-DRD-Dyn-Stokes}
\rho(1-\psi)\left(\partial_t \boldsymbol v+\boldsymbol v\!\cdot\nabla \boldsymbol v\right)=-(1-\psi)\nabla P
+\nabla\!\cdot\!\left[\eta_{\mathrm b}(\phi,\psi)\left(\nabla \boldsymbol v+\nabla \boldsymbol v^{\mathrm T}\right)\right]
+(1-\psi)\hat\mu_\phi\nabla\phi,
\end{equation}
\begin{equation}\label{Eq:theory-DRD-Dyn-phi}
\partial_t[(1-\psi)\phi]+\nabla\!\cdot[(1-\psi)\phi\boldsymbol v]=\nabla\!\cdot\!\left[M_\phi(\phi,\psi)\nabla\hat\mu_\phi\right],
\end{equation}
\end{subequations}
where $\hat\mu_\phi$ is the masked chemical potential in Eq.~\eqref{Eq:theory-DRD-muEq}. As in the DDM block, the inertial term in Eq.~\eqref{Eq:theory-DRD-Dyn-Stokes} is appended in the standard Navier--Stokes form. The system is supplemented by the incompressibility condition in Eq.~\eqref{Eq:theory-DDM-incomp} and by the same set of outer-boundary conditions as in DDM on $\partial\Omega$. 

The DDM and DRD formulations above share the same masked bulk Cahn--Hilliard free energy and the same wall energy localized by $\delta_\epsilon [\psi]$. Their difference is constitutive and appears at finite $\epsilon$. DDM imposes tangential wall friction and local wall relaxation explicitly through $\delta_\epsilon [\psi]$-weighted dissipation terms. DRD instead absorbs the wall resistance into smoothly varying viscosity and mobility fields in the diffuse solid and, if desired, in an intermediate layer. In both cases, the full bulk gradient $\nabla\phi$ is used because $\phi$ is a bulk binary-fluid order parameter, not a tangential surface state variable as discussed in Sec.~\ref{sec:DDM}. 
For the passive overdamped systems, the above equations satisfy the expected Onsager dissipation laws under the stated boundary conditions. In the DDM case, $\dot{\mathcal F}=
-\int_\Omega
\left[2(1-\psi)\eta_{\mathrm b}\,\boldsymbol D:\boldsymbol D
+(1-\psi)M_\phi|\nabla\bar\mu_\phi|^2
+\delta_\epsilon [\psi]\beta|{\bf P}_\epsilon\cdot\boldsymbol v|^2+\delta_\epsilon [\psi]\Gamma_\phi^{-1}\dot\phi_{\mathrm w}^2
\right]dV=-2\Phi \le 0$.  
In the DRD case, $\dot{\mathcal F} = -\int_\Omega
\left[2\eta_{\mathrm b}(\phi,\psi)\boldsymbol D:\boldsymbol D
+M_\phi(\phi,\psi)|\nabla\hat\mu_\phi|^2\right]dV =-2\Phi
\le 0$. 
If the inertial terms in Eqs.~\eqref{Eq:theory-DDM-Dyn-Stokes} and \eqref{Eq:theory-DRD-Dyn-Stokes} are retained, these relations should be written for the sum of the free energy and the corresponding kinetic energy.

\textcolor{black}{\emph{Remark on inertial extensions}. The OVP derivation based on the Rayleighian minimization in this subsection gives the dissipative (overdamped) Stokes part of the dynamics. The inertial terms in Eqs.~\eqref{Eq:theory-DDM-Dyn-Stokes} and \eqref{Eq:theory-DRD-Dyn-Stokes} are standard reversible Navier--Stokes extensions, consistent with the inertial form of Onsager--Machlup theory~\cite{Onsager1953b,Komura2026CPB}. Consequently, when these terms are retained, the passive energy law should be written for the total energy~\cite{WangLiu2022,Komura2026CPB}
\begin{equation}
\mathcal E_{\mathrm{tot}}=\mathcal F+
\int_\Omega \frac{1}{2}\rho(1-\psi)|\boldsymbol v|^2\,dV,
\end{equation}
rather than for the free energy alone. Under periodic, no-slip, or otherwise no-work outer-boundary conditions, and in the absence of active or externally imposed power input, the formal balance becomes $d\mathcal E_{\mathrm{tot}}/dt=-2\Phi$ for the corresponding passive DDM or DRD dissipation functional~\cite{WangLiu2022}. In the low-Reynolds-number limit, relevant to many microscale applications, the kinetic-energy term is negligible and the overdamped free-energy law stated above is recovered.}

\subsection{Hydrodynamics of multicomponent deformable vesicles}\label{sec:MPF-vesicle}

We now specialize the general diffuse-domain framework of Sec.~\ref{sec:DDM} to the dynamics of multicomponent, deformable vesicles, thereby extending the surface phase-separation models of Secs.~\ref{sec:SurfPSD-RigidSurf} and \ref{sec:SurfPSD-DeformSurf}. In contrast to the rigid-solid setting of Sec.~\ref{sec:MPF-2Phase}, the phase field $\psi(\boldsymbol r,t)$ here represents a mobile membrane separating two fluids rather than a fluid-solid boundary; the bulk flow is therefore defined on both sides of the interface, while the vesicle shape evolves through the time dependence of $\psi$. 

A vesicle is a closed lipid-bilayer membrane that, at the continuum level, is modeled as a fluid surface with Helfrich-type bending elasticity and coupling to the surrounding inner and outer fluids~\cite{Seifert1997AdvPhys,Scriven1960,ArroyoDesimone2009PRE}. In sharp-interface description, this leads to coupled bulk Stokes (or Navier--Stokes) equations together with an in-surface Boussinesq--Scriven balance~\cite{Scriven1960,ArroyoDesimone2009PRE}, which serves as the template for the OVP-based diffuse-domain formulation below. Although many vesicle models impose local inextensibility as the large-area-modulus limit, here we retain a finite local area-dilatational elasticity and thus treat the membrane as extensible~\cite{KloppeAland2024}. This choice is physically reasonable because micropipette-aspiration experiments on giant lipid bilayers show a large but finite area-compressibility modulus and measurable tension-induced area expansion, with the low-tension response dominated by the suppression of thermal undulations and the high-tension response by direct stretching of the bilayer~\cite{EvansRawicz1990,Rawicz2000BiophysJ}. By contrast, fluid-phase bilayers do not sustain a static in-plane shear elasticity because lipids rearrange laterally, so tangential shear is modeled dissipatively through membrane fluidity and surface viscosity rather than through a shear-elastic energy~\cite{Scriven1960,Seifert1997AdvPhys,ArroyoDesimone2009PRE,DimovaPoulignyDietrich2000}. Accordingly, the DDM model developed in this subsection combines bulk hydrodynamics, membrane surface viscosity, tangential slip, Helfrich bending elasticity, and finite local area-dilatational elasticity within a single OVP-based diffuse-domain formulation.

\textbf{State variables and constraints} --- 
This subsection extends Sec.~\ref{sec:DDM} to a mixed scalar--vector description of a deformable vesicle. The geometry is represented by the bulk phase field $\psi(\boldsymbol r,t)$, and the surrounding fluid by the bulk velocity $\boldsymbol v(\boldsymbol r,t)$. The membrane carries two surface scalar fields, the areal-compression variable $c_{\mathrm a}(\boldsymbol r,t)$ and the molecular concentration $c_{\mathrm m}(\boldsymbol r,t)$, together with a surface tangential velocity $\boldsymbol V_\parallel(\boldsymbol r,t)$. Following Ref.~\cite{KloppeAland2024}, we interpret $c_{\mathrm a}<1$ and $c_{\mathrm a}>1$ as local dilation and compression, respectively. We retain $\boldsymbol V_\parallel$ as an independent variable to allow viscous in-surface flow and tangential slip relative to the surrounding bulk fluid. This distinction is physically important because continuum and numerical studies show that surface hydrodynamics can substantially affect domain evolution on vesicles~\cite{ArroyoDesimone2009PRE,OlshanskiiQuaini2025PhaseSeparatedVesicles}. The present subsection is therefore conceptually distinct from the nonhydrodynamic phase-separation models of Sec.~III.B and treats bulk flow, surface transport, and membrane viscous dynamics within a single diffuse-domain formulation.

The embedding follows the scalar/vector distinction emphasized in Sec.~\ref{sec:DDM}. The scalar surface fields $c_{\mathrm a}$ and $c_{\mathrm m}$ are embedded isotropically with $\mathcal G=\nabla$, exactly as in Secs.~\ref{sec:SurfPSD-RigidSurf} and \ref{sec:SurfPSD-DeformSurf}. The explicitly tangential membrane velocity $\boldsymbol V_\parallel$, by contrast, is a kinematic surface variable and is restricted to the tangent bundle, either strongly by imposing $\boldsymbol V_\parallel=\boldsymbol P_\epsilon\cdot\boldsymbol V_\parallel$, or weakly through the penalty term specified below. In this subsection, we use $\delta_\epsilon [\psi]\equiv B_4(\psi,\nabla\psi) = \frac{\epsilon}{2} |\nabla\psi|^2+\epsilon^{-1}G(\psi)$. 
The full membrane material velocity used in the surface balance laws is
\begin{equation}\label{Eq:MPF-vesicle-V}
\boldsymbol V= V_n\hat{\boldsymbol n}_\epsilon+\boldsymbol V_\parallel= V_{\psi n}\hat{\boldsymbol n}_\epsilon+\boldsymbol V_\parallel,
\end{equation}
where the normal velocity $V_n\hat{\boldsymbol n}_\epsilon$ of the diffuse membrane is given by the level-set velocity $\boldsymbol V_\psi=V_{\psi n}\hat{\boldsymbol n}_\epsilon=-({\partial_t\psi}/{|\nabla\psi|^2})\nabla\psi$ of $\psi$ in Eq.~\eqref{Eq:SurfPSD-DeformSurf-Vpsi}.  
Since the vesicle separates two fluids rather than a fluid and a solid as discussed in Sec.~\ref{sec:MPF-2Phase}, the bulk flow is incompressible in the usual sense $\nabla\cdot\boldsymbol v=0$. The vesicle phase field $\psi$ obeys the conservative bulk balance law
\begin{equation}\label{Eq:MPF-vesicle-bal-psi}
\partial_t\psi+\nabla\!\cdot(\psi\boldsymbol v)
= -\nabla\!\cdot\boldsymbol J_\psi.
\end{equation}
Combining Eqs.~\eqref{Eq:SurfPSD-DeformSurf-Vpsi},
\eqref{Eq:MPF-vesicle-bal-psi}, and $\nabla\cdot\boldsymbol v=0$ gives, inside the diffuse layer
\begin{equation}\label{Eq:MPF-vesicle-Vpsi}
V_{\psi n} =\frac{\partial_t\psi}{|\nabla\psi|}= \boldsymbol v\!\cdot\!\hat{\boldsymbol n}_\epsilon
-\frac{\nabla\!\cdot\boldsymbol J_\psi}{|\nabla\psi|}.
\end{equation}
Thus, the hydrodynamic normal motion is carried by the bulk velocity $(\boldsymbol v\cdot\hat{\boldsymbol n}_\epsilon)
\hat{\boldsymbol n}_\epsilon$, while the phase-field flux $\boldsymbol J_\psi$ contributes an additional dissipative shape-relaxation velocity. In the material-interface limit
$\boldsymbol J_\psi=\boldsymbol 0$, one recovers
$V_n=V_{\psi n}=\boldsymbol v\cdot\hat{\boldsymbol n}_\epsilon$.

The two membrane scalars satisfy the embedded material balance law in Eq.~\eqref{Eq:DDM-bal-local} of Sec.~\ref{sec:DDM} with the full membrane velocity $\boldsymbol V$ in Eq.~\eqref{Eq:MPF-vesicle-V}:
\begin{subequations}\label{Eq:MPF-vesicle-bal}
\begin{equation}\label{Eq:MPF-vesicle-bal-a}
\partial_t(\delta_\epsilon  c_{\mathrm a})+\nabla\!\cdot(\delta_\epsilon  c_{\mathrm a}\boldsymbol V)=0,
\end{equation}
\begin{equation}\label{Eq:MPF-vesicle-bal-m}
\partial_t(\delta_\epsilon  c_{\mathrm m})+\nabla\!\cdot(\delta_\epsilon  c_{\mathrm m}\boldsymbol V)=-\nabla\!\cdot(\delta_\epsilon \boldsymbol J_{\mathrm m}),
\end{equation}
\end{subequations}
where $\boldsymbol J_{\mathrm m}$ is the nonconvective molecular flux on the membrane and $k_{\mathrm m}=0$ for passive closed-membrane dynamics. Equation~\eqref{Eq:MPF-vesicle-bal-a} treats $c_{\mathrm a}$ as a purely advected areal-compression variable. 

\textbf{Free energy functional} --- The total embedded free energy is taken as
\begin{equation}\label{Eq:MPF-vesicle-F}
\mathcal F[\psi,c_{\mathrm a},c_{\mathrm m}]=
\mathcal F_\psi[c_{\mathrm a},c_{\mathrm m},\psi]+
\mathcal F_{\mathrm m}[c_{\mathrm m},\psi].
\end{equation}
The shell elastic energy is the same diffuse bending/tension functional used in Sec.~\ref{sec:SurfPSD-DeformSurf}, but the membrane tension $\gamma_\psi$ now depends on both $c_{\mathrm a}$ and $c_{\mathrm m}$:
\begin{equation}\label{Eq:MPF-vesicle-Fpsi}
\mathcal F_\psi[c_{\mathrm a},c_{\mathrm m},\psi]=\int_\Omega
\left\{\frac{K_{\mathrm b}(c_{\mathrm m})}{2\epsilon}\left[-\epsilon^{-1}G'(\psi)+6\psi(1-\psi)\kappa_0(c_{\mathrm m})+
\epsilon\nabla^2\psi\right]^2+
\gamma_\psi(c_{\mathrm a},c_{\mathrm m})\delta_\epsilon [\psi]\right\}dV.
\end{equation}
where the bending modulus $K_{\mathrm b}(c_\mathrm{m})$ and the spontaneous curvature $\kappa_0$ can both depend on $c_{\mathrm m}$.
A simple stable constitutive choice for the areal-compression contribution is
\begin{subequations}\label{Eq:MPF-vesicle-gamma}
\begin{equation}\label{Eq:MPF-vesicle-gamma-a}
\gamma_\psi(c_{\mathrm a},c_{\mathrm m}) = \gamma_{\psi0}
+ \frac{K_{\mathrm a}(c_{\mathrm m})}{2} \left(c_{\mathrm a}+c_{\mathrm a}^{-1}-2\right)
\end{equation}
\begin{equation}\label{Eq:MPF-vesicle-Kab}
K_{\mathrm a}(c_{\mathrm m})= K_{\mathrm a0}e^{-c_{\mathrm m}/c_{\mathrm m0}},
\qquad
K_{\mathrm b}(c_{\mathrm m}) = K_{\mathrm b0}e^{-c_{\mathrm m}/c_{\mathrm m0}},
\qquad
\kappa_0(c_{\mathrm m})=\kappa_{00}+\alpha_\kappa c_{\mathrm m},
\end{equation}
\end{subequations}
where $\gamma_{\psi0}$, $K_{\mathrm{a0}}$, $K_{\mathrm{b0}}$, and $\kappa_{00}$ are some reference constants, so that increasing molecular concentration $c_{\mathrm m}$ softens both bending modulus $K_{\mathrm{b}}$ and stretching modulus $K_{\mathrm{a}}$ if $c_{\mathrm m0}>0$, while $\alpha_\kappa>0$ makes the spontaneous curvature increase with molecular abundance. 
We take the convention that $c_{\mathrm a}<1$ corresponds to local area dilation and $c_{\mathrm a}>1$ to compression.  
The corresponding thermodynamic surface tension is: $\sigma_{\mathrm a} = \gamma_\psi-c_{\mathrm a}\partial_{c_{\mathrm a}}\gamma_\psi = \gamma_{\psi0}+K_{\mathrm a}(c_{\mathrm m})(c_{\mathrm a}^{-1}-1)$, which is the diffuse-domain analogue of the sharp-interface dilational tension law used in elastic-surface phase-field models~\cite{KloppeAland2024}. The field $c_{\mathrm a}$ is a purely advected areal-compression variable, while the compression elasticity enters the hydrodynamic force balance through the isotropic surface stress $\sigma_{\mathrm a}\boldsymbol P_\epsilon$.

The molecular free energy $\mathcal F_{\mathrm m}[c_{\mathrm m},\psi]$ is the scalar isotropic specialization of Eq.~\eqref{Eq:DDM-F}:
\begin{equation}\label{Eq:MPF-vesicle-Fm}
\mathcal F_{\mathrm m}[c_{\mathrm m},\psi]=\gamma_{\mathrm L}\int_\Omega \delta_\epsilon [\psi]\left[
\epsilon_{\mathrm m}^{-1}f_{\mathrm m}(c_{\mathrm m})+\frac{\epsilon_{\mathrm m}}{2}|\nabla c_{\mathrm m}|^2\right]dV,
\end{equation}
with $f_{\mathrm m}(c_{\mathrm m})=c_{\mathrm m}\ln\!\left({c_{\mathrm m}}/{c_{\mathrm {m,max}}}\right)+
(c_{\mathrm {m,max}}-c_{\mathrm m})\ln\!\left(1-{c_{\mathrm m}}/{c_{\mathrm {m,max}}}\right)$. 
The energy density $f_{\mathrm m}$ is convex on $0<c_{\mathrm m}<c_{{\mathrm m},\max}$. Thus, $\mathcal F_{\mathrm m}$ alone does not generate an intrinsic passive spinodal instability. In the present multicomponent vesicle model, any demixing or domain formation of $c_{\mathrm m}$ should be understood as an effective instability of the full coupled free energy, induced only when the $c_{\mathrm m}$-dependent elastic and geometric contributions, such as $K_{\mathrm b}(c_{\mathrm m})$, $K_{\mathrm a}(c_m)$, $\kappa_0(c_{\mathrm m})$, and $\gamma_\psi(c_{\mathrm a},c_{\mathrm m})$, overcome the convex mixing and gradient-penalty terms. If these couplings are absent or too weak, Eq.~\eqref{Eq:MPF-vesicle-Fm} describes conserved surface transport with interfacial regularization rather than spontaneous phase separation. In addition, the use of $|\nabla c_{\mathrm m}|^2$ in Eq.~\eqref{Eq:MPF-vesicle-Fm} is deliberate: $c_{\mathrm m}$ is a surface scalar and is therefore embedded by the isotropic R\"atz--Voigt choice $\mathcal G=\nabla$, exactly as in Secs.~\ref{sec:DDM}, \ref{sec:SurfPSD-RigidSurf}, and \ref{sec:SurfPSD-DeformSurf}. 
Using Eqs.~\eqref{Eq:MPF-vesicle-bal-psi}, \eqref{Eq:MPF-vesicle-bal}, and~\eqref{Eq:MPF-vesicle-F}, one obtains the rate of change of the free energy 
\begin{equation}\label{Eq:MPF-vesicle-dFdt}
\dot{\mathcal F}=\int_\Omega\left[\boldsymbol J_\psi\!\cdot\nabla\bar\mu_\psi+\delta_\epsilon \boldsymbol J_{\mathrm m}\!\cdot\nabla\hat\mu_{\mathrm m}-\bar\mu_\psi\nabla\psi\!\cdot\boldsymbol v+\delta_\epsilon \boldsymbol V_\parallel\!\cdot\left(c_{\mathrm a}\nabla_{\mathrm s}\hat\mu_{\mathrm a}+c_{\mathrm m}\nabla_{\mathrm s}\hat\mu_{\mathrm m}\right)
\right]dV.
\end{equation}
Since $\boldsymbol V_\parallel$ is tangential, only the projected gradients of $\hat\mu_{\mathrm a}$ and $\hat\mu_{\mathrm m}$ enter its reversible work. The scalar chemical potentials and the effective geometry chemical potential are defined by
\begin{subequations}\label{Eq:MPF-vesicle-mu}
\begin{equation}\label{Eq:MPF-vesicle-mua}
\delta_\epsilon\hat\mu_{\mathrm a} \equiv \frac{\delta\mathcal F}{\delta c_{\mathrm a}} = \delta_\epsilon \,\partial_{c_{\mathrm a}}
\gamma_\psi(c_{\mathrm a},c_{\mathrm m})= \delta_\epsilon \frac{K_{\mathrm a}(c_{\mathrm m})}{2}\left(1-c_{\mathrm a}^{-2}\right),
\end{equation}
\begin{equation}\label{Eq:MPF-vesicle-mum}
\delta_\epsilon \hat\mu_{\mathrm m} \equiv \frac{\delta\mathcal F}{\delta c_{\mathrm m}}=\gamma_{\mathrm L} \left[
\epsilon_{\mathrm m}^{-1}\delta_\epsilon  f_{\mathrm m}'(c_{\mathrm m})-\epsilon_{\mathrm m}\nabla\! \cdot(\delta_\epsilon \nabla c_{\mathrm m})\right]+
\frac{\delta\mathcal F_\psi}{\delta c_{\mathrm m}},
\end{equation}
\begin{equation}\label{Eq:MPF-vesicle-mupsi}
\bar\mu_\psi\equiv\left.\frac{\delta \mathcal F}{\delta\psi}\right|_{\delta_\epsilon  c_{\mathrm a},\,\delta_\epsilon  c_{\mathrm m}}-
\delta_\epsilon|\nabla\psi|^{-2} \left(c_{\mathrm a}\nabla\psi\!\cdot\nabla\hat\mu_{\mathrm a}
+c_{\mathrm m}\nabla\psi\!\cdot\nabla\hat\mu_{\mathrm m}\right)\simeq \left.\frac{\delta\mathcal F}{\delta\psi}\right|_{c_{\mathrm a},c_{\mathrm m}}
-A_4 (c_{\mathrm a},\psi)\hat{\mu}_{\mathrm a}-A_4 (c_{\mathrm m},\psi)\hat{\mu}_{\mathrm m},
\end{equation}
\end{subequations} 
respectively, with $f_{\mathrm m}'\equiv df_{\mathrm m}/dc_{\mathrm m}$ and $\bar\mu_\psi$ is the hydrodynamic analogue of the shifted geometry chemical potential used in Sec.~\ref{sec:SurfPSD-DeformSurf}. 
The approximate $A_4$-form in Eq.~\eqref{Eq:MPF-vesicle-mupsi}
follows from the standard one-dimensional interfacial-profile assumption $\delta_\epsilon \simeq \epsilon |\nabla\psi|^2$, with $A_4 (c_i,\psi) \equiv -\epsilon\nabla\!\cdot (c_i\nabla\psi)+\epsilon^{-1}c_iG'(\psi)$, and $i=\mathrm a,\mathrm m$. Appendix~\ref{sec:App-ActPolar-scalar-balance} derives this relation from the conserved scalar material balances and shows that $\bar\mu_\psi$ is the effective geometry force entering the constitutive relation of $\boldsymbol J_\psi$.

\textbf{Dissipation functional} --- Consistently with the mixed scalar/vector embedding, we take the total dissipation functional as
\begin{equation}\label{Eq:MPF-vesicle-Phi}
\Phi[\boldsymbol v,\boldsymbol J_\psi,\boldsymbol V_\parallel,\boldsymbol J_{\mathrm m}]=\int_\Omega\left[
\eta_{\mathrm b}(\psi)\boldsymbol D:\boldsymbol D+\frac{|\boldsymbol J_\psi|^2}{2M_\psi}+\delta_\epsilon [\psi]\left( \eta_{\mathrm s}\mathbb{D}_{\mathrm s}(\boldsymbol V):\mathbb{D}_{\mathrm s}(\boldsymbol V)+\frac{\beta}{2}\big|\boldsymbol V_\parallel-\boldsymbol P_\epsilon\!\cdot\!\boldsymbol v\big|^2+\frac{|\boldsymbol J_{\mathrm m}|^2}{2M_{\mathrm m}}+ \frac{\lambda_v}{2}\left|
\hat{\boldsymbol n}_\epsilon\!\cdot\boldsymbol V_\parallel
\right|^2 \right)\right]dV,
\end{equation}
where $\boldsymbol D\equiv\tfrac12(\nabla\boldsymbol v+\nabla\boldsymbol v^{\mathrm T})$ denotes the three-dimensional bulk rate-of-strain tensor associated with the ambient bulk velocity $\boldsymbol v$, and $\mathbb{D}_{\mathrm s}(\boldsymbol V)\equiv \tfrac12(\bbNabla_{\mathrm s}\boldsymbol V +\bbNabla_{\mathrm s}\boldsymbol V^{\mathrm T})=\mathbb{D}_{\mathrm s}(\boldsymbol V_\parallel)+V_n\boldsymbol\kappa_\epsilon$ is the intrinsic projected tangential membrane rate-of-strain tensor associated with the full membrane material velocity $\boldsymbol V$. The use of $\mathbb{D}_{\mathrm s}(\boldsymbol V)$, rather than $\mathbb{D}_{\mathrm s}(\boldsymbol V_\parallel)$, is essential for a Boussinesq–Scriven-type metric-rate surface-viscous model. For simplicity, we use a one-coefficient isotropic surface-viscous dissipation; a fully general Boussinesq–Scriven surface fluid would allow independent shear and dilatational surface viscosities. 
The viscosity $\eta_{\mathrm b}(\psi)$ interpolates between the inner and outer bulk viscosities, $\eta_{\mathrm s}$ is the membrane surface viscosity, and $M_\psi$ and $M_{\mathrm m}$ are mobility constants. The slip term $\tfrac12\beta |\boldsymbol V_\parallel-\boldsymbol P_\epsilon\cdot\boldsymbol v|^2$ measures dissipation associated with relative tangential motion between the membrane flow and the adjacent bulk fluid, with $\beta>0$ denoting the slip coefficient. The final term is a penalty contribution that enforces the approximate constraint $\boldsymbol V_\parallel\cdot\hat{\boldsymbol n}_\epsilon\approx 0$, thereby restricting the surface velocity $\boldsymbol{V}_\parallel$ to remain tangential to the membrane surface; here $\lambda_v>0$ characterizes the strength of this penalty. 
In the tangential no-slip limit $\beta\to\infty$, the constraint gives $\boldsymbol V_\parallel\to \boldsymbol P_\epsilon\cdot \boldsymbol v$, so that the surface tangential velocity is no longer independent. The full single-velocity limit $\boldsymbol V= \boldsymbol v$ is recovered only in the material-interface limit $\boldsymbol J_\psi=0$, or when the phase-field relaxation contribution to the normal velocity is neglected.

\textbf{Rayleighian and dynamic equations} --- Using Eqs.~\eqref{Eq:MPF-vesicle-V}--\eqref{Eq:MPF-vesicle-bal},  \eqref{Eq:MPF-vesicle-dFdt}, \eqref{Eq:MPF-vesicle-Phi}, we minimize the Rayleighian $\mathcal R[\boldsymbol v,\boldsymbol J_\psi,\boldsymbol V_\parallel,\boldsymbol J_{\mathrm m}]
=\dot{\mathcal F} + \Phi - \int_\Omega P\,\nabla\cdot\boldsymbol v\,dV$ with respect to $\boldsymbol J_\psi$, $\boldsymbol J_{\mathrm m}$, $\boldsymbol v$, and $\boldsymbol V_\parallel$. This yields
\begin{subequations}\label{Eq:MPF-vesicle-Dyn}
\begin{equation}\label{Eq:MPF-vesicle-Dyn-bulk}
\rho(\partial_t\boldsymbol v+\boldsymbol v\cdot\nabla\boldsymbol v)
=-\nabla P+\nabla\cdot\!\boldsymbol\sigma_{\mathrm b}^{\mathrm{vis}}+\bar\mu_\psi\nabla\psi+
\delta_\epsilon \beta\,\boldsymbol P_\epsilon\!\cdot\!\left(\boldsymbol V_\parallel-\boldsymbol P_\epsilon\!\cdot\!\boldsymbol v\right)- \delta_\epsilon 
\left(\boldsymbol\sigma_{\mathrm s}^{\mathrm{vis}}:\boldsymbol{\kappa}_\epsilon\right)\hat{\boldsymbol n}_\epsilon,
\end{equation}
\begin{equation}\label{Eq:MPF-vesicle-Dyn-surf} 
\bbNabla_{\mathrm s}\cdot\left(\delta_\epsilon \boldsymbol\sigma_{\mathrm s}^{\mathrm{vis}}
\right)-\delta_\epsilon \beta
\left(\boldsymbol V_\parallel-\boldsymbol P_\epsilon\cdot \boldsymbol v\right) - \delta_\epsilon
\left(c_{\mathrm a}\nabla_{\mathrm s}\hat\mu_{\mathrm a}+c_{\mathrm m}\nabla_{\mathrm s}\hat\mu_{\mathrm m}\right)- \delta_\epsilon\lambda_v(\hat{\boldsymbol n}_\epsilon \cdot\boldsymbol V_\parallel)\hat{\boldsymbol n}_\epsilon=0,
\end{equation}
\begin{equation}\label{Eq:MPF-vesicle-Dyn-ca}
\partial_t\!\left(\delta_\epsilon c_{\mathrm a}\right)+
\nabla\cdot\!\left(\delta_\epsilon c_{\mathrm a}\boldsymbol V\right)=0,
\end{equation}
\begin{equation}\label{Eq:MPF-vesicle-Dyn-cm}
\partial_t\!\left(\delta_\epsilon c_{\mathrm m}\right)+
\nabla\cdot\!\left(\delta_\epsilon c_{\mathrm m}\boldsymbol V\right) = \nabla\cdot\!\left(\delta_\epsilon M_{\mathrm m}\nabla\hat\mu_{\mathrm m}\right),
\end{equation}
\begin{equation}\label{Eq:MPF-vesicle-Dyn-psi}
\partial_t\psi+\nabla\cdot(\psi\boldsymbol v)=
\nabla\cdot(M_\psi\nabla\tilde\mu_\psi),
\end{equation}
\end{subequations}
where $P$ is the usual bulk pressure enforcing the bulk incompressibility $\nabla\!\cdot\!\boldsymbol v=0$, $\boldsymbol\sigma_{\mathrm b}^{\mathrm{vis}}\equiv 2\eta_{\mathrm b}(\psi)\boldsymbol D =\eta_{\mathrm b}(\psi)\left(\nabla\boldsymbol v+\nabla\boldsymbol v^{\mathrm T}\right)$, $\boldsymbol\sigma_{\mathrm s}^{\mathrm{vis}} \equiv 2\eta_{\mathrm s}\mathbb{D}_{\mathrm s}(\boldsymbol V)=\eta_{\mathrm s}
(\bbNabla_{\mathrm s}\boldsymbol V+\bbNabla_{\mathrm s}\boldsymbol V^{\mathrm T})$, $\boldsymbol V$ is given by Eq.~\eqref{Eq:MPF-vesicle-V}, and $\tilde\mu_\psi\equiv \bar\mu_\psi+\delta_\epsilon(\boldsymbol{\sigma}_\mathrm{s}^{\mathrm{vis}}:\boldsymbol{\kappa}_\epsilon)/|\nabla \psi|$. Note that, for a purely local one-component stretch energy $\gamma_{\mathrm a}(c_{\mathrm a})$, the tangential force $-\delta_\epsilon c_{\mathrm a}\nabla_s\hat\mu_{\mathrm a} $ in Eq.~\eqref{Eq:MPF-vesicle-Dyn-surf} may equivalently be written as~\cite{KloppeAland2024}: $\delta_\epsilon\nabla_s\sigma_{\mathrm a}=\bbNabla_s\cdot(\delta_\epsilon\sigma_{\mathrm a}\boldsymbol P_\epsilon)$ with $\sigma_{\mathrm a}=\gamma_{\mathrm a}-c_{\mathrm a} \hat\mu_{\mathrm a}$ for equilibrium interfacial profiles (such as signed-distance profile) where $\boldsymbol P_\epsilon\cdot\nabla \delta_\epsilon \simeq 0$. 
In the present multicomponent model, however, $\gamma_\psi$ may also depend on $c_{\mathrm m}$, and the chemical-potential form is therefore kept here to avoid separating the isotropic surface tension from the remaining compositional and Korteweg contributions. In addition, as in Sec.~\ref{sec:MPF-2Phase}, the Rayleighian minimization gives the overdamped Stokes form. The inertial term has been appended in the standard equal-density Navier–Stokes extension; when inertia is retained, the passive dissipation law should be written for the sum of kinetic energy and free energy.

\textcolor{black}{The inertial term in Eq.~\eqref{Eq:MPF-vesicle-Dyn-bulk} should be interpreted in the same way as in Sec.~\ref{sec:MPF-2Phase}: it is a reversible bulk Navier--Stokes extension appended to an otherwise overdamped OVP derivation. If it is retained, the passive energy balance is written for
\begin{equation}
\mathcal E_{\mathrm{tot}}=\mathcal F+
\int_\Omega \frac{1}{2}\rho|\boldsymbol v|^2\,dV,
\end{equation}
with no-work outer-boundary conditions. Surface inertia is neglected in the present formulation, which is appropriate for the overdamped membrane dynamics considered here. A theory with membrane inertia would require adding a surface kinetic energy and corresponding acceleration terms in the in-surface momentum balance; this is outside the scope of the present work.}

Eq.~\eqref{Eq:MPF-vesicle-Dyn-bulk} is the bulk hydrodynamic balance, Eq.~\eqref{Eq:MPF-vesicle-Dyn-surf} is the in-surface Stokes--Boussinesq-type membrane balance, Eq.~\eqref{Eq:MPF-vesicle-Dyn-psi} governs the conserved vesicle-shape dynamics, Eq.~\eqref{Eq:MPF-vesicle-Dyn-ca} transports the areal-compression variable, and Eq.~\eqref{Eq:MPF-vesicle-Dyn-cm} is the conservative molecular transport law on the extensible membrane. Note that the slip force appears with opposite sign in the bulk balance equation~\eqref{Eq:MPF-vesicle-Dyn-bulk} and surface balance equation~\eqref{Eq:MPF-vesicle-Dyn-surf} because it is an internal momentum exchange. In the no-slip tangential limit $\beta\to\infty$, one has $\boldsymbol V_\parallel=\boldsymbol P_\epsilon\!\cdot\!\boldsymbol v$ and the two-velocity formulation collapses to the single-velocity extensible-membrane limit. 

The system is supplemented by $\nabla\!\cdot\boldsymbol v=0$ and appropriate outer-boundary conditions on $\partial\Omega$. For example, one may impose no-slip or periodic boundary conditions for the bulk flow. Conservation of $\psi$ and $c_{\mathrm m}$ requires $\hat{\boldsymbol n}_\Omega\!\cdot
M_\psi\nabla\tilde\mu_\psi=0$, $\hat{\boldsymbol n}_\Omega\!\cdot
\delta_\epsilon M_{\mathrm m}\nabla\hat\mu_{\mathrm m}=0$. 
The free-energy variation gives the weighted natural condition $\hat{\boldsymbol n}_\Omega\!\cdot\delta_\epsilon \nabla c_{\mathrm m}=0$. Because $\mathcal F_\psi$ contains $\nabla^2\psi$, the boundary conditions for $\psi$ must also remove the fourth-order bending boundary terms. This can be achieved by periodic boundaries, by keeping the diffuse
vesicle layer away from $\partial\Omega$, or by imposing an equivalent split-form or auxiliary-field set of natural boundary conditions for $\psi$ and the bending auxiliary field.

Finally, we give some remarks before ending this subsection.
\begin{enumerate}[label=(\roman*)] 
\item \textcolor{black}{\emph{Sharp-surface counterpart and two-velocity limits.}
The sharp-interface counterpart of the present OVP construction, expressed at the level of the free energy, dissipation functional, constraints, and Rayleighian variations, is summarized in Appendix~\ref{sec:App-Membrane-OVP}. The distinction between the bulk velocity $\boldsymbol v$, the diffuse normal membrane velocity $\boldsymbol V_\psi$, and
the full membrane material velocity $\boldsymbol V$ is central to this correspondence. The bulk velocity $\boldsymbol v$ transports the surrounding fluids and enters the phase-field advection in Eq.~\eqref{Eq:MPF-vesicle-Dyn-psi}. The normal velocity of the diffuse membrane, however, is the level-set velocity $\boldsymbol V_\psi=V_{\psi n}\hat{\boldsymbol n}_\epsilon$, with $V_{\psi n}$ given by Eq.~\eqref{Eq:SurfPSD-DeformSurf-Vpsi}. Thus, the material velocity appearing in the surface balance laws is $\boldsymbol V=\boldsymbol V_\psi+\boldsymbol V_\parallel$.}

In the material-interface limit, $\boldsymbol J_\psi=\boldsymbol 0$, the diffuse membrane is transported normally by the bulk fluid, so that $V_n=V_{\psi n}=\boldsymbol v\cdot \hat{\boldsymbol n}_\epsilon$. When $\boldsymbol J_\psi\neq \boldsymbol 0$, the phase-field flux contributes an additional dissipative normal shape-relaxation velocity. In the tangential
no-slip limit, $\beta\to\infty$, the slip dissipation enforces
$\boldsymbol V_\parallel=\boldsymbol P_\epsilon\cdot\boldsymbol v$. Taking both limits together gives $\boldsymbol V=\boldsymbol v$ in the diffuse layer and, as $\epsilon\to0$, the usual single-velocity material-interface kinematics of sharp-surface vesicle hydrodynamics. The surface-viscous dissipation then corresponds to the Boussinesq--Scriven metric-rate form based on the full surface rate of deformation.

\textcolor{black}{Thus, the finite-$\epsilon$ equations derived here should be viewed as a thermodynamically consistent diffuse-Rayleighian regularization of the corresponding sharp-surface OVP formulation, rather than as an \emph{ad hoc} diffuse PDE ansatz. What remains separate is a full matched-asymptotic analysis of the resulting finite-$\epsilon$ field equations, including higher-order corrections, convergence, and error estimates. }

\item \emph{Extensible versus inextensible membrane formulation.} The field $c_{\mathrm a}$ is an advected areal-compression variable~\cite{KloppeAland2024}.  With the energy in Eq.~\eqref{Eq:MPF-vesicle-gamma-a}, the relaxed state is $c_{\mathrm a}=1$. The large-$K_{\mathrm a}$ limit formally drives $c_{\mathrm a} \to 1$, and the sharp-surface transport law $\dot c_{\mathrm a\Gamma}+c_{\mathrm a\Gamma} \nabla_\Gamma\!\cdot\boldsymbol V_\Gamma=0$ then implies $\nabla_\Gamma\!\cdot\boldsymbol V_\Gamma\to 0$.
Thus, the model approaches local inextensibility as a penalty limit. At finite diffuse thickness, however, this remains a stiff penalty approximation, not the exact Lagrange-multiplier formulation of an inextensible membrane.

\item \emph{Choice of diffuse surface density.} The present subsection uses $\delta_\epsilon =B_4$ to keep the scalar material balances consistent with Sec.~\ref{sec:DDM} and with the deformable-surface construction in Sec.~\ref{sec:SurfPSD-DeformSurf}. The simpler $B_2(\psi)$ choice gives a more compact finite-$\epsilon$ algebra, but then the shifted geometry chemical potential and the material normal transport are not the same finite-$\epsilon$ model. The two choices are asymptotically equivalent only in the sharp-surface limit under the usual one-dimensional profile assumptions. 

\item \emph{Weak normal regularization.} As discussed in Appendix~\ref{app:finiteepsilon-reg}, projected vector-field embeddings may benefit from an additional weak control of the diffuse-layer extension. One may therefore add
\begin{equation}\label{Eq:MPF-vesicle-Phi-n}
\Phi^{(n)}_\parallel=\int_\Omega \delta_\epsilon [\psi]\frac{\eta_n}{2}\big|( \hat{\boldsymbol n}_\epsilon\!\cdot\!\nabla)\boldsymbol V_\parallel\big|^2\,dV,
\qquad
0<\eta_n\ll \eta_{\mathrm s},
\end{equation}
whose variation adds the term $\nabla\cdot\!\left[
\delta_\epsilon [\psi]\eta_n\nabla\boldsymbol V_\parallel\cdot
(\hat{\boldsymbol n}_\epsilon\otimes\hat{\boldsymbol n}_\epsilon) \right]$ 
to the left-hand side of Eq.~\eqref{Eq:MPF-vesicle-Dyn-surf}. Because $c_{\mathrm a}$ is purely advective in the basic model, one may also add a small artificial normal-diffusion flux 
$\boldsymbol J_{\mathrm a}^{(n)}
=
-\,M_{\mathrm a}^{(n)}
(\hat{\boldsymbol n}_\epsilon\otimes\hat{\boldsymbol n}_\epsilon)\cdot\nabla\hat\mu_{\mathrm a}$ with $0<M_{\mathrm a}^{(n)}\ll M_{\mathrm m}$, together with the associated dissipation
\begin{equation}
\Phi_{\mathrm a}^{(n)}
=
\int_\Omega
\delta_\epsilon [\psi]\frac{|\boldsymbol J_{\mathrm a}^{(n)}|^2}{2M_{\mathrm a}^{(n)}}\,dV,
\end{equation}
which adds the term
$\nabla\cdot\!\left[
\delta_\epsilon [\psi]M_{\mathrm a}^{(n)}
(\hat{\boldsymbol n}_\epsilon\otimes\hat{\boldsymbol n}_\epsilon)\cdot\nabla\hat\mu_{\mathrm a}
\right]$
to Eq.~\eqref{Eq:MPF-vesicle-Dyn-ca}. These terms do not change the sharp-interface physics; they only improve control of the diffuse-layer extension at finite $\epsilon$.
\end{enumerate}

\section{Active polar dynamics on multicomponent membranes (multiple scalar \& vector fields)}\label{sec:ActPolar} 

Biological membranes and epithelial surfaces are active, multicomponent interfaces in which in-plane flow, chemical patterning, orientational order, and curvature evolve in a strongly coupled way. At the subcellular scale, actomyosin cortices generate contractile flows that reorganize polarity and cortical tension, as in the anterior--posterior polarization of \emph{C.~elegans} embryos~\cite{Munro2004,Mayer2010}. At larger scales, epithelial sheets and tissue shells exhibit curvature-coupled active flows and tension anisotropies during morphogenetic events such as \emph{Drosophila} germ-band extension and \emph{zebrafish} epiboly~\cite{He2014,Behrndt2012}. Reconstituted lipid and actomyosin systems further show spontaneous flow patterns, curvature--composition coupling, and domain coexistence in vitro~\cite{Kumar2014,Turlier2014,Miao2020}. These observations motivate continuum models that treat the interface not as a passive liquid-liquid interface or elastic sheet but as an active material surface carrying both conserved and nonconserved internal fields.  

Continuum theories of active surfaces and polar gels provide a thermodynamically consistent framework to describe these processes~\cite{Salbreux2017,Khoromskaia2023,NitschkeVoigt2024ActiveNematodynamics}, in which curvature, flow, and internal order are coupled through balance laws and constitutive relations. Onsager's variational principle (OVP) furnishes a compact route to derive such coupled equations systematically by minimizing the Rayleighian~\cite{Xu2021}, $\mathcal{R} = \dot{F} + \Phi - {\mathcal{P}}_{\mathrm{act}}$, under the relevant kinematic constraints. For the passive sector, minimization of the Rayleighian enforces Onsager symmetry and nonnegative dissipation. In the active model below, active stresses and prescribed chemical production are added as external nonequilibrium power inputs, so the total free energy need not decrease.
In this section, we extend the passive hydrodynamic model of multicomponent membranes in Sec.~\ref{sec:MPF-vesicle}. The extension is deliberately formulated as a passive Onsager sector supplemented by active and chemically driven power inputs. We include a tangential polarization field $\boldsymbol p$ representing the coarse-grained orientation of active cellular assemblies; allow passive composition--polarization coupling, including alignment with the surface composition gradient $\nabla_\Gamma c_{\mathrm m}$; and allow curvature--polarization coupling through the deviatoric curvature tensor. Activity is introduced through an active surface stress, or equivalently an active work power. Thus, the passive part retains the Onsager energy--dissipation structure, while the active stress and prescribed production terms represent non-equilibrium driving.  


\textbf{State variables and constraints} --- We retain the membrane phase field $\psi$, the bulk velocity $\boldsymbol v$, the surface scalar fields $c_{\mathrm a}$ and $c_{\mathrm m}$, the tangential membrane velocity $\boldsymbol V_\parallel$, and add the tangential polarization field $\boldsymbol p$. The membrane carries two scalar balance-law fields, $c_{\mathrm a}$ and $c_{\mathrm m}$, and two explicitly tangential vector fields, $\boldsymbol V_\parallel$ and $\boldsymbol p$. Throughout this section, we use the same diffuse surface delta density as in Sec.~\ref{sec:MPF-vesicle}, $\delta_\epsilon[\psi]= B_4(\psi,\nabla\psi)
=\frac{\epsilon}{2}|\nabla\psi|^2+\epsilon^{-1}G(\psi)$. 
The bulk flow is incompressible, $\nabla\!\cdot\boldsymbol v=0$, and the vesicle phase field obeys
\begin{subequations}\label{Eq:ActPolar-bal}
\begin{equation}\label{Eq:ActPolar-bal-psi}
\partial_t\psi+\nabla\!\cdot(\psi\boldsymbol v)
=-\nabla\!\cdot\boldsymbol J_\psi,
\end{equation}
where the full material velocity $\boldsymbol V=V_{\psi n}\hat{\boldsymbol n}_\epsilon+\boldsymbol V_\parallel$ as used in Eq.~\eqref{Eq:MPF-vesicle-V} and \eqref{Eq:MPF-vesicle-Vpsi} of Sec.~\ref{sec:MPF-vesicle} with $V_{\psi n}\hat{\boldsymbol n}_\epsilon
=-({\partial_t\psi}/{|\nabla\psi|^2}) \nabla\psi$ and  $V_{\psi n} = \boldsymbol v\!\cdot\!\hat{\boldsymbol n}_\epsilon-{\nabla\!\cdot\boldsymbol J_\psi}/{|\nabla\psi|}$. 

The scalar surface balances are
\begin{equation}\label{Eq:ActPolar-bal-ca}
\partial_t\!\big[\delta_\epsilon[\psi]c_{\mathrm a}\big]
+\nabla\!\cdot\!\big[\delta_\epsilon[\psi]c_{\mathrm a}\boldsymbol V\big]=0,
\end{equation}
\begin{equation}\label{Eq:ActPolar-bal-cm}
\partial_t\!\big[\delta_\epsilon[\psi]c_{\mathrm m}\big]
+\nabla\!\cdot\!\big[\delta_\epsilon[\psi]c_{\mathrm m}\boldsymbol V\big]=-\nabla\!\cdot\!\big[\delta_\epsilon[\psi]\boldsymbol J_{\mathrm m}\big]+\delta_\epsilon[\psi]k_{\mathrm m}.
\end{equation}
\end{subequations}
Here, $\boldsymbol J_{\mathrm m}$ is the nonconvective membrane flux. A possible prescribed, chemically driven source is
\begin{equation}\label{Eq:ActPolar-kp}
k_{\mathrm m}=-k_{\mathrm d}c_{\mathrm m}
+k_{\mathrm p}^{(a)}\frac{\varepsilon_{\mathrm a}}{\varepsilon_{\mathrm a}+\varepsilon_{\mathrm {a,max}}}
+k_{\mathrm p}^{(H)}\frac{\kappa_+}{\kappa_+ +\kappa_{\mathrm{max}}},\end{equation}
with $\varepsilon_{\mathrm a}\equiv \max \{c_{\mathrm a}^{-1}-1,0\}$, $\kappa_+\equiv \max\{2\mathcal H_\epsilon-\kappa_i,0\}$, $2\mathcal H_\epsilon\equiv\nabla\!\cdot\hat{\boldsymbol n}_\epsilon$, and $k_{\mathrm p}^{(a)}$, $k_{\mathrm p}^{(H)}$, $\varepsilon_{\mathrm {a,max}}$, $\kappa_i$, $\kappa_{\max}$ constant parameters. The source reactive term in Eq.~\eqref{Eq:ActPolar-kp} is not a passive Onsager relaxation. The field $c_{\mathrm a}$ is still treated as a purely advected areal-compression variable; if explicit remodeling or proliferative relaxation of $c_{\mathrm a}$ is needed, one may add a source or dissipative flux to Eq.~\eqref{Eq:ActPolar-bal-ca}.

The polarization field is represented by an ambient vector
$\boldsymbol p(\boldsymbol x,t)$ in the diffuse layer. It can be decomposed as: $\boldsymbol p=\boldsymbol p_\parallel+p_n \hat{\boldsymbol n}_\epsilon$ with the projected tangential component $\boldsymbol p_\parallel \equiv \boldsymbol P_\epsilon\cdot\boldsymbol p$ and the normal component $p_n \equiv \hat{\boldsymbol n}_\epsilon\cdot\boldsymbol p$. Only $\boldsymbol p_\parallel$ has the interpretation of a physical surface polarization in the sharp-interface theory. The normal component $p_n=\hat{\boldsymbol n}_\epsilon\cdot \boldsymbol p$ is controlled weakly by a finite-$\epsilon$ anchoring penalty term in the free energy functional, Eq.~\eqref{Eq:ActPolar-Fp}. The corresponding dissipative rate is $\dot p_n\equiv \partial_t p_n+\boldsymbol V\cdot\nabla p_n$. Thus, the intrinsic polar free-energy terms, active stress, and co-rotational dynamics are written in terms of $\boldsymbol p_\parallel$, while the normal component satisfies an auxiliary relaxation law. In the limit of strong anchoring, $p_n\to 0$, and the formulation reduces to the strongly tangential surface theory.

In this case of weak anchoring, we use $\boldsymbol p_\parallel$ in the variational formulation. Its dissipative co-rotational rate must therefore be projected onto the same tangential state space
\begin{equation}\label{Eq:ActPolar-Pdot}
\dot{\mathbb P}_\parallel \equiv \boldsymbol P_\epsilon\!\cdot\! \left(\dot{\boldsymbol p}_\parallel  
-\boldsymbol\Omega_\parallel \!\cdot\!\boldsymbol p_\parallel
\right),
\end{equation} 
where $\dot{\boldsymbol p}_\parallel \equiv \partial_t\boldsymbol p_\parallel
+\boldsymbol V\!\cdot\!\nabla\boldsymbol p_\parallel$ and 
$\boldsymbol\Omega_\parallel\equiv
\frac12\left(\bbNabla_{\mathrm s}\boldsymbol V_\parallel-\bbNabla_{\mathrm s}\boldsymbol V_\parallel^{\mathrm T}\right)$ is the tangential spin associated with the lateral membrane velocity. 
The normal component of $\boldsymbol V$ is still required in Eq.~\eqref{Eq:ActPolar-Pdot}; under a closest-point extension its explicit convective contribution vanishes, but normal motion still enters through the evolving projector and through the metric-rate tensor defined below. 
The sharp-surface counterpart of Eq.~\eqref{Eq:ActPolar-Pdot} is the standard co-rotational rate used in surface liquid-crystal and active-gel hydrodynamics~\cite{Kruse2004,Khoromskaia2023,AlandWohlgemuth2023,NitschkeVoigt2024ActiveNematodynamics}; it measures reorientation of $\boldsymbol p$ relative to the local co-rotating surface material element. 

\textbf{Free energy functional} --- The total free energy functional is now taken as
\begin{equation}\label{Eq:ActPolar-Ftot}
\mathcal F[\psi,c_{\mathrm a},c_{\mathrm m},\boldsymbol p]
=\mathcal F_\psi[c_{\mathrm a},c_{\mathrm m},\psi]
+\mathcal F_{\mathrm m}[c_{\mathrm m},\psi]
+\mathcal F_{\mathrm p}[\boldsymbol p,c_{\mathrm m},\psi],
\end{equation}
with $\mathcal F_\psi$ and $\mathcal F_{\mathrm m}$ exactly as in Eqs.~\eqref{Eq:MPF-vesicle-Fpsi}--\eqref{Eq:MPF-vesicle-Fm}, and the new polar free energy given by
\begin{equation}\label{Eq:ActPolar-Fp}
\mathcal F_{\mathrm p}[\boldsymbol p,c_{\mathrm m},\psi]
=\int_\Omega\delta_\epsilon[\psi]\left[
-\frac{a_0}{2}|\boldsymbol p_\parallel|^2
+\frac{a_0}{4}|\boldsymbol p_\parallel|^4
+\frac{K_{\mathrm p}}{2}|\bbNabla_{\mathrm s}\boldsymbol p_\parallel|^2-\frac{\chi}{2}\left(\boldsymbol p_\parallel\!\cdot\!\nabla_{\mathrm s}c_{\mathrm m}\right)^2
-\Lambda_{\mathrm{cp}}\boldsymbol q_\parallel:\tilde{\boldsymbol\kappa}_\epsilon
+\frac{\lambda_{\mathrm p}}{2}|\hat{\boldsymbol n}_\epsilon\!\cdot\!\boldsymbol p|^2\right]dV,
\end{equation}
with the traceless alignment tensor $\boldsymbol q_\parallel \equiv \boldsymbol p_\parallel\boldsymbol p_\parallel -\frac12|\boldsymbol p_\parallel|^2\boldsymbol P_\epsilon$, and the deviatoric curvature tensor $\tilde{\boldsymbol \kappa}_\epsilon \equiv \boldsymbol \kappa_\epsilon-\mathcal H_\epsilon \boldsymbol P_\epsilon$. 
The operator $\bbNabla_{\mathrm s}$ projects both the derivative slot and the vector state slot, as required by the tangential state-space convention in Sec.~\ref{sec:DDM}. 
For the passive part to remain bounded at finite wavelength, the composition--polarization gradient coupling is assumed to be sufficiently weak, or supplemented by higher-order regularization, so that the total effective gradient stiffness of $c_{\mathrm{m}}$ remains positive in the absence of active driving. 
The curvature--polarization coupling is nematic in $\boldsymbol p_\parallel$ and favors alignment of the polarization axis with a principal-curvature direction for $\Lambda_{\mathrm{cp}}>0$ and the orthogonal principal direction for $\Lambda_{\mathrm{cp}}<0$. Since $\boldsymbol q_\parallel$ is traceless, $\boldsymbol q_\parallel:\tilde{\boldsymbol\kappa}_\epsilon=\boldsymbol q_\parallel:\boldsymbol\kappa_\epsilon$. The last tangential anchoring term is a finite-$\epsilon$ penalty for the normal component of the ambient extension of $\boldsymbol p$; it imposes the surface tangentiality of $\boldsymbol p$ weakly. 
In the active model, ``phase separation'' should be understood in a coupling-induced or activity-induced patterning sense. The convex molecular free energy $\mathcal F_{\mathrm{m}}$ does not by itself imply passive equilibrium demixing; in the absence of sufficiently strong elastic, curvature, polarization, or active couplings, the $c_{\mathrm{m}}$ equation describes morphogen transport with interfacial regularization.


Using the hydrodynamically coupled balances~\eqref{Eq:ActPolar-bal}, the co-rotational rate~\eqref{Eq:ActPolar-Pdot}, and the same manipulations as in Sec.~\ref{sec:MPF-vesicle}, the free-energy rate becomes
\begin{align}\label{Eq:ActPolar-Fdot}
\dot{\mathcal F}
={}&
\int_\Omega \left(\boldsymbol J_\psi\!\cdot\!\nabla \bar\mu_\psi+ \delta_\epsilon [\psi]\,\boldsymbol J_{\mathrm m}\!\cdot\!\nabla \hat\mu_{\mathrm m}+ \delta_\epsilon [\psi]k_{\mathrm m}\hat\mu_{\mathrm m}\right)\,dV\nonumber\\
&+\int_\Omega \left(-\bar\mu_\psi\,\nabla\psi\!\cdot\!\boldsymbol v + \delta_\epsilon [\psi]c_{\mathrm a}\,\boldsymbol V_\parallel\!\cdot\!\nabla \hat\mu_{\mathrm a} +\delta_\epsilon [\psi]c_{\mathrm m}\,\boldsymbol V_\parallel\!\cdot\!\nabla \hat\mu_{\mathrm m}\right)\,dV\nonumber\\
&+\int_\Omega \left\{\delta_\epsilon \lambda_{\mathrm p} p_n \dot{p}_n -\delta_\epsilon [\psi]\boldsymbol h_{\mathrm p}^{\parallel}\!\cdot\!\dot{\mathbb P}_\parallel + \boldsymbol V_\parallel\!\cdot\!\Big[-\delta_\epsilon\lambda_{\mathrm p}p_n \nabla_s p_n
-\bbNabla_{\mathrm s}\!\cdot
\left(\delta_\epsilon\boldsymbol\sigma_{\mathrm p}^{\mathrm{asy}}\right)+(\bbNabla_{\mathrm s}\boldsymbol p_\parallel)^{\mathrm T}\!\cdot\! \left(\delta_\epsilon\boldsymbol h_{\mathrm p}^{\parallel}\right) \Big]\right\}\,dV.
\end{align}
The scalar chemical potentials $\bar\mu_\psi$, $\hat\mu_{\mathrm a}$, and $\hat\mu_{\mathrm m}$ are still defined by Eqs.~\eqref{Eq:MPF-vesicle-mu}, but  must now be computed from the full free energy in Eq.~\eqref{Eq:ActPolar-Ftot}. In particular,
\begin{equation}\label{Eq:ActPolar-mum}
\delta_\epsilon\hat\mu_{\mathrm m}= \left(\delta_\epsilon\hat\mu_{\mathrm m}\right)_{\mathrm{Sec.~IVB}}+\frac{\delta\mathcal F_{\mathrm p}}{\delta c_{\mathrm m}},
\end{equation}
where the alignment term in Eq.~\eqref{Eq:ActPolar-Fp} contributes ${\delta\mathcal F_{\mathrm p}}/{\delta c_{\mathrm m}} = \nabla\!\cdot \left[\delta_\epsilon\chi \left(\boldsymbol p_\parallel\!\cdot\!\nabla_{\mathrm s} c_{\mathrm m}\right) \boldsymbol p_\parallel \right]$. 
The shifted geometry chemical potential $\bar\mu_\psi$ is also the derivative of the full free energy at fixed conserved scalar densities $\delta_\epsilon c_{\mathrm a}$ and $\delta_\epsilon c_{\mathrm m}$, and at fixed admissible polar state. Thus, it contains the additional $\psi$-variations of $\mathcal F_{\mathrm p}$ through $\delta_\epsilon[\psi]$, $\boldsymbol P_\epsilon$, $\boldsymbol\kappa_\epsilon$, $\tilde{\boldsymbol\kappa}_\epsilon$, and the anchoring term. The polar molecular field conjugate to the projected rate is
\begin{equation}\label{Eq:ActPolar-hp}
\delta_\epsilon\boldsymbol h_{\mathrm p}^{\parallel}
\equiv -\boldsymbol P_\epsilon\!\cdot\!
\frac{\delta\mathcal F}{\delta\boldsymbol p_\parallel} =
\boldsymbol P_\epsilon\!\cdot\!
\left[-\delta_\epsilon\partial_{\boldsymbol p_\parallel}\hat f_{\mathrm p}
+\nabla\!\cdot\left(\delta_\epsilon
\boldsymbol P_\epsilon\!\cdot\!\partial_{\bbNabla_{\mathrm s}\boldsymbol p_\parallel}\hat f_{\mathrm p}\!\cdot\!\boldsymbol P_\epsilon\right)\right],
\end{equation}
where $\hat f_{\mathrm p}$ denotes the bracketed polar free-energy density in Eq.~\eqref{Eq:ActPolar-Fp}. For example, the curvature--polarization term contributes $2\Lambda_{\mathrm{cp}}\tilde{\boldsymbol\kappa}_\epsilon\!\cdot\!\boldsymbol p_\parallel$ when $\psi$ is held fixed.
The antisymmetric reactive stress associated with the co-rotational kinematics is
\begin{equation}\label{Eq:ActPolar-sigmaA}
\boldsymbol\sigma_{\mathrm p}^{\mathrm{asy}}=
\frac12\left(\boldsymbol p_\parallel\boldsymbol h_{\mathrm p}^{\parallel}-\boldsymbol h_{\mathrm p}^{\parallel}\boldsymbol p_\parallel
\right),
\end{equation}
and the polar contribution in the last two terms in Eq.~\eqref{Eq:ActPolar-Fdot} is the diffuse counterpart of the sharp-surface Ericksen/reactive force. 

\textbf{Dissipation functional} --- The passive dissipation is the direct sum of the bulk viscous dissipation, the conserved geometry and morphogen flux dissipations, the membrane viscous and slip dissipations, and the polar rotational dissipation:
\begin{align}\label{Eq:ActPolar-Phi}
\Phi[\boldsymbol v,\boldsymbol J_\psi,\boldsymbol V_\parallel,\boldsymbol J_{\mathrm m},\dot{\mathbb P}_\parallel]=\int_\Omega
\Bigg[&\eta_{\mathrm b}(\psi)\boldsymbol D:\boldsymbol D
+\frac{|\boldsymbol J_\psi|^2}{2M_\psi}+\delta_\epsilon[\psi]
\Bigg(\eta_{\mathrm s}\mathbb D_{\mathrm s}(\boldsymbol V):\mathbb D_{\mathrm s}(\boldsymbol V)+\frac{\beta}{2}
\big|\boldsymbol V_\parallel-\boldsymbol P_\epsilon\!\cdot\!\boldsymbol v\big|^2\nonumber\\
& +\frac{|\boldsymbol J_{\mathrm m}|^2}{2M_{\mathrm m}} 
+\frac{\gamma_{\mathrm{p}}}{2} \left|\dot{\mathbb P}_\parallel
+\nu_{\mathrm p}\mathbb D_{\mathrm s}(\boldsymbol V)\!\cdot\!\boldsymbol p_\parallel\right|^2+   
\frac{\gamma_{\mathrm{p}}^\perp}{2}
\dot{p}_n^2
+\frac{\lambda_v}{2}|\hat{\boldsymbol n}_\epsilon\!\cdot\!\boldsymbol V_\parallel|^2\Bigg)\Bigg]dV .
\end{align}
with $\mathbb D_{\mathrm s}(\boldsymbol V) \equiv \frac12
\left(\bbNabla_{\mathrm s}\boldsymbol V+
\bbNabla_{\mathrm s}\boldsymbol V^{\mathrm T}\right)=
\mathbb D_{\mathrm s}(\boldsymbol V_\parallel)+V_{\psi n}\boldsymbol\kappa_\epsilon$. The surface viscosity and the polar flow-alignment coupling use the full metric rate of the moving membrane, not only the lateral rearrangement rate. The coefficient $\gamma_{\mathrm{p}}^\perp>0$ is the normal-component rotational/friction coefficient. The orientational term proportional to $\gamma_{\mathrm p}$ penalizes reorientation of $\boldsymbol p$ relative to the locally co-rotating surface material element; $\gamma_{\mathrm p}$ is therefore the rotational viscosity, while $\nu_{\mathrm p}$ measures flow alignment. Expanding it gives the usual cross-coupling between the co-rotational polar rate and the surface metric rate,
\begin{equation}
\frac{\gamma_{\mathrm p}}{2}\left|\dot{\mathbb P}_\parallel
+\nu_{\mathrm p}\mathbb D_{\mathrm s}(\boldsymbol V)\!\cdot\!\boldsymbol p_\parallel\right|^2=\frac{\gamma_{\mathrm p}}{2}|\dot{\mathbb P}_\parallel|^2+\gamma_{\mathrm p}\nu_{\mathrm p}\,\mathbb D_{\mathrm s}(\boldsymbol V):\frac12\left(\dot{\mathbb P}_\parallel\boldsymbol p_\parallel+\boldsymbol p_\parallel\dot{\mathbb P}_\parallel\right)+ \frac{\gamma_{\mathrm p}\nu_{\mathrm p}^2}{2}\left|\mathbb D_{\mathrm s}(\boldsymbol V)\!\cdot\!\boldsymbol p_\parallel\right|^2 .
\end{equation} 
The last term in Eq.~\eqref{Eq:ActPolar-Phi} weakly enforces the tangentiality of $\boldsymbol V_\parallel$; it is unnecessary when $\boldsymbol V_\parallel=\boldsymbol P_\epsilon\!\cdot\!\boldsymbol V_\parallel$ is imposed strongly.

\textbf{Active work power} --- Activity enters through the symmetric active surface stress
\begin{equation}\label{Eq:ActPolar-sigmaact}
\boldsymbol\sigma_{\mathrm s}^{\mathrm{act}}=-\zeta\boldsymbol q_\parallel,
\end{equation}
and the corresponding active work power
\begin{equation}\label{Eq:ActPolar-Wact}
{\mathcal P}^{\mathrm{act}}[\boldsymbol V]=-\int_\Omega
\delta_\epsilon[\psi]\,\boldsymbol\sigma_{\mathrm s}^{\mathrm{act}}:\mathbb D_{\mathrm s}(\boldsymbol V)\,dV .
\end{equation}
With this convention, $\zeta<0$ is contractile and $\zeta>0$ is extensile. Because the work is written with the full metric rate $\mathbb D_{\mathrm s}(\boldsymbol V)$, active stresses generate both tangential and curvature-induced normal forces on a curved surface~\cite{Kruse2004,Kruse2005,Julicher2007,Marchetti2013}.

\textbf{Rayleighian and dynamic equations} --- Using Eqs.~\eqref{Eq:ActPolar-bal} and~\eqref{Eq:ActPolar-Pdot}, we minimize $\mathcal R = \dot{\mathcal F} + \Phi - {\mathcal P}^{\mathrm{act}}-\int_\Omega P\,\nabla\!\cdot\!\boldsymbol v\,dV$ with respect to $\boldsymbol J_\psi$, $\boldsymbol J_{\mathrm m}$, $\dot{\mathbb P}_\parallel$, $\boldsymbol V_\parallel$, and $\boldsymbol v$. This yields the dynamic equations
\begin{subequations}\label{Eq:ActPolar-DynEqn}
\begin{align}\label{Eq:ActPolar-DynEqn-bulk}
\rho(\partial_t\boldsymbol v+\boldsymbol v\!\cdot\!\nabla\boldsymbol v) =-\nabla P
+\nabla\!\cdot\!\boldsymbol\sigma_{\mathrm b}^{\mathrm{vis}} +\bar\mu_\psi\nabla\psi +\delta_\epsilon\beta\, \boldsymbol P_\epsilon\!\cdot\!\left(\boldsymbol V_\parallel-\boldsymbol P_\epsilon\!\cdot\!\boldsymbol v
\right)-\delta_\epsilon \left(\boldsymbol{\Sigma}_\mathrm{s}^{\mathrm{sym}}:\boldsymbol\kappa_\epsilon\right)\hat{\boldsymbol n}_\epsilon,
\end{align}
\begin{align}\label{Eq:ActPolar-DynEqn-surf}
\bbNabla_{\mathrm s}\cdot\left(\delta_\epsilon \boldsymbol\Sigma_{\mathrm s}\right) &-\delta_\epsilon\beta
\left(\boldsymbol V_\parallel-\boldsymbol P_\epsilon\cdot\boldsymbol v\right)-\delta_\epsilon
\left(c_{\mathrm a}\nabla_{\mathrm s}\hat\mu_{\mathrm a}
+c_{\mathrm m}\nabla_{\mathrm s}\hat\mu_{\mathrm m}\right)
\nonumber\\
&-(\bbNabla_{\mathrm s}\boldsymbol p_\parallel)^T
\cdot(\delta_\epsilon\boldsymbol h_{\mathrm p}^{\parallel})
+\delta_\epsilon\lambda_{\mathrm p}p_n\nabla_{\mathrm s}p_n
-\delta_\epsilon\lambda_v (\hat{\boldsymbol n}_\epsilon\cdot\boldsymbol V_\parallel)\hat{\boldsymbol n}_\epsilon=0,
\end{align}
\begin{equation}\label{Eq:ActPolar-DynEqn-ca}
\partial_t \left(\delta_\epsilon  c_{\mathrm a}\right)
+\nabla\!\cdot\!\left(\delta_\epsilon  c_{\mathrm a}\boldsymbol V\right)=0,
\end{equation}
\begin{equation}\label{Eq:ActPolar-DynEqn-cm}
\partial_t\!\left(\delta_\epsilon  c_{\mathrm m}\right)
+\nabla\!\cdot\!\left(\delta_\epsilon  c_{\mathrm m}\boldsymbol V\right)= \nabla\!\cdot\!\left(M_{\mathrm m}\delta_\epsilon \nabla\hat\mu_{\mathrm m}\right) +\delta_\epsilon k_{\mathrm m},
\end{equation} 
\begin{equation}\label{Eq:ActPolar-DynEqn-p}
\boldsymbol P_\epsilon\!\cdot\! \left(\partial_t\boldsymbol p_\parallel+\boldsymbol V\!\cdot\!\nabla\boldsymbol p_\parallel-\boldsymbol\Omega_\parallel\!\cdot\!\boldsymbol p_\parallel\right) = -\nu_{\mathrm p}\, \mathbb D_{\mathrm s}(\boldsymbol V)\!\cdot\! \boldsymbol p_\parallel+
\gamma_{\mathrm p}^{-1}\boldsymbol h_{\mathrm p}^{\parallel},
\end{equation}
\begin{equation} 
\gamma_{\mathrm{p}}^\perp \left(
\partial_t p_n+\boldsymbol V\cdot\nabla p_n\right) = -\lambda_{\mathrm{p}}p_n, 
\end{equation}
\begin{equation}\label{Eq:ActPolar-DynEqn-psi}
\partial_t\psi+\nabla\!\cdot(\psi\boldsymbol v)=\nabla\!\cdot\!\left(M_\psi\nabla\tilde\mu_\psi\right),
\end{equation}
\end{subequations}
Here, $\boldsymbol\sigma_{\mathrm b}^{\mathrm{vis}}=2\eta_{\mathrm b}(\psi)\boldsymbol D$, $\boldsymbol{\Sigma}_\mathrm{s}^{\mathrm{sym}} \equiv \boldsymbol\sigma_{\mathrm s}^{\mathrm{vis}}+\boldsymbol\sigma_{\mathrm p}^{\mathrm{irr}}+\boldsymbol\sigma_{\mathrm s}^{\mathrm{act}}$, $\boldsymbol{\Sigma}_\mathrm{s}  \equiv \boldsymbol{\Sigma}_\mathrm{s}^{\mathrm{sym}} +\boldsymbol\sigma_{\mathrm p}^{\mathrm{asy}}$, $\boldsymbol\sigma_{\mathrm s}^{\mathrm{vis}}=2\eta_{\mathrm s}\mathbb D_{\mathrm s}(\boldsymbol V)= \eta_{\mathrm s}(\bbNabla_{\mathrm s}\boldsymbol V+\bbNabla_{\mathrm s}\boldsymbol V^{\mathrm T})$, and $\tilde\mu_\psi\equiv \bar\mu_\psi+\delta_\epsilon(\boldsymbol{\Sigma}_\mathrm{s}^{\mathrm{sym}}:\boldsymbol{\kappa}_\epsilon)/|\nabla \psi|$. The geometry equation retains its previous form, but the shifted shape chemical potential $\bar{\mu}_\psi$ must be computed from the full free energy; it contains additional polar contribution $\left.{\delta \mathcal F_{\mathrm p}}/{\delta\psi} 
\right|_{\delta_\epsilon c_{\mathrm a},\, \delta_\epsilon c_{\mathrm m},\,\boldsymbol p_\parallel,\,p_n}$. Its explicit expression is lengthy because $\tilde{\boldsymbol\kappa}_\epsilon$ depends on gradients of $\psi$, and we do not write it out here. Here
\begin{equation}\label{Eq:ActPolar-sigmad}
\boldsymbol\sigma_{\mathrm p}^{\mathrm{irr}}=\frac{\nu_{\mathrm p}}{2}\left(\boldsymbol p_\parallel\boldsymbol h_{\mathrm p}^{\parallel}+\boldsymbol h_{\mathrm p}^{\parallel}\boldsymbol p_\parallel\right).
\end{equation}
Equation~\eqref{Eq:ActPolar-DynEqn-bulk} is the bulk hydrodynamic balance. As in Sec.~\ref{sec:MPF-2Phase}, the inertial term has been appended in the standard equal-density Navier–Stokes extension. Equation~\eqref{Eq:ActPolar-DynEqn-surf} is the tangential in-surface Stokes--Boussinesq balance with passive, polar, and active surface stresses. Equation~\eqref{Eq:ActPolar-DynEqn-psi} gives the conserved phase-field relaxation of the vesicle geometry, while Eqs.~\eqref{Eq:ActPolar-DynEqn-ca}--\eqref{Eq:ActPolar-DynEqn-cm} transport the scalar surface densities by the full material surface velocity $\boldsymbol V=\boldsymbol V_\psi+\boldsymbol V_\parallel$. Equation~\eqref{Eq:ActPolar-DynEqn-p} evolves the tangential internal order parameter through a projected co-rotational rate. Under a closest-point extension, the explicit normal convective derivative of $\boldsymbol p_\parallel$ vanishes, but normal motion still enters through $V_{\psi n}\boldsymbol \kappa_\epsilon$ in $\mathbb D_{\mathrm s}(\boldsymbol V)$ and through the evolving tangent projector.
If $\boldsymbol p=\boldsymbol 0$, $\zeta=0$, and the  active contributions are removed or made passive, the system reduces to the passive hydrodynamic extensible-shell model of Sec.~\ref{sec:MPF-vesicle}. 
In the tangential no-slip limit $\beta\to\infty$, one obtains $\boldsymbol V_\parallel=\boldsymbol P_\epsilon\!\cdot\!\boldsymbol v$. The full single-velocity limit $\boldsymbol V=\boldsymbol v$ is recovered only when, in addition, the interface is material, $\boldsymbol J_\psi=\boldsymbol 0$.

\textcolor{black}{When the bulk inertial term in Eq.~\eqref{Eq:ActPolar-DynEqn-bulk} is kept, the passive part of the energy balance is again a total-energy balance for $\mathcal E_{\mathrm{tot}}=\mathcal F+\int_\Omega \rho|\boldsymbol v|^2/2\,dV$. The active and chemically driven terms then appear as power inputs. Formally, under no-work outer-boundary conditions,
\begin{equation}
\frac{d\mathcal E_{\mathrm{tot}}}{dt}
= -2\Phi+{\mathcal P}^{\mathrm{act}}+
\int_\Omega \delta_\epsilon k_m\hat\mu_m\,dV,
\end{equation}
up to the same finite-$\epsilon$ regularization assumptions used in deriving the diffuse balance laws. Thus, the passive sector constitutes a thermodynamically consistent dissipative subsystem, whereas the active stress and the imposed chemical production act as sources of nonequilibrium work.}

Finally, we make several remarks before ending this section. 
\begin{enumerate}[label=(\roman*)]
\item \emph{Passive energy structure and active power.}
For $\zeta=0$ and for a passive choice of $k_{\mathrm m}$, the model follows from the Rayleighian and has nonnegative passive dissipation. If $k_{\mathrm m}$ is prescribed as in Eq.~\eqref{Eq:ActPolar-kp}, the free-energy balance contains the chemical power input $\int_\Omega \delta_\epsilon k_{\mathrm m}\hat\mu_{\mathrm m}\,dV$. 
If $\zeta\neq0$, the active stress contributes the active power ${\mathcal P}^{\mathrm{act}}$. Thus, the passive part is thermodynamically consistent, while activity and prescribed production act as external nonequilibrium driving mechanisms.
\item \emph{Weak normal regularization.}
The morphogen field $c_{\mathrm m}$ is embedded through the isotropic scalar operator $\nabla\!\cdot(\delta_\epsilon\nabla\cdot)$, which already regularizes the normal extension. By contrast, $\boldsymbol V_\parallel$ and $\boldsymbol p_\parallel$ use projected tangential gradients, and $c_{\mathrm a}$ is purely advected. For numerical robustness one may add weak normal-extension controls such as
\begin{equation}
\Phi_{\mathrm p}^{(n)}
=
\int_\Omega
\delta_\epsilon
\frac{\eta_{\mathrm p}^{(n)}}{2}
\left|
(\hat{\boldsymbol n}_\epsilon\!\cdot\!\nabla)\boldsymbol p_\parallel
\right|^2 dV,
\qquad
0<\eta_{\mathrm p}^{(n)}\ll K_{\mathrm p},
\end{equation}
and the analogous terms for $\boldsymbol V_\parallel$ and $c_{\mathrm a}$ discussed in Sec.~\ref{sec:MPF-vesicle}. These are finite-$\epsilon$ numerical regularizations and are not part of the intended sharp-interface physics. 
\end{enumerate}

\section{Conclusion and outlook}\label{sec:conclusion}
 
We have developed an Onsager-variational formulation of diffuse-domain methods (DDMs) for microscale fluid--structure interactions (mFSIs) in multicomponent and multiphase flow systems. The framework treats, within a common construction, two classes of problems: dynamics occurring in the vicinity of interfaces or surfaces, and dynamics intrinsically confined to evolving interfaces or surfaces. The central idea is to embed sharp-surface free-energy and dissipation functionals into a regular bulk domain through a diffuse surface delta density, and then derive the diffuse equations by minimizing the Rayleighian. 

A key outcome of Sec.~\ref{sec:DDM} is the separation of surface variables into three operational classes: balance-law fields, internal nonconserved order-parameter fields, and kinematic or constitutive rate variables. Balance-law
fields, including conserved surface densities per unit current area, are represented by weighted variables such as $\delta_\epsilon \boldsymbol a$ and evolve through conservative embedded balance laws. Their transport involves the full material surface velocity, with normal motion entering through the evolution of the surface measure. Internal nonconserved variables require a different treatment. For scalar internal variables the dissipative rate reduces to a material derivative, whereas for explicitly tangential vector or tensor order parameters it must be replaced by an objective rate projected onto the admissible tangent state space. Kinematic variables, such as surface velocities, fluxes, and sources, enter through constraints and dissipation rather than as independent free-energy variables. These distinctions are essential for obtaining thermodynamically consistent passive diffuse-domain models.

For scalar phase separation on rigid and deformable interfaces, and for interfacial hydrodynamics near rigid walls, the present construction recovers established DDM equations with known sharp-interface limits
(Secs.~\ref{sec:SurfPSD} and~\ref{sec:MPF-2Phase})~\cite{RatzVoigt2006,LiLowengrubRatzVoigt2009,AlandLowengrubVoigt2010,LowengrubRatzVoigt2009Vesicles}.
This provides an internal check on the framework and shows that standard DDM models can be recovered from free-energy and dissipation functionals rather than assembled term by term at the level of partial differential equations. The same construction also provides variationally consistent finite-$\epsilon$ diffuse-domain models for extensible multicomponent vesicles with bulk hydrodynamics, surface viscosity, tangential slip, and finite areal compressibility (Sec.~\ref{sec:MPF-vesicle}), and for active polar shells with chemical transport, phase separation, tangential flow, curvature--polarization coupling, and polarization (Sec.~\ref{sec:ActPolar}). In the active model, activity enters through an active work power, or equivalently an active stress, while the passive part retains the Onsager energy-dissipation structure.

The framework offers several advantages beyond thermodynamic consistency. By deriving bulk dynamics, interfacial forces, and surface transport laws from a single Rayleighian, it provides a modular route for constructing coupled bulk/surface models. \textcolor{black}{It also makes explicit which quantities are thermodynamic state variables and which quantities are rates, fluxes, constraints, or active power inputs. Moreover, as discussed in Sec.~\ref{sec:DDM}, under the standard smoothness,  normalization, and normal-extension assumptions, the embedded free energy, dissipation functional, and Rayleighian reduce to their sharp-surface counterparts. The diffuse-domain construction therefore has a well-defined sharp-interface counterpart at the Rayleighian level, including the associated
energy, dissipation, constraints, and variational structure. 
Because the diffuse-domain embedding is posed on fixed regular domains, it is also naturally compatible with standard finite-difference, finite-volume, and finite-element discretizations, as well as adaptive refinement and parallel solvers. This
Rayleighian-level consistency, however, is not a substitute for
problem-specific matched-asymptotic or numerical convergence analysis. The variational route identifies thermodynamically consistent finite-$\epsilon$ equations and their limiting sharp-surface variational structure, whereas asymptotic and numerical analyses are needed to prove PDE-level convergence, quantify finite-$\epsilon$ corrections, and establish accuracy and robustness for each specific coupled model, especially for the new vector- and tensor-valued surface systems.} We summarize the main open issues as follows.
\begin{enumerate}[label=(\roman*)]
\item \emph{PDE-level sharp-interface analysis and state-space justification.} Although the Rayleighian-level reduction gives the corresponding sharp-surface free energy, dissipation functional, and variational equations, a full matched-asymptotic analysis of the newly proposed vesicle and active-surface diffuse equations remains to be carried out. Such an analysis would show directly how the finite-$\epsilon$ field equations recover the sharp traction balances, the transport laws for conserved surface densities, and the projected objective dynamics of class-(ii) tangential order parameters. It would also
identify possible finite-$\epsilon$ correction terms and clarify the assumptions under which the diffuse-layer extensions converge to intrinsic surface fields. A more systematic comparison between ambient bulk extensions and intrinsically
tangential surface state spaces remains desirable, especially when tangentiality is imposed only weakly at finite $\epsilon$.

\item \emph{Finite-$\epsilon$ numerics and bulk-extension control.} The mixed scalar/vector constructions of Secs.~\ref{sec:DDM}, \ref{sec:MPF-vesicle}, and \ref{sec:ActPolar} combine isotropic scalar embeddings with projected tangential embeddings. For scalar fields treated with
the isotropic embedding, the weighted operator $\nabla\!\cdot(\delta_\epsilon \nabla\cdot)$ already regularizes the normal direction. For tangential vector and tensor fields based on projected-gradient embeddings, however, the diffuse-layer extension may require additional weak normal control at finite $\epsilon$. Systematic numerical studies are therefore
needed to quantify accuracy, stiffness, resolution requirements, and convergence with respect to $\epsilon$, and to clarify the role of optional weak normal regularization for fields such as $\boldsymbol V_\parallel$, $\boldsymbol p$, and the purely advected areal-compression field $c_{\mathrm a}$. The tradeoff
between diffuse-interface width and numerical resolution should also be explored more systematically, especially in fully coupled three-dimensional computations where steep gradients can lead to substantial stiffness. More generally, the consequences of different asymptotically equivalent choices of the diffuse surface delta density $\delta_\epsilon$ at finite $\epsilon$ deserve closer study.

\item \emph{Reduced limits and constitutive interpretation.} 
Several limits used in the present models should be understood as formal rather than exact at finite diffuse thickness $\epsilon$. In particular, the large-$K_{\mathrm a}$ limit is a penalty approximation to local membrane inextensibility, not the exact Lagrange-multiplier formulation, and the relation between the two-velocity slip model and the no-slip single-velocity limit deserves a more systematic analysis. Establishing how these diffuse limits connect to the corresponding constrained sharp-interface theories remains an important open problem.

\item \emph{Physics of the new models.}
The vesicle model raises questions about the coupled roles of surface viscosity, tangential slip, finite areal compressibility, and curvature--composition coupling in multicomponent membrane dynamics. The active polar shell model raises related questions concerning contractile and extensile regimes, curvature--polarization coupling, transport-driven instabilities, and defect-mediated morphogenesis on deformable surfaces~\cite{Salbreux2017,Khoromskaia2023}. These issues require both analytical study and targeted DNS simulations, and they will determine which aspects of the present construction are most relevant in specific biological and soft-matter settings.

\item \emph{Parameter identification and comparison with experiment.}
Predictive applications will require constitutive parameters that are often poorly constrained, including surface viscosities, area moduli, slip coefficients, mobilities, and active stresses. For biological systems in particular, meaningful comparison with experiment will demand parameter estimation, sensitivity analysis, and reduced descriptions adapted to measurable observables.

\item \emph{Further applications and extensions of the framework.}
The active polar model should be viewed as one representative member of a broader
class of surface field theories accessible within the same variational
construction. Replacing the tangential polarization vector $\boldsymbol p$ by a
tangential $\boldsymbol Q$ tensor is a natural route to active surface nematic
models on deformable interfaces~\cite{Salbreux2017,NitschkeVoigt2024ActiveNematodynamics},
provided the corresponding state-space projection, objective rate, molecular
field, and sharp-interface reduction are treated consistently. Other directions
include viscoelastic surface rheology, bilayer or multilayer membrane models,
more elaborate reaction networks, and further OVP-based domain-embedding
formulations for interfacial problems~\cite{KloppeAland2024,YuQianWang2025,YuWangZhangQian2025}.
\end{enumerate}
 
In summary, the main contribution of the present work is a reusable model-construction framework rather than a collection of isolated example models. Combining diffuse-domain embedding with Onsager's variational principle provides a modular way to incorporate energetic, dissipative, and active ingredients while preserving the passive energy--dissipation structure and making the relevant geometric assumptions explicit. This framework should be useful for constructing and analyzing models of  vesicles, microfluidic interfaces, subcellular actomyosin dynamics, active tissues, and morphogenetic surface dynamics. For each new coupled vector- or tensor-valued application, the Rayleighian-level sharp-interface correspondence established here should be complemented by problem-specific matched-asymptotic analysis and finite-$\epsilon$ numerical validation.

\section*{Acknowledgements}
We thank Masao Doi for insightful discussions, constructive guidance, and sustained encouragement in the past decade. X.X. thanks Tiezheng Qian and Ping Sheng for longstanding support and training, Shigeyuki Komura for discussions and collaborations on OVP, and Min Gao and Zhenlin Guo for motivating computational modeling problems in complex soft matter and biological systems. X.X. is supported by the National Natural Science Foundation of China (NSFC, Nos. 12374209 and 12131010).  
\section*{DATA AVAILABILITY}

No new data were generated or analyzed in this work.


\appendix
\section{Sharp-interface model for multicomponent active polar membranes}\label{sec:App-Membrane}

\subsection{Differential geometry and kinematics of a moving vesicle surface}\label{sec:surface-geometry}

To connect this appendix section directly to the diffuse-domain formulation of a moving vesicle surface in Secs.~\ref{sec:DDM}, \ref{sec:MPF-vesicle}, and \ref{sec:ActPolar}, we collect here only the sharp-surface identities that are used later. The diffuse-domain embedding has already been formulated in Sec.~\ref{sec:DDM}; in this appendix we work entirely on the smooth, oriented, closed 2D sharp surface $\Gamma(t)$ that is embedded in a 3D (regular) domain $\Omega \subset\mathbb R^3$ (3D Euclidean space).

\textbf{Surface geometry: intrinsic and extrinsic quantities} ---We parametrize $\Gamma(t)$ by an embedding $\boldsymbol X(\boldsymbol\xi,t)$ with surface coordinates $\boldsymbol\xi=(\xi^1,\xi^2)$. The covariant tangent basis is $\boldsymbol e_\alpha\equiv \partial_\alpha\boldsymbol X$, the unit normal is $\hat{\boldsymbol n}_\Gamma\equiv (\boldsymbol e_1\times\boldsymbol e_2)/|\boldsymbol e_1\times\boldsymbol e_2|$. The induced metric (the first fundamental form) is $g_{\alpha\beta}\equiv  \boldsymbol{e}_\alpha \cdot \boldsymbol{e}_\beta=g_{\beta\alpha}$, with inverse $g^{\alpha\beta}=g^{\beta\alpha}$ satisfying $g^{\alpha\gamma}g_{\gamma\beta}=\delta^\alpha_{\ \beta}$ and determinant $g\equiv  \det(g_{\alpha\beta})$. The dual (contravariant) basis $\boldsymbol e^\alpha$ is defined by $\boldsymbol{e}^{\alpha}\cdot \boldsymbol{e}_\beta=\delta^\alpha_{\ \beta}$, equivalently $\boldsymbol e^\alpha=g^{\alpha\beta}\boldsymbol e_\beta$. The tangential projector is the symmetric tensor: ${\boldsymbol P}_\Gamma  \equiv  \boldsymbol I-{\boldsymbol{\hat{n}}}_\Gamma \otimes{\boldsymbol{\hat{n}}}_\Gamma
= \boldsymbol e_\alpha\otimes \boldsymbol e^\alpha = \boldsymbol e^\alpha\otimes \boldsymbol e_\alpha$, which projects ambient vectors in $\mathbb{R}^3$ onto the tangent plane of the surface $\Gamma(t)$. Throughout, Latin indices ($i,j,k,\ell\in\{1,2,3\}$) label ambient Cartesian components and Greek indices ($\alpha,\beta,\gamma\in\{1,2\}$) label surface coordinate components; repeated indices are summed (Einstein summation convention). 

\textbf{Surface connection and curvature} --- Since $\{\boldsymbol e_1,\boldsymbol e_2,\hat{\boldsymbol n}_\Gamma\}$ spans $\mathbb R^3$ at each point, both $\partial_\alpha \boldsymbol e_\beta$ and $\partial_\alpha \hat{\boldsymbol n}_\Gamma$ admit unique tangential--normal decompositions:
\begin{equation}\label{Eq:AppA-GW}
\partial_\alpha \boldsymbol e_\beta=\Gamma^{\gamma}_{\alpha\beta}\,\boldsymbol e_\gamma-\kappa_{\alpha\beta}\hat{\boldsymbol n}_\Gamma, \qquad 
\partial_\alpha\hat{\boldsymbol n}_\Gamma=\kappa_{\alpha\beta}\,\boldsymbol e^\beta=\kappa_{\alpha}^{\ \beta}\,\boldsymbol e_\beta,
\end{equation}
with $\partial_\alpha \equiv \partial/\partial\xi^\alpha$. The first is the Gauss formula, and the second is the Weingarten equation, obtained by differentiating $\hat{\boldsymbol n}_\Gamma\cdot \boldsymbol e_\beta=0$ and $\hat{\boldsymbol n}_\Gamma\cdot \hat{\boldsymbol n}_\Gamma=1$ with respect to $\xi^\alpha$. Similarly, we can also derive $\partial_\alpha \boldsymbol e^\beta=-\Gamma^{\beta}_{\alpha\gamma}\,\boldsymbol e^\gamma-\kappa^{\beta}_{\alpha}\hat{\boldsymbol n}_\Gamma$. Here $\Gamma^\gamma_{\alpha\beta}=\Gamma^\gamma_{\beta\alpha}$ are the Christoffel symbols of the second kind, \emph{i.e.}, the coefficients of the Levi--Civita connection associated with the surface metric; the Christoffel symbols of the first kind are
$\Gamma_{\alpha\beta\gamma}\equiv g_{\gamma\delta}\Gamma^\delta_{\alpha\beta}
=\boldsymbol e_\gamma\cdot \partial_\alpha \boldsymbol e_\beta
=\tfrac12\left(\partial_\alpha g_{\beta\gamma}+\partial_\beta g_{\alpha\gamma}-\partial_\gamma g_{\alpha\beta}\right)$, with $\Gamma^\gamma_{\alpha\beta}=g^{\gamma\delta}\Gamma_{\alpha\beta\delta}$.
The symmetric tensor $\kappa_{\alpha\beta}=\kappa_{\beta\alpha}$ is the second fundamental form, and the mixed tensor $\kappa_\alpha^{\ \beta}\equiv g^{\beta\gamma}\kappa_{\alpha\gamma}$ is the shape operator (Weingarten map, up to sign convention), whose eigenvalues $\kappa_1$ and $\kappa_2$ are the principal curvatures. The mean curvature and Gaussian curvature are given, respectively, by
${\mathcal H}_\Gamma\equiv \tfrac12\,\mathrm{tr}(\kappa_\alpha^{\ \beta})
=\tfrac12\,g^{\alpha\beta}\kappa_{\alpha\beta}
=\tfrac12\,\nabla_{\Gamma}\cdot \hat{\boldsymbol n}_\Gamma
=\tfrac12(\kappa_1+\kappa_2)$, and $\kappa_{\mathrm G}\equiv \det(\kappa_\alpha^{\ \beta})
={\det(\kappa_{\alpha\beta})}/{\det(g_{\alpha\beta})}
=\kappa_1\kappa_2$, 
where $\nabla_{\Gamma}\cdot$ is defined below; with this sign convention, a sphere with outward normal has $\kappa_1=\kappa_2={\mathcal H}_\Gamma>0$. Differentiating $g_{\alpha\beta}=\boldsymbol e_\alpha\cdot \boldsymbol e_\beta$ ($g^{\alpha\beta}=\boldsymbol e^\alpha\cdot \boldsymbol e^\beta$) with respect to $\xi^\gamma$ and using Eq.~\eqref{Eq:AppA-GW} gives the metric-compatibility condition
$\partial_\gamma g_{\alpha\beta}
=\Gamma^\delta_{\gamma\alpha}g_{\delta\beta}
+\Gamma^\delta_{\gamma\beta}g_{\alpha\delta}$ ($\partial_\gamma g^{\alpha\beta}
=-\Gamma^\alpha_{\gamma\delta}g^{\delta\beta}-\Gamma^\beta_{\gamma\lambda}g^{\lambda\alpha}$), or equivalently, $g_{\alpha\beta;\gamma}=0$ ($g^{\alpha\beta}_{\,\,\,\,;\gamma}=0$), where $;$ denotes the covariant derivative of a surface tensor. Together with Eqs.~\eqref{Eq:AppA-GW}, these relations complete the geometric framework for an embedded surface: intrinsic derivatives are determined by $g_{\alpha\beta}$ and $\Gamma^\gamma_{\alpha\beta}$, whereas the extrinsic geometry enters through $\kappa_{\alpha\beta}$ and $\hat{\boldsymbol n}_\Gamma$.

\textbf{Surface differential operators} --- Surface derivatives are defined intrinsically. For a general surface field $\boldsymbol{a}_\Gamma(\boldsymbol{\xi},t)$ of arbitrary rank $m$, the covariant surface derivative of its mixed tensor component ($s$ covariant components, $r$ contra-variant components and $m=r+s$) is given by
\begin{equation}
(a_\Gamma)^{\alpha_1\cdots \alpha_r}{}_{\beta_1\cdots \beta_{\mathrm s};\gamma}=
\partial_\gamma (a_\Gamma)^{\alpha_1\cdots \alpha_r}{}_{\beta_1\cdots \beta_{\mathrm s}}
+\sum_{i=1}^{r}\Gamma^{\alpha_i}_{\gamma\mu}\,
(a_\Gamma)^{\alpha_1\cdots \mu \cdots \alpha_r}{}_{\beta_1\cdots \beta_{\mathrm s}}
-\sum_{j=1}^{s}\Gamma^{\mu}_{\gamma\beta_j}\,
(a_\Gamma)^{\alpha_1\cdots \alpha_r}{}_{\beta_1\cdots \mu \cdots \beta_{\mathrm s}}.
\end{equation}
Surface derivatives of surface fields can also be defined conveniently using the tangential projector ${\boldsymbol P}_\Gamma $.
\begin{subequations}
For example, for a scalar surface field $a_\Gamma$ and its extension $a$, we define
\begin{equation}
\nabla_\Gamma a_\Gamma
\equiv  e^\alpha \partial_\alpha a_\Gamma=\boldsymbol P_\Gamma \cdot \nabla a\big|_{\Gamma},
\qquad
\Delta_\Gamma a_\Gamma
\equiv \frac{1}{\sqrt g}\partial_\alpha
\left(\sqrt g\,g^{\alpha\beta}\partial_\beta a_\Gamma\right)=\nabla_\Gamma\cdot\nabla_\Gamma a_\Gamma. 
\end{equation}
For a tangential vector field $\boldsymbol a_\Gamma=a^\alpha e_\alpha$ and its extension $\boldsymbol a$, we define the intrinsic projected gradient and projected rough Laplacian as
\begin{equation}
\bbNabla_\Gamma \boldsymbol a_\Gamma
\equiv \boldsymbol P_\Gamma\cdot(\nabla \boldsymbol a)\cdot \boldsymbol P_\Gamma\big|_\Gamma, 
\qquad
\bbDelta_\Gamma \boldsymbol a_\Gamma
\equiv \bbNabla_\Gamma\cdot\bbNabla_\Gamma \boldsymbol a_\Gamma,
\end{equation} 
with $(\bbNabla_\Gamma \boldsymbol a_\Gamma)^\alpha{}_\beta
=a^\alpha{}_{;\beta}$. 
For a right-tangential flux $\boldsymbol J_{\mathrm{a}\Gamma}$ carrying a final flux index, the intrinsic divergence is
\begin{equation} 
(\bbNabla_\Gamma\cdot \boldsymbol J_{\mathrm{a}\Gamma})^{\alpha_1\cdots\alpha_r}
\equiv (\boldsymbol J_{\mathrm{a}\Gamma})^{\alpha_1\cdots\alpha_r\beta}{}_{;\beta}.
\end{equation}
For a fully tangential rank-two surface stress $\boldsymbol a_\Gamma$, the intrinsic projected covariant divergence and ambient surface divergence are
\begin{equation} 
(\bbNabla_\Gamma\cdot\boldsymbol a_\Gamma)^\alpha
=a_\Gamma^{\alpha\beta}{}_{;\beta},
\qquad
\nabla_\Gamma\cdot\boldsymbol a_\Gamma
=\bbNabla_\Gamma\cdot\boldsymbol a_\Gamma
-(\boldsymbol a_\Gamma:\boldsymbol \kappa_\Gamma)\hat {\boldsymbol n}_\Gamma.
\end{equation}    
\end{subequations}
Here, $\boldsymbol{a}(\boldsymbol{x}(\boldsymbol{\xi},t),t)$ is a smooth normal extension of $\boldsymbol{{a}}_\Gamma(\boldsymbol{\xi},t)$ from $\Gamma$ to a tubular neighborhood $\boldsymbol{x}(\boldsymbol{\xi},t)=\boldsymbol{X}(\boldsymbol{\xi},t)+{\mathcal{D}}\boldsymbol{\hat{n}}$ (we typically use the zero-normal extension $\partial_{\mathcal{D}} \boldsymbol{a}=0$) with $\boldsymbol{{a}}_\Gamma(\boldsymbol{\xi},t)=\boldsymbol{a}(\boldsymbol{X}(\boldsymbol{\xi},t),t)$, and we use the chain rule $\partial_\alpha a_{\Gamma i}=\boldsymbol{e}_\alpha\cdot \nabla a_i$. Note that a surface vector/tensor field may have nonzero normal components; by contrast, a surface-tangential vector/tensor field has only tangential components (all normal components vanish). 

The membrane material velocity is defined and decomposed as $\boldsymbol V_\Gamma(\boldsymbol{\xi},t)\equiv d\boldsymbol{X}/dt \equiv\left.\partial_t\boldsymbol{X}(\boldsymbol{\xi},t)\right|_{\boldsymbol{\xi}} =\boldsymbol V_{\Gamma\parallel} + V_{\Gamma n}\,{\boldsymbol{\hat{n}}}$, with the surface tangential velocity ${\boldsymbol{V}}_{\Gamma\parallel}\equiv {\boldsymbol P}_\Gamma  \cdot {\boldsymbol{V}_{\Gamma}}$ and $V_{\Gamma n}=\boldsymbol{V}_{\Gamma} \cdot \boldsymbol{\hat{n}}$. 
The material time derivative of the surface field $\boldsymbol{a}_\Gamma(\boldsymbol{X}(\boldsymbol{\xi},t),t)$ is defined via $\dot{\boldsymbol{a}}_\Gamma\equiv d{\boldsymbol{a}}_\Gamma/dt\equiv \partial_t\boldsymbol{a}_\Gamma|_{\boldsymbol{\xi}}=\partial_t \boldsymbol{a}_\Gamma+{\boldsymbol{V}}_{\Gamma\parallel} \cdot\nabla_\Gamma \boldsymbol{a}_\Gamma$. 
For scalar fields, the material derivative is a dissipative rate, while for tangential vector and tensor internal variables, this material derivative is not itself an admissible dissipative rate; it must be projected onto the tangent bundle and, when appropriate, replaced by an objective/co-rotational rate.  

The area of the surface element is given by $dA=\sqrt{g}\,d\xi^1 d\xi^2$; using $\dot{\boldsymbol{e}}_\alpha=\partial_t\boldsymbol{e}_\alpha|_{\boldsymbol{\xi}}=\partial_\alpha {\boldsymbol{V}}_\Gamma$, the identity $\partial g/\partial g_{\alpha\beta}=g\,g^{\alpha\beta}$, and differentiating $g_{\alpha\beta}=\boldsymbol{e}_\alpha\cdot \boldsymbol{e}_\beta$ over time at fixed $\xi$, we obtain~\cite{Seifert1997AdvPhys,ArroyoDesimone2009PRE}
\begin{equation}\label{Eq:SI-area-rate-local}
\frac{1}{dA}\frac{d(dA)}{dt}
= \nabla_\Gamma\cdot \boldsymbol V_\Gamma
= \nabla_\Gamma\cdot \boldsymbol V_{\Gamma\parallel} + 2{\mathcal H}_\Gamma\,V_{\Gamma n}, 
\end{equation} 
which is purely kinematic: surface area changes only through tangential convergence 
$\nabla_\Gamma\cdot\boldsymbol V_{\Gamma\parallel}$ and through normal motion coupled to curvature ($2{\mathcal H}_\Gamma V_{\Gamma n}$). Local surface inextensibility (area incompressibility) is therefore $\nabla_\Gamma\cdot{\boldsymbol{V}_\Gamma}=0$, which reduces to $\nabla_\Gamma \cdot{\boldsymbol{V}}_{\Gamma\parallel}=0$ on a flat surface and requires in-plane convergence to compensate normal motion on curved surfaces.

\textbf{Surface integral theorems} --- On a smooth material surface patch $S(t)\subset\Gamma(t)$ with outward co-normal $\hat{\boldsymbol\nu}_S$, the surface divergence theorem is~\cite{Aris1962Book,Federer1969GMT}: $\int_{S}\nabla_\Gamma\!\cdot\!\boldsymbol{a}_\Gamma\,dA=\int_{\partial S}\boldsymbol{a}_\Gamma\!\cdot\!\hat{\boldsymbol\nu}_S\,d\ell$ for right-tangential $\boldsymbol a_\Gamma\cdot\hat{\boldsymbol n}_\Gamma=0$. 

If the surface patch $S(t)$ is convected by the velocity $\boldsymbol V_\Gamma$, the surface Reynolds transport theorem (SRTT) reads
\begin{equation}\label{Eq:SI-SRTT}
\frac{d}{dt}\int_{ S(t)} \boldsymbol{a}_\Gamma\,dA
=\int_{S(t)} \left(\dot{\boldsymbol{a}}_\Gamma+\boldsymbol{a}_\Gamma\,\nabla_\Gamma\cdot\boldsymbol V_\Gamma\right)dA
=\int_{S(t)} \left(\dot{\boldsymbol{a}}_\Gamma+\boldsymbol{a}_\Gamma\,\nabla_\Gamma\cdot\boldsymbol V_{\Gamma\parallel}+2{\mathcal H}_\Gamma \boldsymbol{a}_\Gamma V_{\Gamma n}\right)dA.
\end{equation} 
where we have used the surface material derivative $\dot{\boldsymbol{a}}_\Gamma$ and the local area-rate identity \eqref{Eq:SI-area-rate-local}. For a closed surface with $S(t)=\Gamma(t)$ and $\partial\Gamma=\varnothing$, taking $\boldsymbol{a}_\Gamma \to 1$ gives the change rate of the area of the closed membrane surface $\Gamma(t)$: $\frac{d}{dt}\int_{\Gamma(t)} dA=\int_{\Gamma(t)} 2{{\mathcal H}_\Gamma} V_{\Gamma n}\,dA$,  since $\int_{\Gamma(t)}\nabla_\Gamma\cdot\boldsymbol V_{\Gamma\parallel}\,dA=0$ on a closed surface. Similarly, the volume enclosed by $\Gamma(t)$ satisfies $\frac{d}{dt}\int_{\Omega_-(t)} dV=\int_{\Gamma(t)} V_{\Gamma n}\,dA$. 

The SRTT also provides a systematic route to the surface balance equations. For the surface field $\boldsymbol{a}_\Gamma$, let $\boldsymbol J_{\mathrm{a}\Gamma}$ be its surface flux (tangent to $\Gamma$, \emph{i.e.}, $\boldsymbol J_{\mathrm{a}\Gamma}\cdot\hat{\boldsymbol n}_\Gamma=0$), defined relative to the moving membrane, and let $\boldsymbol{k}_{\mathrm{a}\Gamma}$ be its surface source (production) rate per unit area. Then, the global (integral) and local balance equations read
\begin{subequations}\label{Eq:AppA-surface-balance}
\begin{equation}\label{Eq:AppA-surface-balance-global}
\frac{d}{dt}\int_{S(t)} \boldsymbol{a}_\Gamma\,dA =-\int_{\partial S(t)}\boldsymbol J_{\mathrm{a}\Gamma}\cdot \hat{\boldsymbol{\nu}}_S\,ds +\int_{S(t)} \boldsymbol k_{\mathrm{a}\Gamma}\,dA,
\end{equation}
\begin{equation}\label{Eq:AppA-surface-balance-local}
\dot{\boldsymbol{a}}_\Gamma+\boldsymbol{a}_\Gamma\, \nabla_\Gamma\cdot \boldsymbol V_\Gamma   
=-\bbNabla_\Gamma\!\cdot\boldsymbol J_{\mathrm{a}\Gamma}+ \boldsymbol k_{\mathrm{a}\Gamma},
\end{equation}
\end{subequations}
respectively, where $\hat{\boldsymbol{\nu}}_S$ is the outward co-normal vector along the boundary curve $\partial S(t)$, $\nabla_\Gamma\cdot \boldsymbol V_\Gamma$ expands as Eq.~\eqref{Eq:SI-area-rate-local}, and we have used the SRTT in Eq.~\eqref{Eq:SI-SRTT}. Equation~\eqref{Eq:AppA-surface-balance-local} is the sharp counterpart of the conservative embedded balance in Sec.~\ref{sec:DDM}; it will be used repeatedly below. In particular, if $\boldsymbol{a}_\Gamma\equiv  1$ and $\boldsymbol{k}_{\mathrm{a}\Gamma}=0$, then local inextensibility is $\nabla_\Gamma\!\cdot\!\boldsymbol V_\Gamma=\nabla_\Gamma\cdot \boldsymbol V_{\Gamma\parallel} + 2{\mathcal H}_\Gamma\,V_{\Gamma n}=0$.

\subsection{Sharp-interface model: OVP formulation consistent with Sec.~V} \label{sec:App-Membrane-OVP}

In Sec.~\ref{sec:MPF-vesicle}, we provide a passive \emph{extensible}-surface model with the areal-compression field $c_{\mathrm a\Gamma}$, the chemical concentration $c_{\mathrm m\Gamma}$, and the tangential velocity $\boldsymbol V_{\Gamma\parallel}$, including the scalar membrane free energies, passive viscous surface dynamics, and the bulk hydrodynamics inside and outside the closed membrane surface. In Sec.~\ref{sec:ActPolar}, we further add the surface tangential polarization $\boldsymbol p_\Gamma$ and the active surface stress. The sharp model presented in this appendix is the material-interface, strong-tangency limit of Sec.~\ref{sec:ActPolar}. Specifically, we set $V_{\Gamma n}=\boldsymbol v_{\mathrm b}\cdot\hat{\boldsymbol n}_\Gamma$ and we impose the sharp tangency constraint $\boldsymbol p_\Gamma\cdot\hat{\boldsymbol n}_\Gamma=0$. 
A convenient sharp/diffuse correspondence is
\begin{align}\label{Eq:AppA-dictionary}
&c_{\mathrm a}\to c_{\mathrm a\Gamma},\qquad
c_{\mathrm m}\to c_{\mathrm m\Gamma},\qquad
\boldsymbol{p} \to \boldsymbol{p}_\Gamma,\qquad
\delta_\epsilon[\psi]\,dV\to dA, \qquad
\boldsymbol P_\epsilon\to\boldsymbol P_\Gamma,\nonumber \\
&\hat{\boldsymbol n}_\epsilon\to\hat{\boldsymbol n}_\Gamma,\qquad
\boldsymbol\kappa_\epsilon\to \boldsymbol\kappa_\Gamma,\quad
\,\,\,\,\,\,\,\nabla_{\mathrm s}\to\nabla_\Gamma,\qquad
\bbNabla_{\mathrm s}\to\bbNabla_\Gamma.
\end{align}
Equivalently, at the level of conservative measures, $\delta_\epsilon c_{\mathrm a}\,dV\to c_{\mathrm a\Gamma}\,dA$, $\delta_\epsilon c_{\mathrm m}\,dV\to c_{\mathrm m\Gamma}\,dA$,
and similarly for $\boldsymbol p$.
Correspondingly, the conjugate variables map as $\hat\mu_{\mathrm a}\to \hat\mu_{\mathrm a\Gamma}$, $\hat\mu_{\mathrm m}\to \hat\mu_{\mathrm m\Gamma}$, $\boldsymbol h_{\mathrm p}^\parallel\to\boldsymbol h_{p\Gamma}$, and $\boldsymbol\sigma_{\mathrm s}^{\mathrm{act}}=-\zeta\boldsymbol q_\parallel \to \boldsymbol\sigma^{\mathrm{act}}_\Gamma =-\zeta \boldsymbol q_\Gamma$. This is exactly the sharp-interface reduction stated in Sec.~\ref{sec:DDM}: the conservative embedded variable is $\delta_\epsilon {\boldsymbol{a}}$, and the sharp integral and local balance are Eqs.~\eqref{Eq:DDM-SI-bal}. 

\textbf{State variables and constraints} ---
We let the moving surface $\Gamma(t)$ carry the scalar fields $c_{\mathrm a\Gamma}$ and $c_{\mathrm m\Gamma}$, the tangential polarization $\boldsymbol p_\Gamma$ with $\boldsymbol p_\Gamma\!\cdot\!\hat{\boldsymbol n}_\Gamma=0$, and the surface material velocity
$\boldsymbol V_\Gamma=\boldsymbol V_{\Gamma\parallel}+V_{\Gamma n}\hat{\boldsymbol n}_\Gamma$.
The surrounding inner and outer bulk fluids occupy $\Omega_-(t)$ and $\Omega_+(t)$ and have velocities $\boldsymbol v^\pm$ and stresses $\boldsymbol\sigma_{\mathrm b}^\pm
=2\eta_{\mathrm b}^\pm\boldsymbol D_{\mathrm b}^\pm-P^\pm\boldsymbol I$, with $\boldsymbol D_{\mathrm b}^\pm
= \frac12\left(\nabla\boldsymbol v^\pm+\nabla\boldsymbol v^{\pm\mathrm T}\right)$. 

To retain tangential slip explicitly, the sharp-surface kinematics must \emph{not} impose the full no-slip condition $\boldsymbol V_\Gamma=\boldsymbol v^\pm|_{\Gamma(t)}$. Instead, only the normal kinematic condition is imposed,
\begin{equation}\label{Eq:AppA-kinematic-slip}
V_{\Gamma n}=\boldsymbol v_{\mathrm b}\!\cdot\!\hat{\boldsymbol n}_\Gamma,
\qquad
\boldsymbol v_{\mathrm b}\equiv \boldsymbol v^+|_{\Gamma(t)}=\boldsymbol v^-|_{\Gamma(t)},
\qquad
\boldsymbol v_{b\parallel}\equiv \boldsymbol P_\Gamma\!\cdot\!\boldsymbol v_{\mathrm b},
\end{equation}
while the surface tangential velocity $\boldsymbol V_{\Gamma\parallel}$ is kept as an independent variable. This is the sharp-interface counterpart of the two-velocity diffuse model in Sec.~\ref{sec:MPF-vesicle}.

The two scalar balances are the direct sharp-surface counterparts of Eqs.~\eqref{Eq:ActPolar-bal-ca} and \eqref{Eq:ActPolar-bal-cm}:
\begin{subequations}\label{Eq:AppA-balances}
\begin{equation}\label{Eq:AppA-bal-ca}
\dot c_{\mathrm a\Gamma}+c_{\mathrm a\Gamma}\,\nabla_\Gamma\!\cdot\!\boldsymbol V_\Gamma=0,
\end{equation}
\begin{equation}\label{Eq:AppA-bal-cm}
\dot c_{\mathrm m\Gamma}+c_{\mathrm m\Gamma}\,\nabla_\Gamma\!\cdot\!\boldsymbol V_\Gamma
=-\nabla_\Gamma\!\cdot\!\boldsymbol J_{\mathrm m\Gamma}+k_{\mathrm m\Gamma},
\end{equation}
\end{subequations}
where the nonconvective flux $\boldsymbol J_{\mathrm m\Gamma}$ is tangential and 
$k_{\mathrm m\Gamma} \equiv -k_{\mathrm d}c_{\mathrm m\Gamma}+k_{\mathrm p}(c_{\mathrm a\Gamma},\mathcal H_\Gamma)$
is the sharp counterpart of the source used in Sec.~\ref{sec:ActPolar}.
For the polarization, the sharp-surface co-rotational rate is taken along the \emph{full} material motion of the surface:
\begin{equation}\label{Eq:AppA-Pdot}
\dot{\mathbb P}_{\Gamma\parallel} \equiv \boldsymbol P_\Gamma\!\cdot\!\left(
\dot{\boldsymbol p}_\Gamma -\boldsymbol\Omega_{\Gamma\parallel}\!\cdot\!\boldsymbol p_\Gamma
\right),
\end{equation}
where $\dot{\boldsymbol p}_\Gamma \equiv \partial_t\boldsymbol p_\Gamma+\boldsymbol V_\Gamma\!\cdot\!\nabla_\Gamma\boldsymbol p_\Gamma$ is the material derivative along $\boldsymbol V_\Gamma$ and $\boldsymbol\Omega_{\Gamma\parallel}\equiv \frac12\left(\bbNabla_\Gamma\boldsymbol V_{\Gamma\parallel}-\bbNabla_\Gamma\boldsymbol V_{\Gamma\parallel}^{\mathrm T}\right)$. In a closest-point extension, $\dot{\boldsymbol p}_\Gamma$ corresponds to $\dot{\boldsymbol{p}}=\partial_t\boldsymbol p+\boldsymbol V\!\cdot\!\nabla\boldsymbol p$ in the diffuse-domain model, so Eq.~\eqref{Eq:AppA-Pdot} is the sharp-interface counterpart of Eq.~\eqref{Eq:ActPolar-Pdot}.

\textbf{Free energy functional} --- We write the sharp-surface free energy as
$\mathcal F_\Gamma=\mathcal F_{\Gamma,\mathrm{sh}}+\mathcal F_{\Gamma,\mathrm c}+\mathcal F_{\Gamma,\mathrm p}$ with
\begin{align}\label{Eq:AppA-F}
\mathcal F_{\Gamma,\mathrm{sh}}
&=
\int_{\Gamma(t)}\!\left[
\frac12K_{\mathrm b}(c_{\mathrm m\Gamma})\left(2\mathcal H_\Gamma-\kappa_0(c_{\mathrm m\Gamma})\right)^2
+\gamma(c_{\mathrm a\Gamma},c_{\mathrm m\Gamma})
\right]dA,
\nonumber\\
\mathcal F_{\Gamma,\mathrm c}
&=
\gamma_{\mathrm L}\int_{\Gamma(t)}\!\left[
\epsilon_{\mathrm m}^{-1}f_{\mathrm m}(c_{\mathrm m\Gamma})
+\frac{\epsilon_{\mathrm m}}2|\nabla_\Gamma c_{\mathrm m\Gamma}|^2
\right]dA,
\nonumber\\
\mathcal F_{\Gamma,\mathrm p}
&=
\int_{\Gamma(t)}\!\left[
-\frac{a_0}{2}|\boldsymbol p_\Gamma|^2
+\frac{a_0}{4}|\boldsymbol p_\Gamma|^4
+\frac{K_{\mathrm p}}2|\bbNabla_\Gamma\boldsymbol p_\Gamma|^2
-\frac\chi2\left(\boldsymbol p_\Gamma\!\cdot\!\nabla_\Gamma c_{\mathrm m\Gamma}\right)^2
- \Lambda_{\mathrm{cp}} \boldsymbol{q}_\Gamma:\boldsymbol{\tilde\kappa}_\Gamma\right]dA,
\end{align}
with $\boldsymbol q_\Gamma \equiv \boldsymbol p_\Gamma\boldsymbol p_\Gamma -\frac12|\boldsymbol p_\Gamma|^2\boldsymbol P_\Gamma$, and $\boldsymbol{\tilde\kappa}_\Gamma \equiv \boldsymbol{\kappa}_\Gamma-{\mathcal H}_\Gamma \boldsymbol{P}_\Gamma$ being the deviatoric curvature tensor. 
This is the sharp-interface counterpart of Eqs.~\eqref{Eq:MPF-vesicle-Fpsi}, \eqref{Eq:MPF-vesicle-Fm}, and \eqref{Eq:ActPolar-Fp}, with $\gamma_\psi$ renamed simply as $\gamma$ in the sharp theory.
For consistency with Eq.~\eqref{Eq:MPF-vesicle-gamma-a}, we take
\begin{equation}
\gamma(c_{\mathrm a\Gamma},c_{\mathrm m\Gamma})=\gamma_0+
\frac{K_{\mathrm a}(c_{\mathrm m\Gamma})}{2} \left(c_{\mathrm a\Gamma}+c_{\mathrm a\Gamma}^{-1}-2\right).
\end{equation}

Using Eqs.~\eqref{Eq:AppA-balances}, the surface transport identity~\eqref{Eq:AppA-surface-balance}, and integration by parts on the closed surface, the free-energy rate can be written as
\begin{align}\label{Eq:AppA-Fdot}
\dot{\mathcal F}_\Gamma={}&
\int_{\Gamma(t)}\!\hat{\mu}_{\Gamma n}V_{\Gamma n}\,dA +\int_{\Gamma(t)}\!\boldsymbol V_{\Gamma\parallel}\!\cdot\!\Big[
- \bbNabla_\Gamma\!\cdot\! \boldsymbol\sigma_{\mathrm p\Gamma}^{\mathrm{asy}}  
+c_{\mathrm a\Gamma} \nabla_\Gamma\hat{\mu}_{\mathrm a\Gamma} + c_{\mathrm m\Gamma}\nabla_\Gamma\hat{\mu}_{\mathrm m\Gamma} +(\bbNabla_\Gamma\boldsymbol p_\Gamma)^{\mathrm T}\!\cdot\!\boldsymbol h_{\mathrm p\Gamma}
\Big]dA \nonumber\\
&\qquad
+\int_{\Gamma(t)}\!\left(
\boldsymbol J_{\mathrm m\Gamma}\!\cdot\!\nabla_\Gamma\hat{\mu}_{\mathrm m\Gamma}+k_{\mathrm m\Gamma}\hat{\mu}_{\mathrm m\Gamma}
\right)dA -\int_{\Gamma(t)}\!\boldsymbol h_{\mathrm p\Gamma}\!\cdot\!\dot{\mathbb P}_{\Gamma\parallel}\,dA,
\end{align}
where $\delta\mathcal F_\Gamma/\delta\boldsymbol X=f_{\Gamma}^{\perp}\hat{\boldsymbol n}_\Gamma$ denotes the variational surface-force density generated by $\mathcal F_\Gamma$, and
\begin{equation}\label{Eq:AppA-mun}
\hat{\mu}_{\Gamma n}
\equiv
f_{\Gamma}^{\perp}
-2\mathcal H_\Gamma\left(c_{\mathrm a\Gamma}\hat{\mu}_{\mathrm a\Gamma}+c_{\mathrm m\Gamma}\hat{\mu}_{\mathrm m\Gamma}\right)
\end{equation}
is the normal driving force associated with shape relaxation. 
The quantity $f_\Gamma^\perp$ includes the normal variations of all terms in $\mathcal F_\Gamma$, including the polar distortion and curvature--polarization coupling contributions. Particularly, a pure tension energy $\mathcal F_\Gamma=\int_\Gamma \gamma\,dA$ gives $f_\Gamma^\perp=2\mathcal H_\Gamma\gamma$. 

Let $\hat{f}_\Gamma$ denote the total surface free-energy density in Eq.~\eqref{Eq:AppA-F}. The scalar chemical potentials are defined, as in Sec.~\ref{sec:DDM}, by
$\hat{\mu}_{\alpha\Gamma}=\partial_{c_{\alpha\Gamma}}\hat{f}_\Gamma-\nabla_\Gamma\!\cdot\!\boldsymbol\pi_{\alpha\Gamma}$ with
$\boldsymbol\pi_{\alpha\Gamma}\equiv \partial_{\nabla_\Gamma c_{\alpha\Gamma}}\hat{f}_\Gamma$ for $\alpha\in\{\mathrm a,\mathrm m\}$. Since there is no gradient term in the advected areal-compression variable $c_{\mathrm a\Gamma}$, the associated chemical potential is
\begin{equation}\label{Eq:AppA-mua}
\hat\mu_{\mathrm a\Gamma}
=
\partial_{c_{\mathrm a\Gamma}}
\gamma(c_{\mathrm a\Gamma},c_{\mathrm m\Gamma})
=
\frac{K_{\mathrm a}(c_{\mathrm m\Gamma})}{2}
\left(1-c_{\mathrm a\Gamma}^{-2}\right).
\end{equation}
The corresponding thermodynamic dilational tension is $\sigma_{\mathrm a\Gamma} =\gamma-c_{\mathrm a\Gamma}\hat\mu_{\mathrm a\Gamma}=\gamma_0+K_{\mathrm a}(c_{\mathrm m\Gamma})\left(c_{\mathrm a\Gamma}^{-1}-1\right)$. 
For the morphogen field,
$\boldsymbol\pi_{\mathrm m\Gamma}
= \gamma_{\mathrm L}\epsilon_{\mathrm m}\nabla_\Gamma c_{\mathrm m\Gamma}
-\chi(\boldsymbol p_\Gamma\!\cdot\!\nabla_\Gamma c_{\mathrm m\Gamma})\boldsymbol p_\Gamma$,
so that
\begin{equation}\label{Eq:AppA-mum}
\hat{\mu}_{\mathrm m\Gamma} = \partial_{c_{\mathrm m\Gamma}}\hat{f}_\Gamma-\nabla_\Gamma\!\cdot\!\boldsymbol\pi_{\mathrm m\Gamma}.
\end{equation}
Equation~\eqref{Eq:AppA-mun} is the intrinsic sharp-surface analogue of the shifted diffuse geometry chemical potential $\bar\mu_\psi$ in Eq.~\eqref{Eq:MPF-vesicle-mupsi} that appears in the reduced conserved-geometry formulation of Secs.~\ref{sec:MPF-vesicle} and~\ref{sec:ActPolar}. The extra term $-2\mathcal H_\Gamma\left(c_{\mathrm a\Gamma}\hat\mu_{\mathrm a\Gamma}+c_{\mathrm m\Gamma}\hat\mu_{\mathrm m\Gamma}\right)$ is the sharp-surface consequence of transporting the conserved surface densities $c_{\mathrm a\Gamma}$ and $c_{\mathrm m\Gamma}$ by a moving area element. 
Equation~\eqref{Eq:AppA-mum} is the intrinsic sharp-surface counterpart of the morphogen chemical potential obtained from the full active free energy in Eq.~\eqref{Eq:ActPolar-Ftot}; in particular, it includes the curvature-, tension-, and polarization-alignment contributions. For the polarization, we define the distortion tensor and molecular field by
\begin{equation}\label{Eq:AppA-hp}
\boldsymbol h_{\mathrm p\Gamma}
=-\boldsymbol{P}_\Gamma\cdot \frac{\delta \mathcal F_\Gamma}{\delta \boldsymbol p_\Gamma}
=\boldsymbol{P}_\Gamma\cdot \left[
a_0\boldsymbol p_\Gamma
-a_0|\boldsymbol p_\Gamma|^2\boldsymbol p_\Gamma
+\chi(\boldsymbol p_\Gamma\!\cdot\!\nabla_\Gamma c_{\mathrm m\Gamma})\nabla_\Gamma c_{\mathrm m\Gamma}
+K_{\mathrm p}\bbDelta_\Gamma \boldsymbol p_\Gamma+ 2\Lambda_{\mathrm{cp}}\tilde{\boldsymbol\kappa}_\Gamma\cdot \boldsymbol p_\Gamma\right],
\end{equation}
where $\hat f_{\mathrm p\Gamma}$ denotes the integrand in $\mathcal F_{\Gamma,\mathrm p}$ and $\bbDelta_\Gamma\equiv \bbNabla_\Gamma\cdot\bbNabla_\Gamma$ is projected rough Laplacian on tangential vector fields. 

In the intrinsic sharp-surface formulation, $\boldsymbol h_{\mathrm p\Gamma}$ is tangential by construction. 
The antisymmetric reactive stress associated with the co-rotational kinematics is
\begin{equation}\label{Eq:AppA-sigmaA}
\boldsymbol\sigma_{\mathrm p\Gamma}^{\mathrm{asy}}
\equiv
\frac12\left(\boldsymbol p_\Gamma\boldsymbol h_{\mathrm p\Gamma}-\boldsymbol h_{\mathrm p\Gamma}\boldsymbol p_\Gamma\right).
\end{equation}

\textbf{Dissipation functional} --- The passive sharp-surface dissipation is the hydrodynamically coupled sharp-surface analogue of Eq.~\eqref{Eq:ActPolar-Phi}:
\begin{equation}\label{Eq:AppA-Phi}
\Phi_\Gamma
=
\sum_{\pm}\int_{\Omega_\pm(t)}\!\eta_{\mathrm b}^\pm\,\boldsymbol D_{\mathrm b}^\pm\!:\!\boldsymbol D_{\mathrm b}^\pm\,dV+
\int_{\Gamma(t)}\!\left[
\eta_{\mathrm s}\, \mathbb{D}_{\Gamma}(\boldsymbol V_\Gamma)\!:\!\mathbb{D}_{\Gamma}(\boldsymbol V_\Gamma)
+\frac{\beta}{2}\big|\boldsymbol V_{\Gamma\parallel}-\boldsymbol v_{b\parallel}\big|^2
+\frac{|\boldsymbol J_{\mathrm m\Gamma}|^2}{2M_{\mathrm m}}
+\frac{\gamma_{\mathrm p}}{2}\left|\dot{\mathbb P}_{\Gamma\parallel}+\nu_{\mathrm p}\,\mathbb{D}_{\Gamma}(\boldsymbol V_\Gamma)\!\cdot\!\boldsymbol p_\Gamma\right|^2
\right]dA.
\end{equation}
Here, $\mathbb{D}_{\Gamma}(\boldsymbol V_\Gamma)\equiv \frac12\left(\bbNabla_\Gamma\boldsymbol V_{\Gamma}+\bbNabla_\Gamma\boldsymbol V_{\Gamma}^{\mathrm T}\right) =\mathbb{D}_{\Gamma}(\boldsymbol V_{\Gamma\parallel})+V_{\Gamma n}\boldsymbol\kappa_\Gamma$,  $\mathbb{D}_{\Gamma}(\boldsymbol V_{\Gamma\parallel})\equiv \frac12\left(\bbNabla_\Gamma\boldsymbol V_{\Gamma\parallel}+\bbNabla_\Gamma\boldsymbol V_{\Gamma\parallel}^{\mathrm T}\right)$, the tangential slip coefficient is denoted by the same symbol $\beta$ as in Eq.~\eqref{Eq:MPF-vesicle-Phi}. The use of $\mathbb{D}_{\Gamma}(\boldsymbol V_\Gamma)$ gives the full Boussinesq–Scriven metric-rate model. 


\textbf{Active work power} --- Activity enters through the symmetric active stress $\boldsymbol\sigma^{\mathrm{act}}_\Gamma=-\zeta\,\boldsymbol q_\Gamma$, and the corresponding active work power
\begin{equation}\label{Eq:AppA-Pact}
\mathcal P^{\mathrm{act}}_\Gamma[\boldsymbol V_\Gamma]=-\int_{\Gamma(t)} \boldsymbol\sigma^{\mathrm{act}}_\Gamma\!:\!\mathbb{D}_{\Gamma}(\boldsymbol V_\Gamma)\,dA.
\end{equation}
The traceless tensor $\boldsymbol q_\Gamma\equiv \boldsymbol p_\Gamma\boldsymbol p_\Gamma-\tfrac12|\boldsymbol p_\Gamma|^2\boldsymbol P_\Gamma$ is the sharp-surface analogue of Eq.~\eqref{Eq:ActPolar-sigmaact}.

\textbf{Rayleighian and dynamic equations} --- Using Eqs.~\eqref{Eq:AppA-balances}, we  minimize the sharp-interface Rayleighian $\mathcal R_\Gamma
= \dot{\mathcal F}_\Gamma+\Phi_\Gamma-\mathcal P^{\mathrm{act}}_\Gamma
- \sum_{\pm}\int_{\Omega_\pm(t)} \!P^\pm\,\nabla\!\cdot\!\boldsymbol v^\pm\,dV$ with respect to $\boldsymbol v^\pm$, $\boldsymbol V_{\Gamma}$, $\boldsymbol J_{\mathrm m\Gamma}$, and $\dot{\mathbb P}_{\Gamma\parallel}$, subject to the normal kinematic condition~\eqref{Eq:AppA-kinematic-slip}. This yields the hydrodynamically coupled sharp-interface system
\begin{subequations}\label{Eq:AppA-DynEqns}
\begin{equation}\label{Eq:AppA-bulk}
-\nabla\!\cdot\! \boldsymbol\sigma_{\mathrm b}^\pm=0,
\qquad
\nabla\!\cdot\!\boldsymbol v^\pm=0,
\qquad \text{in }\Omega_\pm(t),
\end{equation}
\begin{equation}\label{Eq:AppA-normal}
\hat{\boldsymbol n}_\Gamma\!\cdot\! \llbracket \boldsymbol\sigma_{\mathrm b} \rrbracket\!\cdot\!\hat{\boldsymbol n}_\Gamma
-\hat{\mu}_{\Gamma n}-
\boldsymbol\Sigma_\Gamma^{\mathrm{sym}}:\boldsymbol\kappa_\Gamma=0,
\end{equation}
\begin{equation}\label{Eq:AppA-tanbulk}
\boldsymbol P_\Gamma\!\cdot\!\left( \llbracket \boldsymbol\sigma_{\mathrm b} \rrbracket\!\cdot\!\hat{\boldsymbol n}_\Gamma\right)
+\beta\left(\boldsymbol V_{\Gamma\parallel}-\boldsymbol v_{b\parallel}\right)=0,
\end{equation} 
\begin{equation}\label{Eq:AppA-forcebalance}
\bbNabla_\Gamma\!\cdot\! \boldsymbol{\Sigma}_\Gamma  
-c_{\mathrm a\Gamma}\nabla_\Gamma\hat{\mu}_{\mathrm a\Gamma}
-c_{\mathrm m\Gamma}\nabla_\Gamma\hat{\mu}_{\mathrm m\Gamma}-(\bbNabla_\Gamma\boldsymbol p_\Gamma)^{\mathrm T}\!\cdot\!\boldsymbol h_{\mathrm p\Gamma}
-\beta\left(\boldsymbol V_{\Gamma\parallel}-\boldsymbol v_{b\parallel}\right)=0,
\end{equation}
\begin{equation}\label{Eq:AppA-bal-ca2}
\dot c_{\mathrm a\Gamma}+c_{\mathrm a\Gamma}\,\nabla_\Gamma\!\cdot\!\boldsymbol V_\Gamma=0,
\end{equation}
\begin{equation}\label{Eq:AppA-bal-cm2}
\dot c_{\mathrm m\Gamma}+c_{\mathrm m\Gamma}\,\nabla_\Gamma\!\cdot\!\boldsymbol V_\Gamma
=M_{\mathrm m}\Delta_\Gamma \hat{\mu}_{\mathrm m\Gamma}+k_{\mathrm m\Gamma},
\end{equation}
\begin{equation}\label{Eq:AppA-constitutive}
\dot{\mathbb P}_{\Gamma\parallel}=
-\nu_{\mathrm p}\,\mathbb{D}_{\Gamma}(\boldsymbol V_\Gamma)\!\cdot\!\boldsymbol p_\Gamma
+\gamma_{\mathrm p}^{-1}\boldsymbol h_{\mathrm p\Gamma},
\end{equation}
\begin{equation}\label{Eq:AppA-kinematic2}
\partial_t\boldsymbol X=V_{\Gamma n}\hat{\boldsymbol n}_\Gamma+\boldsymbol V_{\Gamma\parallel},
\qquad
V_{\Gamma n}=\boldsymbol v_{\mathrm b}\!\cdot\!\hat{\boldsymbol n}_\Gamma,
\qquad
\boldsymbol v_{b\parallel}=\boldsymbol P_\Gamma\!\cdot\!\boldsymbol v_{\mathrm b},
\end{equation}
\end{subequations}
where $\boldsymbol\sigma_\Gamma^{\mathrm{vis}}\equiv 2\eta_{\mathrm s}\mathbb{D}_{\Gamma}(\boldsymbol V_\Gamma)$, 
$\boldsymbol{\Sigma}_\Gamma^{\mathrm{sym}}\equiv \boldsymbol\sigma_\Gamma^{\mathrm{vis}}+\boldsymbol\sigma_{\mathrm p\Gamma}^{\mathrm{irr}}+\boldsymbol\sigma^{\mathrm{act}}_\Gamma$, 
$\boldsymbol{\Sigma}_\Gamma\equiv \boldsymbol{\Sigma}_\Gamma^{\mathrm{sym}}+\boldsymbol\sigma_{\mathrm p\Gamma}^{\mathrm{asy}}$, 
$\boldsymbol\sigma_{\mathrm p\Gamma}^{\mathrm{irr}}
\equiv \frac{\nu_{\mathrm p}}{2}\left(
\boldsymbol p_\Gamma\boldsymbol h_{\mathrm p\Gamma}
+\boldsymbol h_{\mathrm p\Gamma}\boldsymbol p_\Gamma
\right)$, and $ \llbracket \boldsymbol\sigma_{\mathrm b} \rrbracket\equiv \boldsymbol\sigma_{\mathrm b}^+-\boldsymbol\sigma_{\mathrm b}^-$. 
Note that it is useful to keep Eqs.~\eqref{Eq:AppA-tanbulk} and \eqref{Eq:AppA-forcebalance} separate: the slip term is an \emph{internal momentum-exchange term} between the bulk fluid and the surface phase, and therefore appears with opposite signs in the bulk and surface tangential balances. Adding those two equations eliminates the internal drag and gives the compact total tangential traction balance
\begin{equation}\label{Eq:AppA-total-tan-balance}
\boldsymbol P_\Gamma\!\cdot\!\left( \llbracket \boldsymbol\sigma_{\mathrm b} \rrbracket\!\cdot\!\hat{\boldsymbol n}_\Gamma\right)
+\bbNabla_\Gamma\!\cdot\!\boldsymbol{\Sigma}_\Gamma
-c_{\mathrm a\Gamma}\nabla_\Gamma\hat{\mu}_{\mathrm a\Gamma}
-c_{\mathrm m\Gamma}\nabla_\Gamma\hat{\mu}_{\mathrm m\Gamma}
-(\bbNabla_\Gamma\boldsymbol p_\Gamma)^{\mathrm T}\!\cdot\!\boldsymbol h_{\mathrm p\Gamma}=0.
\end{equation}
Combining Eq.~\eqref{Eq:AppA-total-tan-balance} with Eq.~\eqref{Eq:AppA-normal} and using the ambient surface divergence identity, 
$\nabla_\Gamma\cdot\boldsymbol\Sigma_\Gamma
=\bbNabla_\Gamma\cdot\boldsymbol\Sigma_\Gamma
-(\boldsymbol\Sigma_\Gamma: \boldsymbol\kappa_\Gamma)\hat{\boldsymbol n}_\Gamma$, we obtain the compact vector traction-jump form
\begin{equation}\label{Eq:AppA-total-traction-balance}
\boldsymbol  \llbracket \boldsymbol\sigma_{\mathrm b} \rrbracket\!\cdot\!\hat{\boldsymbol n}_\Gamma
+\nabla_\Gamma\!\cdot\!\boldsymbol{\Sigma}_\Gamma
-\hat{\mu}_{\Gamma n}\hat{\boldsymbol n}_\Gamma 
-c_{\mathrm a\Gamma}\nabla_\Gamma\hat{\mu}_{\mathrm a\Gamma}
-c_{\mathrm m\Gamma}\nabla_\Gamma\hat{\mu}_{\mathrm m\Gamma}
-(\bbNabla_\Gamma\boldsymbol p_\Gamma)^{\mathrm T}\!\cdot\!\boldsymbol h_{\mathrm p\Gamma}=0.
\end{equation}
Equation~\eqref{Eq:AppA-total-traction-balance} contains no explicit $\beta$ because the slip force has canceled between the separate bulk and surface momentum balances. The slip effect is therefore \emph{not absent}; it is simply hidden once one writes only the total traction-jump closure. Equation~\eqref{Eq:AppA-total-traction-balance} is the sharp-interface counterpart of the coupled diffuse-domain force balances in Sec.~\ref{sec:ActPolar}, while Eqs.~\eqref{Eq:AppA-bal-ca2}--\eqref{Eq:AppA-constitutive} are the sharp-surface counterparts of the transport and polarization laws for $c_{\mathrm a\Gamma}$, $c_{\mathrm m\Gamma}$, and $\boldsymbol p_\Gamma$.
In the no-slip limit $\beta\to\infty$, Eqs.~\eqref{Eq:AppA-tanbulk} and \eqref{Eq:AppA-forcebalance} enforce $\boldsymbol V_{\Gamma\parallel}=\boldsymbol v_{b\parallel}$ and the formulation collapses to the usual single-velocity sharp-interface hydrodynamics.

Finally, two remarks are in order.
\begin{enumerate}[label=(\roman*)]
\item If one intentionally wants the \emph{reduced} nonhydrodynamic formulation in which bulk flow effects are neglected, it becomes necessary to introduce an additional conservative shape-relaxation closure. This leads to a distinct reduced theoretical framework that is not equivalent to, and should not be mixed with the model proposed in the present work.

\item If one wants the strictly inextensible membrane limit instead of the extensible $c_{\mathrm a\Gamma}$-based formulation, then $c_{\mathrm a\Gamma}$ should be eliminated from the state space and replaced by the local inextensibility constraint $\nabla_\Gamma\!\cdot\!\boldsymbol V_\Gamma = 0$, supplemented by a Lagrange-multiplier field representing the membrane tension. This results in a distinct reduced theoretical framework, which should not be superimposed on the present extensible formulation.
\end{enumerate} 

\section{Auxiliary derivations for deformable membrane models}\label{sec:App-ActPolar-deriv}

This appendix collects several algebraic derivations utilized in the variational formulations of the dynamics of deformable membranes, as developed in the diffuse-domain models in Secs.~\ref{sec:SurfPSD-DeformSurf}, \ref{sec:MPF-vesicle}, \ref{sec:ActPolar}, as well as in the sharp-interface model discussed in Appendix~\ref{sec:App-Membrane-OVP}. 
Specifically, we introduce the diffuse surface delta density in the form $\delta_\epsilon[\psi] = B_4(\psi,\nabla \psi)
= \frac{\epsilon}{2}\,|\nabla \psi|^2 + \epsilon^{-1} G(\psi)$, and decompose the membrane velocity as $\boldsymbol V = \boldsymbol V_\psi + \boldsymbol V_\parallel$, where $\boldsymbol V_\psi = V_{\psi n}\,\hat{\boldsymbol n}_\epsilon = -({\partial_t \psi}/{|\nabla \psi|^2})\,\nabla \psi$, $\hat{\boldsymbol n}_\epsilon = -{\nabla \psi}/{|\nabla \psi|}$, and $\boldsymbol P_\epsilon = \boldsymbol I - \hat{\boldsymbol n}_\epsilon \hat{\boldsymbol n}_\epsilon$.  
All integrations by parts in the domain $\Omega$ below assume either periodic boundary conditions, or that the diffuse layer remains away from the boundary $\partial\Omega$, or boundary conditions that remove the corresponding outer-boundary terms. 

\subsection{Scalar balance laws and shifted geometry chemical potential}
\label{sec:App-ActPolar-scalar-balance}

\emph{Scalar balance laws} --- Let $c_i$ denote a scalar surface balance-law field, with $i=\mathrm a,\mathrm m$.  The local embedded material balance is
\begin{equation}
\label{Eq:AppAct-bal-start}
\partial_t(\delta_\epsilon c_i)+\nabla\!\cdot(\delta_\epsilon c_i\boldsymbol V)
=
-\nabla\!\cdot(\delta_\epsilon\boldsymbol J_i)+\delta_\epsilon k_i ,
\end{equation}
where $\boldsymbol J_i$ is the nonconvective surface flux and $k_i$ is a source.  In Sec.~\ref{sec:ActPolar}, $\boldsymbol J_{\mathrm a}=\boldsymbol0$, $k_{\mathrm a}=0$, while $\boldsymbol J_{\mathrm m}$ and $k_{\mathrm m}$ are generally nonzero.

We first expand the normal-motion part.  Since
$\delta_\epsilon=B_4$, one has
$\partial_t\delta_\epsilon=\epsilon\nabla\psi\!\cdot\!\nabla(\partial_t\psi)+\epsilon^{-1}G'(\psi)\partial_t\psi$. Thus, $\partial_t(\delta_\epsilon c_i)=\delta_\epsilon\partial_t c_i +\epsilon c_i\nabla\psi\!\cdot\!\nabla(\partial_t\psi)
+\epsilon^{-1}c_iG'(\psi)\partial_t\psi$. For a standard one-dimensional diffuse-interface profile, we have $\delta_\epsilon=B_4 \simeq \epsilon |\nabla\psi|^2$ in the diffuse layer. Hence, 
$\delta_\epsilon c_i\boldsymbol V_\psi\simeq
-\epsilon c_i\nabla\psi\,\partial_t\psi$, and therefore $\nabla\!\cdot (\delta_\epsilon c_i \boldsymbol V_\psi) \simeq -\epsilon\nabla\!\cdot (c_i\nabla\psi) \partial_t\psi-\epsilon c_i\nabla\psi\!\cdot\!\nabla(\partial_t\psi)$.
The two terms proportional to $\nabla(\partial_t\psi)$ cancel, giving
\begin{equation}
\label{Eq:AppAct-A4-block}
\partial_t(\delta_\epsilon c_i)+\nabla\!\cdot(\delta_\epsilon c_i\boldsymbol V_\psi) \simeq \delta_\epsilon\partial_t c_i
+A_4(c_i,\psi)\partial_t\psi,
\qquad
A_4(c_i,\psi)\equiv
-\epsilon\nabla\!\cdot(c_i\nabla\psi)+\epsilon^{-1}c_iG'(\psi).
\end{equation}
Substituting $\boldsymbol V=\boldsymbol V_\psi+\boldsymbol V_\parallel$ into Eq.~\eqref{Eq:AppAct-bal-start} gives the finite-$\epsilon$ scalar balance used in Secs.~\ref{sec:MPF-vesicle} and \ref{sec:ActPolar},
\begin{equation}\label{Eq:AppAct-scalar-expanded}
\delta_\epsilon\partial_t c_i
+A_4(c_i,\psi)\partial_t\psi
+\nabla\!\cdot(\delta_\epsilon c_i\boldsymbol V_\parallel)
=
-\nabla\!\cdot(\delta_\epsilon\boldsymbol J_i)+\delta_\epsilon k_i .
\end{equation}
The reduced identity
$\delta_\epsilon\partial_t c_i+A_4(c_i,\psi)\partial_t\psi
=-\nabla\!\cdot(\delta_\epsilon\boldsymbol J_i)$
is recovered from Eq.~\eqref{Eq:AppAct-scalar-expanded} when $\boldsymbol V_\parallel=\boldsymbol0$ and $k_i=0$.
The assumptions behind Eq.~\eqref{Eq:AppAct-A4-block} are: (i) $\psi$ is locally an equilibrium signed-distance profile on the $O(\epsilon)$ normal scale, so that $B_4\simeq\epsilon|\nabla\psi|^2$; (ii) the flux $\boldsymbol J_i$ is interpreted as the nonconvective flux relative to the material surface velocity $\boldsymbol V$.

\emph{Shifted geometry chemical potential} ---
Equations~\eqref{Eq:AppAct-bal-start}--\eqref{Eq:AppAct-scalar-expanded}
also give the scalar-balance contribution to the shifted geometry
chemical potential used in Eq.~\eqref{Eq:MPF-vesicle-mupsi}.  Let $m_i\equiv \delta_\epsilon c_i$ ($i=\{\mathrm a,\mathrm m\}$) denote the conserved diffuse areal densities in Secs.~\ref{sec:MPF-vesicle}
and~\ref{sec:ActPolar}.  At fixed $\psi$, $\delta\mathcal F/\delta m_i=\hat\mu_i$, since $\delta\mathcal F/\delta c_i=\delta_\epsilon\hat\mu_i$. Hence, the
part of the free-energy rate associated with the scalar balance-law
fields can be written as
\begin{equation}\label{Eq:AppAct-scalar-dotFsc}
\dot{\mathcal F}_{\mathrm sc} =\int_\Omega\left[\hat{\mu}_\psi^{(m)}\partial_t\psi
+\sum_{i}\hat\mu_i\,\partial_t m_i\right]dV,
\qquad
\hat{\mu}_\psi^{(m)}\equiv\left.\frac{\delta\mathcal F}{\delta\psi}
\right|_{\{m_i\},\,{\mathrm other\ states}}.    
\end{equation} 
The fixed-$m_i$ chemical potential, $\mu_\psi^{(m)}$, is related to the fixed-$c_i$ chemical potential by 
\begin{equation}
\hat{\mu}_\psi^{(m)} \;=\; \left.\frac{\delta\mathcal F}{\delta\psi}\right|_{\{c_i\},\,{\mathrm other\ states}}
\;-\; \sum_i A_4(c_i,\psi)\,\hat\mu_i
\;+\; \epsilon\sum_i c_i\,\nabla\psi\cdot\nabla\hat\mu_i.
\end{equation}
On the other hand, using Eq.~\eqref{Eq:AppAct-bal-start}, and integrating by parts gives
\begin{equation}
\sum_i\int_\Omega\hat\mu_i\left[-\nabla\!\cdot(\delta_\epsilon c_i\boldsymbol V_\psi)\right]dV=-\int_\Omega\delta_\epsilon |\nabla\psi|^{-2}\left(
\sum_i c_i\,\nabla\psi\!\cdot\!\nabla\hat\mu_i\right)\partial_t\psi\,dV,
\end{equation}
where
$\boldsymbol V_\psi=-|\nabla\psi|^{-2}\partial_t\psi\,\nabla\psi$
has been used. Therefore, the effective geometry force multiplying
$\partial_t\psi$ is
\begin{equation}\label{Eq:AppAct-scalar-mubar1}
\bar\mu_\psi=\hat{\mu}_\psi^{(m)}-\delta_\epsilon |\nabla\psi|^{-2}
\sum_{i}c_i\,\nabla\psi\!\cdot\!\nabla\hat\mu_i.    
\end{equation}  
Furthermore, by invoking the same one-dimensional-profile approximation employed in Eq.~\eqref{Eq:AppAct-A4-block}, namely
$\delta_\epsilon\simeq\epsilon|\nabla\psi|^2$, Eq.~\eqref{Eq:AppAct-scalar-mubar1} can be approximated by the second expression in Eq.~\eqref{Eq:MPF-vesicle-mupsi} as
\begin{equation}
\bar\mu_\psi \;\simeq\; \left.\frac{\delta\mathcal F}{\delta\psi}\right|_{\{c_i\},\,{\mathrm other\ states}}
\;-\; \sum_{i} A_4(c_i,\psi)\,\hat\mu_i.
\end{equation}

\subsection{Polar contribution to the free-energy rate: diffuse and sharp forms}
\label{sec:App-ActPolar-Fdot-polar}

We first record the polar part of the free-energy rate in the diffuse-domain model and then give its intrinsic sharp-surface counterpart.  Only the terms involving the polar rate and the tangential velocity are displayed in this subsection; the scalar-balance and shape-variation terms are already absorbed into $\hat\mu_i$ and $\bar\mu_\psi$ in the main text.

\emph{Diffuse-Domain Model} --- The polar molecular field is defined by $\delta_\epsilon\boldsymbol h_{\mathrm p}^{\parallel} \equiv -\boldsymbol P_\epsilon\cdot\delta\mathcal F/\delta\boldsymbol p_\parallel$. Thus, the free-energy rate contribution from the explicit time dependence of $\boldsymbol p_\parallel$ is $\dot{\mathcal F}_{\mathrm p}=-\int_\Omega\delta_\epsilon \boldsymbol h_{\mathrm p}^{\parallel}\!\cdot\! \partial_t\boldsymbol p_\parallel\,dV$.
The projected co-rotational rate is $\dot{\mathbb P}_\parallel
=\boldsymbol P_\epsilon\cdot\left(\partial_t\boldsymbol p_\parallel
+\boldsymbol V\cdot\nabla\boldsymbol p_\parallel
-\boldsymbol\Omega_\parallel\cdot\boldsymbol p_\parallel
\right)$ with  $\boldsymbol\Omega_\parallel=\frac12\left( \bbNabla_{\mathrm s}\boldsymbol V_\parallel- \bbNabla_{\mathrm s}\boldsymbol V_\parallel^{\mathrm T}\right)$. 
Since $\boldsymbol h_{\mathrm p}^{\parallel}$ is tangential, contraction with $\boldsymbol h_{\mathrm p}^{\parallel}$ gives
$-\boldsymbol h_{\mathrm p}^{\parallel}\cdot\partial_t\boldsymbol p_\parallel=-\boldsymbol h_{\mathrm p}^{\parallel}\cdot\dot{\mathbb P}_\parallel +\boldsymbol V_\parallel \cdot (\nabla\boldsymbol p_\parallel)^{\mathrm T}\cdot\boldsymbol h_{\mathrm p}^{\parallel}
-\boldsymbol h_{\mathrm p}^{\parallel}\cdot (\boldsymbol\Omega_\parallel \cdot\boldsymbol p_\parallel)$,
where the explicit normal convective contribution vanishes under the closest-point extension.  
After integration by parts, the polar contribution in free-energy rate becomes
\begin{equation} \label{Eq:AppAct-Fdot-polar-diff}
\dot{\mathcal F}_{\mathrm p}= \int_\Omega
\left\{-\delta_\epsilon \boldsymbol h_{\mathrm p}^{\parallel}\cdot\dot{\mathbb P}_\parallel + \boldsymbol V_\parallel\cdot \left[ -\bbNabla_{\mathrm s}\cdot
(\delta_\epsilon\boldsymbol\sigma_{\mathrm p}^{\mathrm{asy}})
+ (\bbNabla_{\mathrm{s}}\boldsymbol p_\parallel)^{\mathrm T}\cdot
(\delta_\epsilon\boldsymbol h_{\mathrm p}^{\parallel})\right]\right\}dV,
\end{equation}
where $\boldsymbol\sigma_{\mathrm p}^{\mathrm{asy}} \equiv
\frac12 \left(\boldsymbol p_\parallel\boldsymbol h_{\mathrm p}^{\parallel} - \boldsymbol h_{\mathrm p}^{\parallel}\boldsymbol p_\parallel \right)$, for which $-\boldsymbol h_{\mathrm p}^{\parallel}\cdot (\boldsymbol\Omega_\parallel\cdot\boldsymbol p_\parallel) = \boldsymbol\sigma_{\mathrm p}^{\mathrm{asy}}:
\bbNabla_{\mathrm s}\boldsymbol V_\parallel$.

\emph{Sharp-Interface Model} --- The corresponding sharp-interface calculation is identical after replacing $\delta_\epsilon dV\to dA$,  $\hat{\boldsymbol n}_\epsilon\to \hat{\boldsymbol n}_\Gamma$, $\boldsymbol P_\epsilon\to\boldsymbol P_\Gamma$, $\bbNabla_{\mathrm s}\to\bbNabla_\Gamma$, 
$\boldsymbol p_\parallel\to\boldsymbol p_\Gamma$, and $\boldsymbol V\to\boldsymbol V_\Gamma$.  On $\Gamma(t)$, the projected co-rotational rate is
$\dot{\mathbb P}_{\Gamma\parallel} = \boldsymbol P_\Gamma\cdot (\dot{\boldsymbol p}_\Gamma-\boldsymbol\Omega_{\Gamma\parallel}\cdot\boldsymbol p_\Gamma)$, where $\dot{\boldsymbol p}_\Gamma=\partial_t\boldsymbol p_\Gamma+\boldsymbol V_\Gamma\cdot\nabla_\Gamma \boldsymbol p_\Gamma$ and $\boldsymbol\Omega_{\Gamma\parallel}
=\frac12(\bbNabla_\Gamma \boldsymbol V_{\Gamma\parallel} -\bbNabla_\Gamma\boldsymbol V_{\Gamma\parallel}^{\mathrm T})$. Using the definition $\boldsymbol\sigma_{\mathrm p\Gamma}^{\mathrm{asy}}
\equiv \frac12 \left(\boldsymbol p_\Gamma\boldsymbol h_{\mathrm p\Gamma} - \boldsymbol h_{\mathrm p\Gamma}\boldsymbol p_\Gamma
\right)$, we obtain
\begin{equation}\label{Eq:AppAct-Fdot-polar-sharp}
\dot{\mathcal F}_{\Gamma,\mathrm p}
= \int_{\Gamma(t)} \left\{-\boldsymbol h_{\mathrm p\Gamma}\cdot\dot{\mathbb P}_{\Gamma\parallel}+ \boldsymbol V_{\Gamma\parallel}\cdot\left[-\bbNabla_\Gamma\cdot\boldsymbol\sigma_{\mathrm p\Gamma}^{\mathrm{asy}} + (\bbNabla_\Gamma\boldsymbol p_\Gamma)^{\mathrm T}\cdot \boldsymbol h_{\mathrm p\Gamma}\right]\right\}dA.
\end{equation}

\subsection{Variations of surface viscous and polar dissipation functionals and active work power} \label{sec:App-ActPolar-metric-var}

We next evaluate the variations of the surface viscous dissipation functional and the surface polar dissipation functional, as well as of the active power input.

\emph{Diffuse-Domain Model} --- The contributions from surface viscosity and polar flow alignment to the dissipation functional, along with the active work power, are combined in the following part of the Rayleighian:
\begin{equation}\label{Eq:AppAct-var-Rvpa}
\mathcal R_{\mathrm{vpa}}= \int_\Omega\delta_\epsilon\left[\eta_{\mathrm s} \mathbb D_{\mathrm s} (\boldsymbol V):\mathbb D_{\mathrm s}(\boldsymbol V) + \frac{\gamma_{\mathrm p}}{2} \left|
\dot{\mathbb P}_\parallel +\nu_{\mathrm p}\mathbb D_{\mathrm s}(\boldsymbol V)\!\cdot\!\boldsymbol p_\parallel
\right|^2+\boldsymbol\sigma_{\mathrm s}^{\mathrm{act}}:\mathbb D_{\mathrm s} (\boldsymbol V)\right]dV.
\end{equation}
Variation of the full Rayleighian $\mathcal{R}$ in the main text with respect to $\dot{\mathbb P}_\parallel$ gives $\gamma_{\mathrm p}\boldsymbol R_{\mathrm p}=\boldsymbol h_{\mathrm p}^{\parallel}$ with $\boldsymbol R_{\mathrm p}\equiv \dot{\mathbb P}_\parallel +\nu_{\mathrm p}\mathbb D_{\mathrm s}(\boldsymbol V)\!\cdot\!\boldsymbol p_\parallel$. Using it, the term $\gamma_{\mathrm p}\boldsymbol R_{\mathrm p}\cdot \left[\nu_{\mathrm p}\delta \mathbb D_{\mathrm s}(\boldsymbol V)\cdot\boldsymbol p_\parallel\right]$ in the first variation of $\mathcal R_{\mathrm{vpa}}$ gives $\frac12 \gamma_{\mathrm p}\nu_{\mathrm p}\left(
\boldsymbol R_{\mathrm p}\boldsymbol p_\parallel+\boldsymbol p_\parallel \boldsymbol R_{\mathrm p} \right):\delta\mathbb D_{\mathrm s}=\boldsymbol\sigma_{\mathrm p}^{\mathrm{irr}}:\delta\mathbb D_{\mathrm s}$, where the symmetric irreversible polar stress $\boldsymbol\sigma_{\mathrm p}^{\mathrm{irr}}$ is defined by
\begin{equation}\label{Eq:AppAct-sigmairr-full}
\boldsymbol\sigma_{\mathrm p}^{\mathrm{irr}}
= \frac{\nu_{\mathrm p}}{2}\left(\boldsymbol p_\parallel\boldsymbol h_{\mathrm p}^{\parallel}
+\boldsymbol h_{\mathrm p}^{\parallel} \boldsymbol p_\parallel\right).
\end{equation} 
Therefore, the first variation of $\mathcal R_{\mathrm{vpa}}$ is given by
\begin{equation}\label{Eq:AppAct-var-dRvpa1}
\delta\mathcal R_{\mathrm{vpa}}=\int_\Omega
\delta_\epsilon \left[
\boldsymbol\Sigma_{\mathrm s}^{\mathrm{sym}}:
\delta\mathbb D_{\mathrm s}(\boldsymbol V)+ \gamma_{\mathrm p} \left(\dot{\mathbb P}_\parallel +\nu_{\mathrm p}\mathbb D_{\mathrm s}(\boldsymbol V)\!\cdot\!\boldsymbol p_\parallel\right)\cdot\delta \dot{\mathbb P}_\parallel\right] \,dV,
\end{equation}
with the stress tensors given by
\begin{equation}\label{Eq:AppAct-total-metric-stress}
\boldsymbol\Sigma_{\mathrm s}^{\mathrm{sym}}\equiv
 \boldsymbol\sigma_{\mathrm s}^{\mathrm{vis}}
+\boldsymbol\sigma_{\mathrm p}^{\mathrm{irr}}
+\boldsymbol\sigma_{\mathrm s}^{\mathrm{act}},
\qquad
\boldsymbol\sigma_{\mathrm s}^{\mathrm{vis}}
=2\eta_s\mathbb D_{\mathrm s}(\boldsymbol V),
\qquad
\boldsymbol\sigma_{\mathrm s}^{\mathrm{act}}=-\zeta\boldsymbol q_\parallel.
\end{equation} 
Since $\mathbb D_{\mathrm s}(\boldsymbol V)
=\mathbb D_{\mathrm s}(\boldsymbol V_\parallel)+V_{\psi n}\boldsymbol\kappa_\epsilon$ and $V_{\psi n}=\boldsymbol v\!\cdot\!\hat{\boldsymbol n}_\epsilon
-\nabla\!\cdot\!\boldsymbol J_\psi/|\nabla\psi|$, we have
\begin{equation}\label{Eq:AppAct-total-metric-var}
\delta\mathbb D_{\mathrm s}(\boldsymbol V)=
\mathbb D_{\mathrm s}(\delta\boldsymbol V_\parallel)
+(\delta V_{\psi n})\boldsymbol\kappa_\epsilon,
\quad
\delta_{\boldsymbol v}V_{\psi n}=\delta\boldsymbol v\!\cdot\!\hat{\boldsymbol n}_\epsilon,
\quad
\delta_{\boldsymbol J_\psi}V_{\psi n}
=-\frac{\nabla\!\cdot\delta\boldsymbol J_\psi}{|\nabla\psi|}.
\end{equation}
Moreover, using the relation $\boldsymbol\Sigma_{\mathrm s}^{\mathrm{sym}}:\mathbb D_{\mathrm s}(\delta\boldsymbol V_\parallel)
=\boldsymbol\Sigma_{\mathrm s}^{\mathrm{sym}}:\bbNabla_{\mathrm s}\delta\boldsymbol V_\parallel$ from the symmetry and tangentiality of $\boldsymbol\Sigma_{\mathrm s}^{\mathrm{sym}}$, and integrating by parts, the first term of Eq.~\eqref{Eq:AppAct-var-dRvpa1} becomes
\begin{equation}\label{Eq:AppAct-var-dRvpa2}
\delta \mathcal R_{\mathrm{vpa}}=
\int_\Omega\left\{-\left[\bbNabla_{\mathrm s}\!\cdot(\delta_\epsilon\boldsymbol\Sigma_{\mathrm s}^{\mathrm{sym}})
\right]\!\cdot\delta\boldsymbol V_\parallel+ 
\delta_\epsilon
(\boldsymbol\Sigma_{\mathrm s}^{\mathrm{sym}}:\boldsymbol\kappa_\epsilon)
\hat{\boldsymbol n}_\epsilon\!\cdot\!\delta\boldsymbol v +
\nabla\!\left[\frac{\delta_\epsilon
(\boldsymbol\Sigma_{\mathrm s}^{\mathrm{sym}}:\boldsymbol\kappa_\epsilon)}{|\nabla\psi|}\right]\!\cdot\delta\boldsymbol J_\psi \right\}\,dV.
\end{equation}  
Note that the last term is present only if the full metric rate uses $V_{\psi n}=\boldsymbol v\cdot\hat{\boldsymbol n}_\epsilon
-\nabla\cdot\boldsymbol J_\psi/|\nabla\psi|$. 
Then, the variation of the full Rayleighian $\mathcal R$ with respect to $\boldsymbol J_\psi$ gives $\boldsymbol J_\psi=-M_\psi\nabla\left[\bar\mu_\psi+{\delta_\epsilon (\boldsymbol\Sigma_{\mathrm s}^{\mathrm{sym}}:\boldsymbol\kappa_\epsilon)}/{|\nabla\psi|}\right]$. Consequently,
\begin{equation}
\partial_t\psi+\nabla\!\cdot(\psi\boldsymbol v)
=
\nabla\!\cdot\left\{
M_\psi\nabla
\left[
\bar\mu_\psi+
\frac{\delta_\epsilon
(\boldsymbol\Sigma_{\mathrm s}^{\mathrm{sym}}:\boldsymbol\kappa_\epsilon)}
{|\nabla\psi|}
\right]\right\}.
\end{equation} 
 
\emph{Sharp-Interface Model} --- The sharp-interface counterpart is obtained by replacing
$\mathbb D_{\mathrm s}(\boldsymbol V)$ with
$\mathbb D_\Gamma(\boldsymbol V_\Gamma)
=\mathbb D_\Gamma(\boldsymbol V_{\Gamma\parallel})
+V_{\Gamma n}\boldsymbol\kappa_\Gamma$ and $\mathcal R_{\mathrm{vpa}}$ in Eq.~\eqref{Eq:AppAct-var-Rvpa} by
\begin{equation}\label{Eq:AppAct-var-sharp-Rvpa}
\mathcal R_{\Gamma,\mathrm{vpa}}= \int_{\Gamma(t)}  \left[\eta_{\mathrm s} \mathbb D_{\Gamma} (\boldsymbol V_\Gamma):\mathbb D_{\Gamma}(\boldsymbol V_\Gamma) 
+ \frac{\gamma_{\mathrm p}}{2} \left|
\dot{\mathbb P}_{\Gamma\parallel} +\nu_{\mathrm p}\mathbb D_{\Gamma}(\boldsymbol V_\Gamma)\!\cdot\!\boldsymbol p_\Gamma
\right|^2+\boldsymbol\sigma_{\Gamma}^{\mathrm{act}}:\mathbb D_{\Gamma} (\boldsymbol V_\Gamma)\right]dA.
\end{equation}
Variation of the full Rayleighian $\mathcal{R}_\Gamma$ in the appendix section~\ref{sec:App-Membrane-OVP} with respect to $\dot{\mathbb P}_{\Gamma\parallel}$ gives 
$\gamma_{\mathrm p}\boldsymbol R_{\mathrm p\Gamma}
=\boldsymbol h_{\mathrm p\Gamma}$ with
$\boldsymbol R_{\mathrm p\Gamma}
=\dot{\mathbb P}_{\Gamma\parallel}
+\nu_{\mathrm p}\mathbb D_\Gamma(\boldsymbol V_\Gamma)\cdot\boldsymbol p_\Gamma$,
and the symmetric irreversible polar stress is given by
\begin{equation}\label{Eq:AppAct-sharp-sigmairr}
\boldsymbol\sigma_{\mathrm p\Gamma}^{\mathrm{irr}}=\frac{\nu_{\mathrm p}}{2}
\left(\boldsymbol p_\Gamma\boldsymbol h_{\mathrm p\Gamma}+\boldsymbol h_{\mathrm p\Gamma}\boldsymbol p_\Gamma\right).
\end{equation}   
Similarly to Eq.~\eqref{Eq:AppAct-total-metric-var} in the above DDM models, we have
\begin{equation}\label{Eq:AppAct-sharp-metric-var}
\delta\mathbb D_\Gamma(\boldsymbol V_\Gamma)
=\mathbb D_\Gamma(\delta\boldsymbol V_{\Gamma\parallel})
+(\delta V_{\Gamma n})\boldsymbol\kappa_\Gamma,
\quad
\delta V_{\Gamma n}=\delta\boldsymbol v_{\mathrm b}\!\cdot\!\hat{\boldsymbol n}_\Gamma.
\end{equation}
Note that the sharp normal kinematic condition is
$V_{\Gamma n}=\boldsymbol v_{\mathrm b}\cdot\hat{\boldsymbol n}_\Gamma$. There is no independent sharp-interface analogue of the $\boldsymbol J_\psi$ variation; the sharp hydrodynamic model treats the interface as material in the normal direction.  
Then, the sharp-interface counterpart of the variation in Eq.~\eqref{Eq:AppAct-var-dRvpa2} is
\begin{equation}\label{Eq:AppAct-sharp-metric-vars}
\delta\mathcal R_{\Gamma,\mathrm{vpa}}=
\int_{\Gamma(t)} \left[-(\bbNabla_\Gamma\cdot \boldsymbol\Sigma_\Gamma)\cdot
\delta\boldsymbol V_{\Gamma\parallel} + 
(\boldsymbol\Sigma_\Gamma:\boldsymbol\kappa_\Gamma) \delta V_{\Gamma n}\right]\,dA,
\end{equation}
where the stress tensors are given by
\begin{equation}
\boldsymbol\Sigma_\Gamma\equiv \boldsymbol\sigma_{\Gamma}^{\mathrm{vis}}+
\boldsymbol\sigma_{\mathrm p\Gamma}^{\mathrm{irr}}+ \boldsymbol\sigma_{\Gamma}^{\mathrm{act}},
\qquad
\boldsymbol\sigma_{\Gamma}^{\mathrm{vis}}=2\eta_s\mathbb D_\Gamma(\boldsymbol V_\Gamma),
\qquad
\boldsymbol\sigma_{\Gamma}^{\mathrm{act}}=
-\zeta\boldsymbol q_\Gamma.
\end{equation}  
 
\textcolor{black}{\section{Technical details supporting the condensed diffuse-domain construction}\label{app:secII-technical}}

\textcolor{black}{This appendix collects the technical definitions and auxiliary derivations in Sec.~\ref{sec:DDM}. }

\textcolor{black}{\subsection{Surface tangentiality: admissible state spaces and projected gradients}\label{app:tangential-state-spaces}}

\textcolor{black}{A surface field $\boldsymbol f_\Gamma(\boldsymbol X,t)$ of tensor rank $r$, defined for $\boldsymbol X \in \Gamma(t)$, is represented in the diffuse interfacial layer by a smooth ambient field $\boldsymbol f(\boldsymbol x,t)$, with $\boldsymbol x\in\Omega$, such that $\boldsymbol f|_{\psi=1/2} = \boldsymbol f_\Gamma$. This extension is not unique. For tensorial surface fields, one must specify not only the extension away from $\Gamma(t)$, but also the admissible tangential state space in which the diffuse variational problem is posed. }

\textcolor{black}{For a rank-$r$ tensor field $\boldsymbol f$ with Cartesian components $f_{i_1\cdots i_r}$, the $m$-th tensor slot is tangential if $\hat n_{\epsilon,i_{\mathrm m}}
f_{i_1\cdots i_{m}\cdots i_r}=0$ for $m=1,\ldots,r$. Thus, $\boldsymbol f$ is called right-tangential if only the last slot is tangential, left-tangential if only the first slot is tangential, and fully tangential if all slots are tangential. We denote the slotwise projection onto the fully tangential rank-$r$ state space by
\begin{equation}\label{Eq:DDM-rank-r-projection}
\left(\boldsymbol P_\epsilon^{(r)}\cdot \boldsymbol f\right)_{i_1\cdots i_r}
\equiv P_{\epsilon,i_1k_1}\cdots P_{\epsilon,i_rk_r}
f_{k_1\cdots k_r}.
\end{equation}
A strongly imposed fully tangential field satisfies $\boldsymbol f=\boldsymbol P_\epsilon^{(r)}\cdot\boldsymbol f$. Additional algebraic constraints, such as symmetry and surface tracelessness for a nematic $\boldsymbol Q$ tensor, are independent of tangentiality and must be imposed separately. Tangentiality may be imposed strongly by using the projected variables as the state variables, or weakly by anchoring or penalty terms. Specifically, for a second-rank tensor, right-, left-, and fully tangentiality reduce, respectively, to $\boldsymbol f\cdot\hat{\boldsymbol n}_\epsilon=\boldsymbol 0$ 
(equivalently, $\boldsymbol f=\boldsymbol f\cdot\boldsymbol P_\epsilon$), $\hat{\boldsymbol n}_\epsilon\cdot\boldsymbol f=\boldsymbol 0$ (equivalently, $\boldsymbol f=\boldsymbol P_\epsilon\cdot\boldsymbol f$), and $\hat{\boldsymbol n}_\epsilon\cdot\boldsymbol f =\boldsymbol f\cdot\hat{\boldsymbol n}_\epsilon =\boldsymbol 0$ (equivalently, $\boldsymbol f=\boldsymbol P_\epsilon\cdot\boldsymbol f\cdot\boldsymbol P_\epsilon$). By contrast, the weaker condition $\hat{\boldsymbol n}_\epsilon\cdot\boldsymbol f\cdot\hat{\boldsymbol n}_\epsilon=0$ removes only the normal--normal component and does not define a tangential tensor.}

\textcolor{black}{Let $\boldsymbol f_\parallel\equiv\boldsymbol P_\epsilon^{(r)}\!\cdot\boldsymbol f$ denote the fully tangential projection of a rank-$r$ tensor field.  If tangentiality is imposed strongly, then $\boldsymbol f_\parallel=\boldsymbol f$.  If tangentiality is imposed only weakly, the projected gradient acts on $\boldsymbol f_\parallel$, not on the unconstrained ambient field.  We distinguish the projection of the derivative direction from the projection of the tensor state slots. Two projected gradients are then relevant:
\begin{equation}\label{Eq:DDM-gradient}
\nabla_{\mathrm s} \boldsymbol{f}_\parallel \equiv (\boldsymbol{P}_\epsilon\cdot \nabla) \boldsymbol{f}_\parallel=(\nabla \boldsymbol{f}_\parallel)\cdot \boldsymbol{P}_\epsilon, \qquad
\bbNabla_{\mathrm s} \boldsymbol{f}_\parallel \equiv \boldsymbol P_\epsilon^{(r)}\cdot(\nabla \boldsymbol{f}_\parallel)\cdot \boldsymbol{P}_\epsilon = \boldsymbol P_\epsilon^{(r)} \cdot(\nabla_{\mathrm s} \boldsymbol{f}_\parallel).
\end{equation} 
Equivalently, $(\nabla_{\mathrm s}\boldsymbol f_\parallel)_{i_1\cdots i_r j}=P_{\epsilon,jk}\partial_k(f_\parallel)_{i_1\cdots i_r}$, whereas $(\bbNabla_{\mathrm s}\boldsymbol f_\parallel)_{i_1\cdots i_r j}=P_{\epsilon,i_1k_1}\cdots P_{\epsilon,i_rk_r}P_{\epsilon,jk}\partial_k(f_\parallel)_{k_1\cdots k_r}$.  Thus, $\nabla_{\mathrm s}$ projects only the derivative slot, whereas $\bbNabla_{\mathrm s}$ also projects the state slots.  For a tensor field $\boldsymbol F$ carrying a designated last flux or divergence index, the corresponding projected divergences are
\begin{equation}
\nabla_{\mathrm s}\!\cdot\boldsymbol F\equiv\boldsymbol P_\epsilon:\nabla\boldsymbol F,
\qquad
\bbNabla_{\mathrm s}\!\cdot\boldsymbol F\equiv\boldsymbol P_\epsilon^{(r-1)}\!\cdot(\nabla_{\mathrm s}\!\cdot\boldsymbol F),
\end{equation}
equivalently, $(\nabla_{\mathrm s}\!\cdot\boldsymbol F)_{i_1\cdots i_{r-1}}\equiv P_{\epsilon,jk}\,\partial_k F_{i_1\cdots i_{r-1}j}$, and $(\bbNabla_{\mathrm s}\!\cdot\boldsymbol F)_{i_1\cdots i_{r-1}}\equiv P_{\epsilon,i_1k_1}\cdots P_{\epsilon,i_{r-1}k_{r-1}}(\nabla_{\mathrm s}\!\cdot\boldsymbol F)_{k_1\cdots k_{r-1}}$, respectively. Note that the double-dot operator, hereafter, denotes (i) the contraction over the last two indices when acting on a pair of tensors of different ranks, and (ii) the full contraction over all indices when acting on a pair of tensors of equal rank, respectively.
The first definition is meaningful as a surface divergence when the last, flux-carrying slot is right-tangential. The second definition is required when the remaining output slots are also constrained to be tangential.}

\textcolor{black}{The same state-space convention applies to variational conjugates and dissipative rates. If a tangential state field is imposed strongly, then admissible variations
are tangential, and only the correspondingly projected chemical potential, molecular field, or generalized force is thermodynamically conjugate to those variations. Normal components are then constraint forces or finite-$\epsilon$
penalty forces. They should not be included in an intrinsic dissipative rate unless the model intentionally treats the normal component as a physical ambient degree of freedom. For scalar surface fields, by contrast, there is no tensor-slot tangentiality. The choice between a projected surface gradient $\nabla_{\mathrm s}c=\boldsymbol P_\epsilon \cdot\nabla c$ and the full ambient gradient $\nabla c$ is therefore a modeling and numerical choice rather than a state-space constraint. In particular, the isotropic R\"atz--Voigt scalar embedding used below employs the full gradient in the diffuse layer; under a closest-point extension it has the same sharp-interface surface-gradient limit while providing normal regularization at finite $\epsilon$.
In addition, when needed in the analysis, we use the standard closest-point, or constant-along-normal, extension $\hat{\boldsymbol n}_\epsilon\cdot\nabla\boldsymbol f=\boldsymbol 0$, understood componentwise in a fixed ambient Cartesian basis. This is an extension convention, not an additional surface evolution law. For tensorial surface fields, it must be combined with the appropriate slotwise tangentiality constraint.} 

\textcolor{black}{\subsection{Internal nonconserved variables and projected objective rates}\label{app:internal-nonconserved-rates}}

\textcolor{black}{If a class-(ii) variable--an internal nonconserved order parameter $\boldsymbol b_\Gamma$--is carried by a surface balance-law density $\rho_\Gamma$, so that $\boldsymbol m_\Gamma\equiv\rho_\Gamma\boldsymbol b_\Gamma$ is itself a balance-law field, then the diffuse-domain embedding can be obtained by writing separate conservative balances for $\rho$ and $\boldsymbol m=\rho\boldsymbol b$ in the form of Eq.~\eqref{Eq:DDM-bal-local}.  The dynamic equation for $\boldsymbol b$ is
\begin{equation}\label{Eq:DDM-intensive-from-density-diffuse}
\delta_\epsilon\rho\dot{\boldsymbol b}=-\nabla\!\cdot(\delta_\epsilon\boldsymbol J_m)+\delta_\epsilon k_m
-\boldsymbol b[-\nabla\!\cdot(\delta_\epsilon\boldsymbol J_\rho)+\delta_\epsilon k_\rho],
\end{equation}
with $\dot{\boldsymbol b}=\partial_t\boldsymbol b+\boldsymbol V\!\cdot\nabla\boldsymbol b$.  For a purely advected carrier, $\boldsymbol J_\rho=0$ and $k_\rho=0$, this reduces to
\begin{equation}\label{Eq:DDM-intensive-from-density-diffuse-simple}
\rho\dot{\boldsymbol b}=-\delta_\epsilon^{-1}\nabla\!\cdot(\delta_\epsilon\boldsymbol J_m)+k_m .
\end{equation}
This derivation explains why the natural transport law and dissipative rate of an internal nonconserved order parameter are written in terms of a material rate rather than in conservative divergence form. Moreover, if the surface moves and the flow field is present, one must subsequently replace the material rate $\dot{\boldsymbol{b}}$ by the appropriate objective/co-rotational rate $\dot{\mathbb B}$. Note that the prefactor $\delta_\epsilon $ should not be built into the definition of the rate itself; instead, $\delta_\epsilon $ belongs in the measure weight of the dissipation functional. This carrier-balance interpretation is justified only when the carrier density $\rho_\Gamma$ is physically identifiable and obeys its own balance law; in general, it is neither automatic nor unique.}

\textcolor{black}{The dissipative rate $\dot{\mathbb B}$ associated with an internal nonconserved order parameter $\boldsymbol b$ depends on its tensorial character and on the surface motion. For a scalar internal variable,
\begin{equation}\label{Eq:DDM-Bdot-scalar}
\dot{\mathbb B}=\dot b\equiv\partial_t b+\boldsymbol V\!\cdot\nabla b .
\end{equation}
For a tangential vector $\boldsymbol b_\parallel$, we use the projected co-rotational rate
\begin{equation}\label{Eq:DDM-Bdot-vector}
\dot{\mathbb B}=\boldsymbol P_\epsilon\!\cdot\left(\dot{\boldsymbol b}_\parallel-\boldsymbol\Omega_\parallel\!\cdot\boldsymbol b_\parallel\right),
\end{equation}
where $\dot{\boldsymbol b}_\parallel=\partial_t\boldsymbol b_\parallel+\boldsymbol V\!\cdot\nabla\boldsymbol b_\parallel$ and $\boldsymbol\Omega_\parallel = \frac12\boldsymbol P_\epsilon\cdot \left[
\nabla_{\mathrm s}\boldsymbol V_\parallel -
(\nabla_{\mathrm s}\boldsymbol V_\parallel)^{\mathrm T}\right]\cdot\boldsymbol P_\epsilon=\frac12 \left[
\bbNabla_{\mathrm s}\boldsymbol V_\parallel -
(\bbNabla_{\mathrm s}\boldsymbol V_\parallel)^{\mathrm T}\right]$.  
For a fully tangential second-rank tensor $\boldsymbol b_\parallel$, the same co-rotational correction acts slotwise:
\begin{equation}\label{Eq:DDM-Bdot-tensor2}
\dot{\mathbb B}=\boldsymbol P_\epsilon\!\cdot\left(\dot{\boldsymbol b}_\parallel-\boldsymbol\Omega_\parallel\!\cdot\boldsymbol b_\parallel+\boldsymbol b_\parallel\!\cdot\boldsymbol\Omega_\parallel\right)\!\cdot\boldsymbol P_\epsilon .
\end{equation}
For higher-rank tensors, the infinitesimal rotation acts on each tensor index, with the appropriate sign convention for that slot, and the result is projected back onto the same admissible tangential state space.  Other objective or flow-alignment rates may be used as constitutive alternatives, but the dissipative rate and its conjugate molecular field must be projected onto the same admissible state space. For the fixed-surface case considered in the section~\ref{sec:DDM}, $\boldsymbol V=\boldsymbol 0$ and $\partial_t\boldsymbol P_\epsilon=\boldsymbol 0$, so the dissipative rate reduces to $\dot{\mathbb B}=\partial_t\boldsymbol b_\parallel$.
The more general moving-surface objective rates $\dot{\mathbb B}$ are introduced only in Secs.~\ref{sec:SurfPSD-DeformSurf}, \ref{sec:MPF-vesicle}, and \ref{sec:ActPolar}, where they are actually needed.}

\textcolor{black}{\subsection{Finite-$\epsilon$ regularization and implementation notes}\label{app:finiteepsilon-reg}}

\textcolor{black}{The conservative variable in a diffuse-domain balance law is $\delta_\epsilon a_\parallel$, not the bulk extension $a_\parallel$ alone.  In numerical implementations where $a_\parallel$ must be evaluated pointwise away from the surface~\cite{RatzVoigt2006}, it is common to introduce a small positive floor $\delta_{\mathrm{reg}}$ with the same physical dimension as $\delta_\epsilon$, replacing $\delta_\epsilon[\psi]\to\delta_\epsilon[\psi]+\delta_{\mathrm{reg}}$.  This floor should be small relative to the characteristic magnitude $O(\epsilon^{-1})$ of $\delta_\epsilon$ inside the diffuse layer, so that it regularizes division by $\delta_\epsilon $ without changing the intended sharp-interface limit.}

\textcolor{black}{The fixed-surface derivation in Sec.~\ref{sec:DDM} assumes that $\psi$, $\delta_\epsilon$, and $\boldsymbol P_\epsilon$ are time independent.  If the surface evolves by velocity transport or through a dissipative evolution of $\psi$, then $\delta_\epsilon$ and $\boldsymbol P_\epsilon$ also evolve; the corresponding $\psi$-dependent contributions must be included in the free energy, dissipation functional, and kinematic identities.  In that moving-surface setting, the dissipative rate $\dot{\mathbb B}$ of a tangential internal order parameter must be replaced by the
appropriate projected material or objective rate. This is also why the deformable-surface models in Secs.~\ref{sec:SurfPSD-DeformSurf}, \ref{sec:MPF-vesicle}, and \ref{sec:ActPolar} explicitly retain the shifted geometry chemical potentials and material scalar balances.}

\textcolor{black}{Scalar balance-law fields treated with the isotropic R\"atz--Voigt embedding use the full-gradient regularization generated by the weighted bulk operator $\nabla\!\cdot(\delta_\epsilon\nabla\cdot)$~\cite{RatzVoigt2006,LiLowengrubRatzVoigt2009}.  Therefore, no additional weak normal diffusion is usually needed for the scalar phase-separation models considered in Sec.~\ref{sec:SurfPSD-RigidSurf}.  When full-gradient regularization is absent, weak normal control can be useful as a numerical device~\cite{NestlerVoigt2024,KloppeAland2024}.  For a purely advected or source-driven scalar balance-law field $a$, one possible normal-extension regularization is
\begin{equation}\label{Eq:DDM-weak-normal-scalar}
\partial_t(\delta_\epsilon a)=\cdots+
\nabla\!\cdot\left[\delta_\epsilon M_{\mathrm{a}}^{(n)}(\hat{\boldsymbol n}_\epsilon\otimes\hat{\boldsymbol n}_\epsilon)\nabla a\right],
\end{equation}
where the dots denote the physical balance-law terms and $M_{\mathrm{a}}^{(n)}$ is chosen small compared with the physical tangential mobility.  For a tangential vector or tensor field $\boldsymbol f$, one may instead add
\begin{equation}\label{Eq:DDM-weak-normal-vector}
\Phi_f^{(n)}=\int_\Omega\delta_\epsilon\frac{\eta_n}{2}|(\hat{\boldsymbol n}_\epsilon\!\cdot\nabla)\boldsymbol f|^2\,dV,
\end{equation}
whose variation contributes $-\nabla\!\cdot[\delta_\epsilon\eta_n(\hat{\boldsymbol n}_\epsilon\otimes\hat{\boldsymbol n}_\epsilon)\nabla\boldsymbol f]$ componentwise to the corresponding Euler--Lagrange equation or force balance.  These terms are finite-$\epsilon$ numerical regularizations that weakly promote a constant-along-normal extension; they are not part of the intended sharp-interface physics.}


\bibliographystyle{iopart-num.bst}
\bibliography{references}

\end{document}